\documentclass[%
 reprint,
showpacs,preprintnumbers,
 amsmath,amssymb,
 aps,
pra,
floatfix,
hidelinks,
]{revtex4-1}
\usepackage{graphicx}
\usepackage[english]{babel}
\usepackage{blindtext}
\usepackage{dcolumn}
\usepackage{bm}
\usepackage{hyperref}
\usepackage{mathtools}
\usepackage{amsmath}
\usepackage[mathlines]{lineno}
\usepackage{xcolor}
\hypersetup{colorlinks,citecolor={blue!50!black},urlcolor={blue!80!black}}
\usepackage[margin={-0.2cm,0cm},position=above,justification=raggedright,singlelinecheck=false]{subfig}
\usepackage[figurename=FIG.]{caption}
\begin{document}
\title{Quantum Vortex States in Bose Hubbard Model With Rotation}
\author{Mukesh P. Khanore}
\email{mukeshkhanore@gmail.com}
\author{Bishwajyoti Dey}
\email{bdey@physics.unipune.ac.in}
\affiliation{Department of Physics, SP Pune University,  Pune 411007, India.}%
\date{\today}
\newcommand{\iu}{{i\mkern1mu}}
\begin{abstract}
{We study quantum vortex states of strongly interacting bosons in a two-dimensional rotating optical lattice. The system is modeled by Bose-Hubbard Hamiltonian with rotation. We consider lattices of different geometries, such as, square, rectangular and triangular. Using numerical exact diagonalization method we show how the rotation introduces vortex states of different ground-state symmetries and the  transition between  these states at discrete rotation frequencies. We show how the geometry of the lattice plays  crucial role in determining the maximum number of vortex states as well as the general characteristics of these states such as, the average angular momentum $<L_z>$, the current at the perimeter of the lattice, phase winding, the relation between the maximum phase difference, the maximum current and also the saturation of the current between the two neighboring lattice points. The  effect of  the two- and three-body interactions between the particles, both attractive and repulsive, also depends on the geometry of the lattice as the current flow or the lattice current depends on the interactions.  We also consider the effect of the spatial inhomogeneity introduced by the presence of an additional confining harmonic trap potential. It is shown that the  curvature of the trap potential and the position of the minimum of the trap potential with respect to the axis of rotation or the center of the lattice have significant effect on the general  characteristics these  vortex states. } 
\end{abstract}
\maketitle
\section{\label{sec:l1} INTRODUCTION}
The study of strongly interacting bosons in a lattice using a Hubbard Hamiltonian was introduced by Fisher et al   \cite{fisher1989} where they studied the competing effects of repulsive two-body interaction and the pinning by lattice potential and predicted the existence of superfluid-Mott insulator (SF-MI) quantum phase transition. Such a transition was experimentally realized in a gas of ultracold atoms in optical lattice by Greiner et al \cite{Greiner2002}. It is well known that the interactions between the bosons can be made attractive or repulsive by varying the scattering length through Feshbach resonance mechanism. In contrast to the usual Bose Hubbard model with repulsive interaction between bosons,  theoretical \cite{sakmann_sacl2010,jack_y2005, khanore_bdey2014, khanore_bdey2015, wladimir_m2015, zin_cost2008} as well as experimental \cite{mark_hldjbdn2012} studies on attractive Bose Hubbard models have gained importance recently. Recent studies of  Bose Hubbard model with three-body interaction between bosons have shown several interesting results \cite{sowinski_crdtlm2015,sowinski2012,safavi-naini_scrs2012}.  An important feature  of the ultracold atoms in optical lattice is  the  ability  to  control  the geometry of the underlying optical lattice potential and even the possibility of implementation of a more complex unit cell. Optical lattice is created by interference of several laser beams and the atoms in optical lattice feel a potential proportional to the intensity of light filed.  The height of the lattice potential can be tuned by changing the laser intensity. Study of the Bose-Hubbard model on different lattice geometry, in particular the triangular lattice and the Kagome lattice is of recent interest \cite{arwas_vc2014,thomas2017}.  The studies of strongly interaction bosons on optical lattice superimposed with additional harmonic trap potential have also shown very interesting properties. The harmonic trap potential introduces space inhomogeneity in the system.   Study of SF-MI quantum phase transition of bosons in optical lattice with superimposed harmonic trap has been reported in \cite{jaksch_bcgz1998}. Studies on the effect of the trap potential on the SF-MI transition in Bose-Hubbard model have shown very different behavior where the Mott transition does not occur via a traditional quantum phase transition \cite{batrouni_rsrmdt2002}. For double-well trap potential the SF-Mott transition shows many characteristic features like, plateaus in local integrated density etc. \cite{thonhauser2007}. Investigation of the ground state properties of the attractive Bose-gas confined on two-dimensional square optical lattice and superimposed wine-bottle-bottom or Mexican hat trap potential have shown the coexistence of Mott and superfluid domains \cite{khanore_bdey2015}. Similarly, the rotation can be introduced in Bose gases in optical lattice  by rotating the optical lattice. Realization of rotating Bose gases in optical lattice  have opened wide the opportunity to explore many properties of atomic clouds, especially collective excitations such as vortices, solitons, mode coupling, nonlinear interferometry \cite{fetter2009}, fractional quantum Hall effect \cite{bhat_kch2007}, etc. It has been shown that for deep optical lattice, the vortex lattice of rotating Bose-Einstein condensate (BEC) gets pinned to the optical lattice and assumes the geometry and symmetry of the optical lattice \cite{dey2014}. The study of Bose-Einstein condensates (BECs) in a random potential has also received much attention as in this system, it is possible to display in a controlled way the interplay between interaction and disorder. Very recently we have shown that disorder can be used to induce vortex lattice melting in rotating  BEC \cite{mithun2016} and that this melting is a two-step process via loss of positional and orientational order \cite{mithun2018}. These studies are directly related to the melting of the Abrikosov vortex lattice in type-II superconductors, where the melting process has been studied extensively to explore its role on the critical current of the superconductor  \citep{pastoriza1994,paltiel2000,lindemann2004,kierfeld2000}. The two-step vortex lattice melting have also been recently experimentally observed  in type-II superconductors \citep{ganguli2016,ganguli2015}.\\
Even though the superfluidity and vortex properties have been explored analytically  in details in continuum rotating BECs through  Gross-Pitaevskii equation \cite{fetter2009}, there have been very few reports of studies of vortex formation by strongly interacting bosons in a rotating optical lattice using  Bose-Hubbard type Hamiltonian. Wu et al  \cite{wu_chjzs2004} considered the single vortex structure  of the strongly interacting bosons in rotating optical lattice and studied the nature of the vortex core near the superfluid-Mott insulator (SF-MI) transition within the mean field theory of Bose-Hubbard model. Bhat et al \cite{bhat_hc2006,bhat_ps2006} studied the  vortex states of strongly interacting bosons in rotating square optical lattice using a modified Bose-Hubbard model with a complex site-dependent hoping term. They showed that in contrast to the superfluid where the vortex refers to quantized circulation, in lattice system, the circulation is the current flow due to inter-site hopping  establishes phase relationship across the system much like the phase of the superfluid order parameter. Due to the discrete rotational symmetry of the lattice the average angular momentum $<L_z>$ is not quantized\cite{bhat_hc2006,bhat_ps2006,bhat_kch2007}. On the other hand, the quasi-angular momentum of the ground state, which is the generator of the discrete rotation, change  by integer value for symmetry-incommensurate filling  with increasing rotational frequency.  It is associated with changes in the phase winding $\Theta_{\rm cf}$ which jumps by $2\pi$ each time the quasi angular momentum  of the ground state changes \cite{bhat_hc2006,peden2007,bhat_kch2007}. The change in the phase winding of the ground state denote quantum phase transitions and is associated with vortex entering the system. For symmetry-incommensurate filling, the signature of the vortex entry is the crossing of energy levels of the ground and the first  excited state. Near the SF-MI transition, several abrupt structural phase transitions has been observed due to the pinning of the vortices by optical lattice \cite{goldbaum2008}. Vignolo et al \cite{vignolo2007} studied the dynamic response of the system as the vortex moves in the periodic potential of the optical lattice which allows to measure the vortex properties, like, the mass, pinning potential etc. \\
In this paper we examine the effects of the nature of the confining potentials, the rotating optical lattice potential and the trap potential, on the vortex states of the strongly interacting bosons at zero temperature. In particular we consider the  effect of the change in lattice  geometry provided by the optical lattice pinning potential and the spatial inhomogeneity introduced by the harmonic trap potential, on the vortex states of interacting bosons. 
We consider optical lattices of three different geometries, the square, rectangular and triangular. Further, we also consider the effects of the nature and the range of the onsite  interaction potential between the bosons on the vortex states of the system.  Accordingly, we consider bosons interacting via the onsite  attractive as well as repulsive two- and three-body interactions and also the off site nearest neighbor (NN) two-body interactions.  We show that the competition between the various parameters gives many interesting new quantum vortex states  in the system. We show how the geometry of the lattice plays a crucial role in determining the average angular momentum $<L_z>$, the current at the perimeter of the lattice, phase winding, the relation between the maximum phase difference and the maximum current and also the saturation of the current between the two neighboring lattice points. The  effect of  the two- and three-body interactions between the particles  also depends on the geometry of the lattice as the current flow or the lattice current depends on the interactions. We also show the spatial inhomogeneity introduced by the harmonic trap potential give rise to the spatial asymmetry in site dependent number density, current as well as the  phase difference between neighboring sites and its effect on the vortex states of the system. 
\section{\label{sec:l2} 2D BOSE-HUBBARD MODEL WITH ROTATION AND HARMONIC TRAP}
We consider neutral bosons confined in a 2D deep optical lattice potential and an additional harmonic trap potential which rotates about the $z$-axis. The bosons interacts via effective two- and three-body  onsite interaction  potentials  as well nearest neighbor (NN) two-body interaction potential. The Hamiltonian of the system as obtained  via the tight binding and the lowest band approximation can be written as 
\begin{eqnarray}\label{bhm1}
  \hat H &=& {-t}\sum_{<i,j>}^{}({\hat a_i} {\hat a_j^\dagger}+{\hat a_i^\dagger} {\hat a_j})- \iu\hbar\Omega\sum_{<i,j>}^{}K_{ij}({\hat a_i} {\hat a_j^\dagger}-{\hat a_i^\dagger} {\hat a_j}) \nonumber \\
&+& \sum_{i}^{} \left[ \frac{U}{2} \left[ {\hat n_i}({\hat n_i}-1) \right] + \frac{W}{6} \left[ {\hat n_i}({\hat n_i}-1)({\hat n_i}-2)\right]\right] \nonumber \\
&+& \sum_{i}^{} {\epsilon_i} {\hat n_i}+U_{nn}\sum_{<i,j>}^{} \hat{n_i} \hat{n_j}
\end{eqnarray}
where ${\hat a_i^\dagger} ({\hat a_i}) $ creates (annihilates) a particle at site $i$, ${\hat n_i}={\hat a_i^\dagger}{\hat a_i}$ gives number of particles at site $i$, $t$ is the hopping amplitude, $\Omega$, $U$, $W$ and 
$U_{nn}$ denote the strengths of rotation, the onsite two- and three-body interactions and the nearest-neighbor (NN) two-body interactions respectively. The co-rotating harmonic trap potential $\epsilon_i$ is given by \cite{jack_y2005}.
\begin{equation}\label{bhm2}
 \epsilon_i = \lambda\Big[i-\frac{1}{2}(c - \Delta)\Big]^2
\end{equation} 
Here,  $\lambda$ is the measure of the curvature of the trap, $c \equiv (x_0, y_0)$ is the center of lattice  and $\Delta \equiv (\Delta_x, \Delta_y)$ is the 
displacement of the minimum position of the trap from the the center of lattice.  $\epsilon_i$ represent an energy offset of each lattice site \cite{jaksch_bcgz1998}. The rotation induces hopping between neighboring sites and is governed by the site-dependent hopping parameter $\Omega K_{ij}$  given by  \cite{bhat_hc2006}
\begin{equation}\label{bhm3}
K_{ij} = \beta (r_i r_j /d^2)\sin{\alpha_{ij}}
\end{equation} 
where $r_i (r_j)$ is the distance from the axis of rotation to the $i^{th}(j^{th})$ site, $\alpha_{ij}$ is the angle between by $i^{th}$ and $j^{th}$ sites with respect to center of axis of rotation, $\beta$ is a dimensionless constant characterizing the lattice geometry and depth, $d$ is the lattice constant. $K_{ij}$ also depends on the lattice geometry.
The  site-dependent hopping parameter $\Omega K_{ij}$ and the site-dependent amplitude of the trap potential $\epsilon_i$ makes the model represented by the Hamiltonian in Eq. (\ref{bhm1}), an inhomogeneous Bose-Hubbard model. This is in contrast to the usual (homogeneous) Bose-Hubbard model where the hopping parameter and the two-body interaction parameter are site-independent. It may be mentioned here that by adding the two hopping terms the  Hamiltonian (Eq. (\ref{bhm1})) can be rewritten in terms of a complex site-dependent hopping parameter $t_{ij}=t + i\hbar\Omega K_{ij}$ \cite{bhat_ps2006}.
\section{\label{sec:l3} NUMERICAL METHOD AND GENERAL CHARACTERISTICS OF VORTEX STATES}	
We consider 2D lattice with different lattice geometries and filling factor less than one with open boundary conditions. We use numerical exact diagonalization method to calculate the eigenvalue and eigenvectors of the ground state of the system. We also calculate the first excited state energy of the system to show the crossing of energy levels between the ground state and the first excited state for each vortex entry  for symmetry-incommensurate filling. Using the eigenvectors of the ground state we calculate various quantities, such as,  site number density and normalized variance  for ground state, inter-site current, average angular momentum for ground state, condensate fraction of ground-state and phase winding on perimeter of the lattice for ground state, which are then used to characterize the vortex states of the system.
The site number density is the expectation value of number operator $\hat{n_i}=<\hat{a_i}^\dagger \hat{a_i} >$, the variation of which over sites shows the presence of a vortex in the rotating lattice system.  The presence of a  vortex in the  lattice shows a depletion of the site number density from the periphery towards the central region of the lattice where the vortex is located. The normalized variance  gives information about the system phase, such as, superfluid or Mott insulator and is defined as
\begin{equation}\label{bhm4}
\nu \equiv \frac{\sum_{i} ( <\hat{n_i}^2>-<\hat{n_i}>^2)}{\sum_{i}<\hat{n_i}>} 
 \end{equation} 
$\nu=1$ represent coherent state, $\nu>1$ the phase-squeezed state, $\nu<1$  the number-squeeze state, and $\nu=0$  single Fock state. 
The expectation value of $J_{ij}$ represents the current between sites $i$ and $j$ and the total current in the lattice boundary is an important  observable to verify the particle hole symmetry in the system.  In the rotating frame it is defined as \cite{bhat_hc2006}
\begin{eqnarray}\label{bhm5}
<J_{ij}>&=&-\frac{1}{\iu \hbar}\Big<[\hat{n_i},\hat{H_{ij}}]\Big> \nonumber 
 = \frac{\iu t}{\hbar}<\hat{a_i}\hat{a_j}^\dagger -\hat{a_i}^\dagger \hat{a_j}>\\
 &-&\Omega K_{ij}<\hat{a_i}\hat{a_j}^\dagger+\hat{a_i}^\dagger \hat{a_j}>
 \end{eqnarray} 
$\hat{H}_{ij}$ in [Eq.(\ref{bhm5})] is the Hamiltonian associated with sites $i$ and $j$. The total current is conserved at each site $i$ since the sum of $<J_{ij}>$ over all nearest neighbor $j$ is zero. The total current on the lattice boundary  $C$  is
\begin{equation}\label{bhm6}
\Lambda_C=(\hbar d/E_R) \sum_{<i,j>\in C}^{} <\hat{J_{ij}}>
\end{equation}
where all sum over $C$ are taken with same sign convention as the helicity of $\Omega$ (i.e. anticlockwise), $E_R = \hbar^2 / Md^2$ is the recoil energy in terms of which the energies of the system are scaled and $d$ is the lattice constant. Another important observable is  the average angular momentum, defined as
\begin{equation}\label{bhm7}
<L_z>=-\frac{d E_0}{d \Omega}.
\end{equation}
where $E_0$ is the ground state energy of the system. The variation of $< L_z >$ with rotational strength $\Omega$ shows discrete jumps at particular values of $\Omega$ where the vortex enters the system accompanied by the crossing of energy levels between $E_0$ and $E_1$ as the ground state symmetry changes abruptly. $< L_z >$ also gives information about rotational symmetry of system. Similarly, the condensate fraction and phase winding are associated with superfluid nature of system, as the condition for the existence of  a Bose-Einstein condensate is the macroscopic occupation of the single particle density mode with the largest eigenvalue \cite{penrose1956}. For lattice system the one body density matrix is given by
\begin{equation}\label{bhm8}
\rho_{ji}=<\Psi_0 |\hat{a_i}^\dagger\hat{a_j}| \Psi_0>
\end{equation}
where $\Psi_0$ is the ground state of the system [Eq.(\ref{bhm1})]. The ratio of the largest eigenvalue of the density matrix [Eq.(\ref{bhm8})] and total number of particles represents condensate fraction  and the associated eigenvector represents the condensate wave function. From the condensate wave function we calculate the phase of the condensate and  the phase winding around the perimeter $C$ of the condensate wave function $\Theta_{cf}$, which when divided by $2\pi$ gives the vorticity of the rotating system.
\section{\label{sec:l4} RESULTS}
\subsection{\label{sec:l4-1} RELATION OF THE PARAMETER  $\beta$ WITH COORDINATION NUMBER FOR DIFFERENT LATTICE GEOMETRIES}
\begin{figure}[!t]
\center
\vspace{-10pt}
     \subfloat[]{\includegraphics[scale=.7,keepaspectratio,trim={0 1.9cm 0 2.5cm},clip]{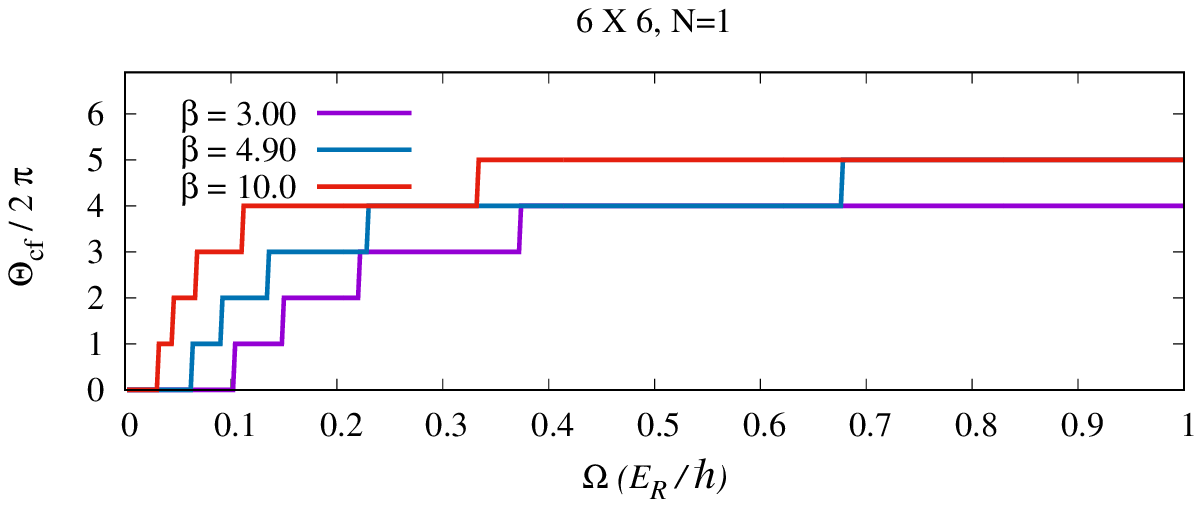}} \\
\vspace{-10pt}
     \subfloat[]{\includegraphics[scale=.7,keepaspectratio,trim={0 1.9cm 0cm 2.5cm},clip]{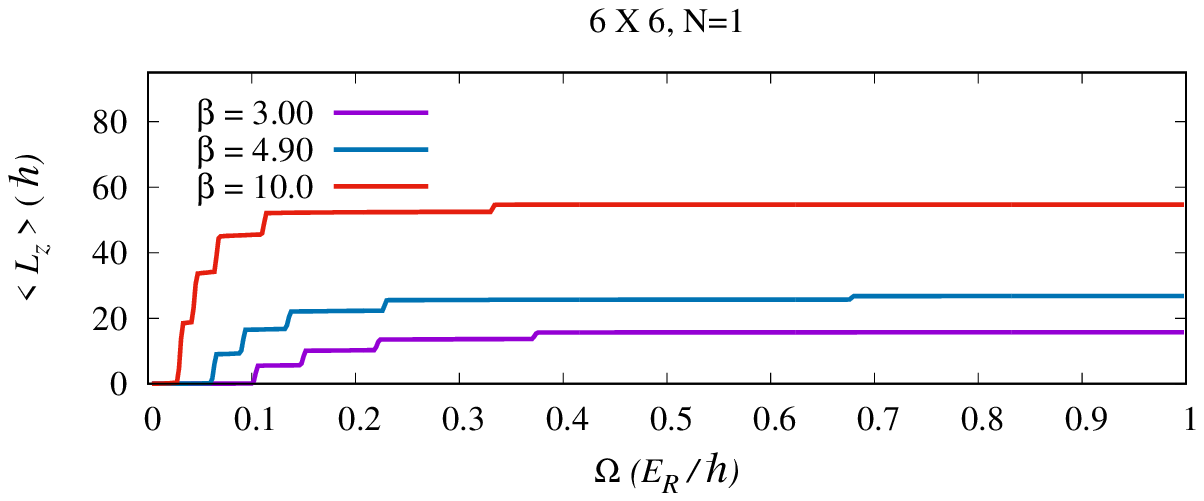}}\\
\vspace{-10pt}
     \subfloat[]{\includegraphics[scale=.7,keepaspectratio,trim={0 1.9cm 0 2.5cm},clip]{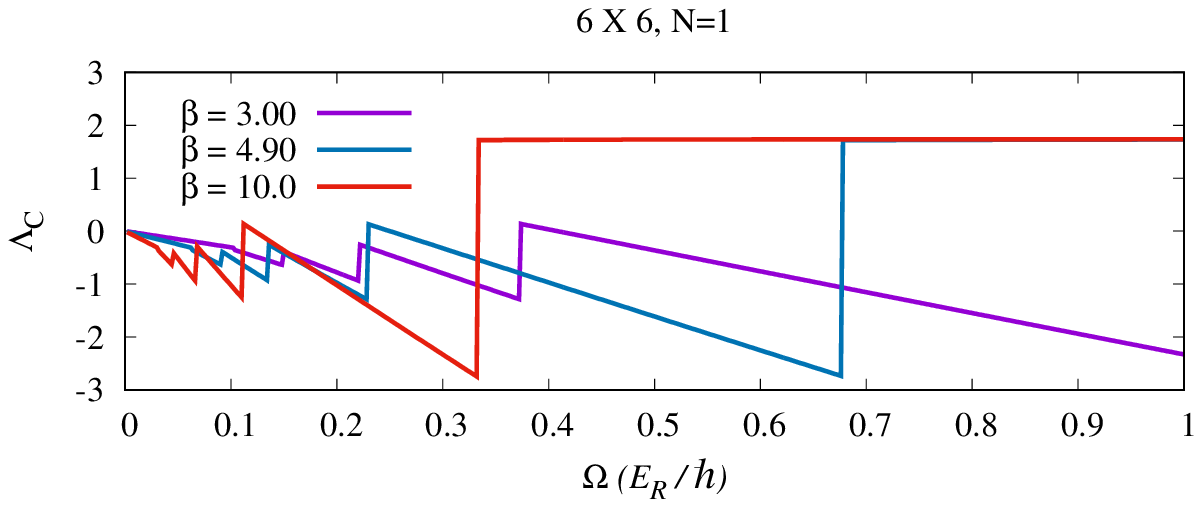}}
\vspace{-5pt}
     \caption{ $6 \times 6$ lattice with $N=1$ and for different values of $\beta$ in units of $t / E_R$. (a) phase winding $\Theta_{cf}/2 \pi$, (b) average angular momentum $<L_z>$, and (c) sum of boundary current $\Lambda_C $. }\label{fig:1}
\end{figure}
The parameter $\beta$ in Eq. (\ref{bhm3}) can be numerically estimated from the overlap integrals for finite numbers of sites \cite{bhat_ps2006}.  However, we observed  that the parameter $\beta$ has an interesting relationship with the coordination number $Z$ of the lattice.  This allows us to calculate the optimal value of the parameter $\beta$ for lattice of different symmetries in terms of its coordination number $Z$.  For a square lattice of size $L \times L$, the four-fold rotational symmetry allows maximum $L - 1$ quantized vortices within the lowest band. The vortices enters the system one after another as the lattice is rotated with increasing frequency $\Omega$  and the phase winding of the condensate wave-function increases in steps of $2\pi$ to a maximum of $2\pi \times (L-1)$. The maximum phase difference between neighboring sites is $\pi/2$.  We calculate the phase winding by varying the rotational frequency  for different values of the parameter $\beta$ and find the optimum value of $\beta$ for which all the allowed number of vortices enters the system and the phase winding remain constant. We find that this happens when $\beta \geq Z$ and any further increase in the value of $\beta$ do not produce any change in phase winding.  This is shown in Fig. \ref{fig:1} for two-dimensional square lattices ($Z=4$) of sizes  $6 \times 6$, where it can be seen  that for $\beta \geq Z$ we get maximum value of phase winding $(L-1)2 \pi$, maximum number of vortices $L-1$ and accordingly $L-1$ jumps in $<L_z>$. For 
$\beta < 4$, we do not get the right number ($L-1$) of vortices and the right value of the phase winding. This makes the connection of the coordination number $Z$ with parameter $\beta$. For a square lattice we therefore take the optimum value of $\beta = 4.9 \ t/E_{R} d, t / E_R = 1$. This is very close to the numerical estimation of $\beta$ from the overlap integrals as $\beta =4.93 \ t/E_{R} d, t / E_R = 1$ \cite{bhat_ps2006}. \\
 We follow the same logic to find $\beta$ value for the  triangular lattice for which the coordination number is $Z=6$. In contrast to the square lattice,  the triangular lattice has three-fold rotational symmetry. As shown in Fig. \ref{fig:6}, we observe similar dependence of $\beta \geq Z$ for the triangular lattice and accordingly  we choose the optimum value of the parameters $\beta =6.6 \ t/E_{R} d, t / E_R = 1$ for the triangular lattice.  
\subsection{\label{sec:l4-2} VORTEX NUMBER, PHASE WINDING, BOUNDARY VALUE CURRENT AND PHASE DIFFERENCE FOR LATTICES OF DIFFERENT GEOMETRIES }
\begin{figure}[!t]
\vspace{-10pt}
     \subfloat[]{\includegraphics[scale=.7,keepaspectratio,trim={0 2cm 0cm 2.5cm},clip]{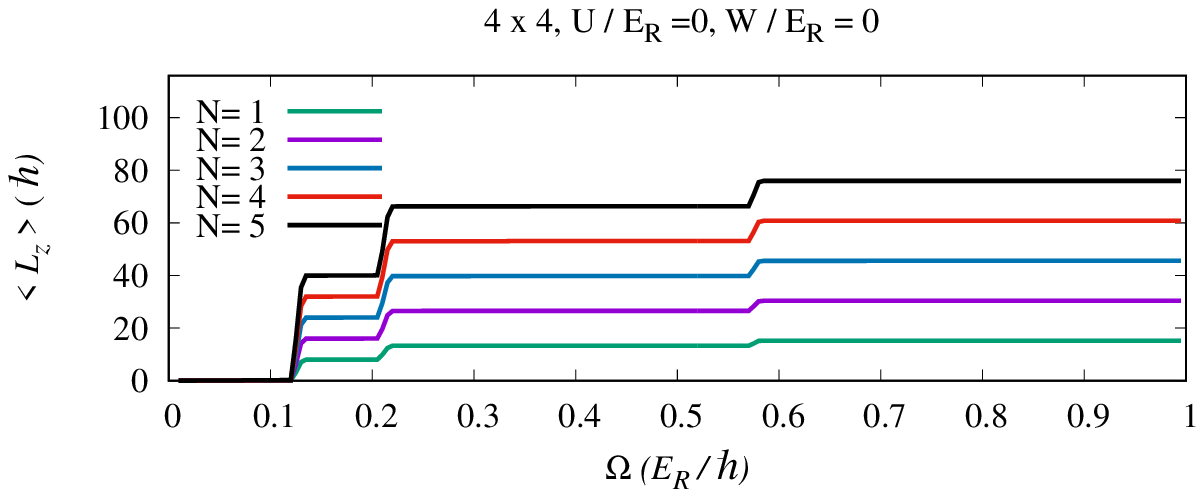}}\\
\vspace{-10pt}
     \subfloat[]{\includegraphics[scale=.7,keepaspectratio,trim={0 2cm 0 2.5cm},clip]{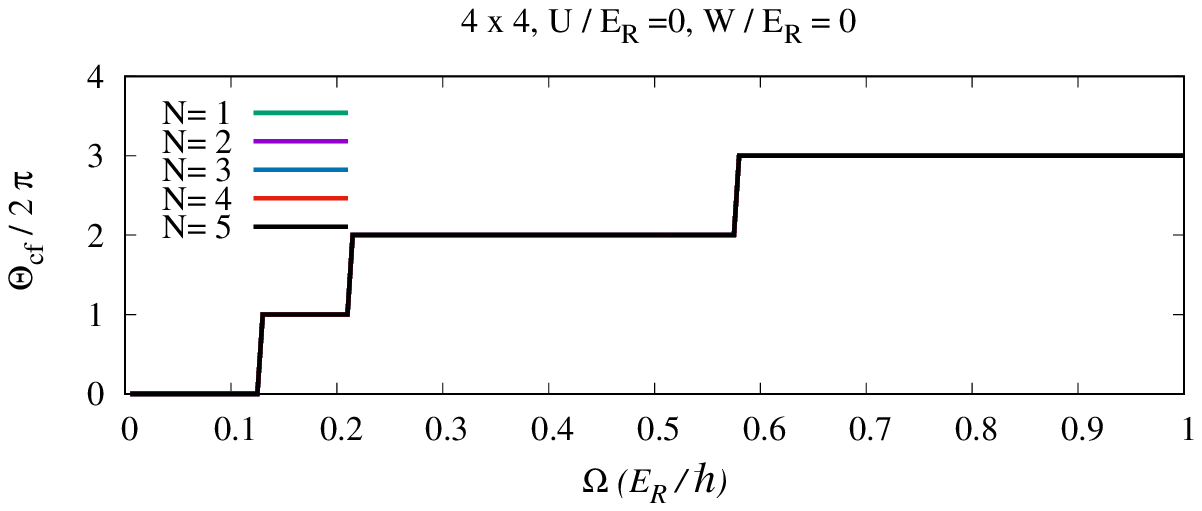}}\\
\vspace{-10pt}
     \subfloat[]{\includegraphics[scale=.7,keepaspectratio,trim={0 2cm 0 2.5cm},clip]{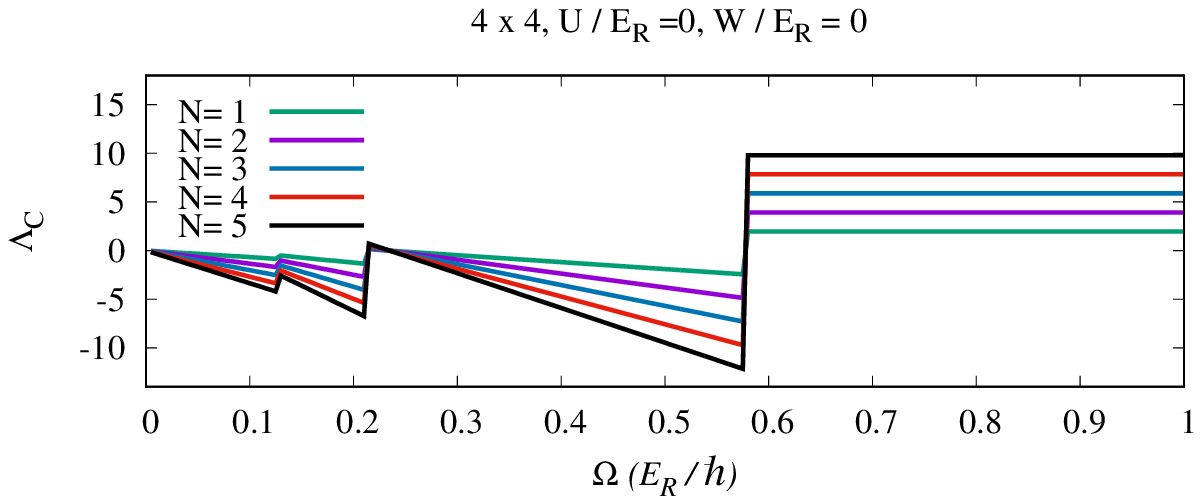}}
\vspace{-5pt}
      \caption{ $4 \times 4$ lattice for different values of $N$ without interactions $U / E_R =0$, (a) average angular momentum $<L_z>$, (b) phase winding $\Theta_{cf}/2 \pi$, and (c) sum of boundary current $\Lambda_C $.}\label{fig:2}
\end{figure}
\begin{figure}[!t]
     \subfloat[]{\includegraphics[scale=.7,keepaspectratio,trim={0 2cm 0cm 2.5cm},clip]{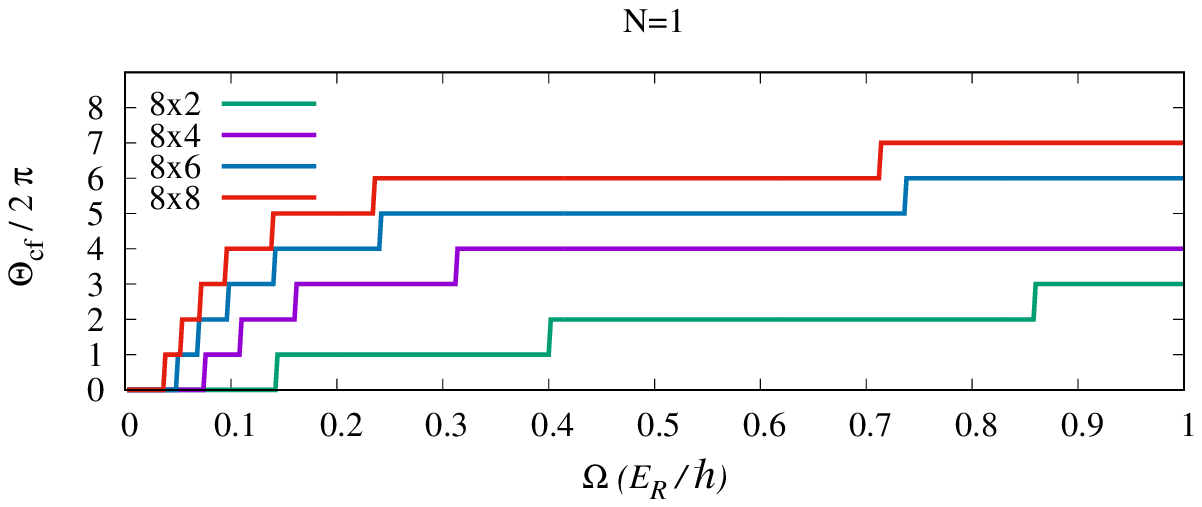}} \\
\vspace{-10pt}
     \subfloat[]{\includegraphics[scale=.7,keepaspectratio,trim={0 0cm 0 0cm},clip]{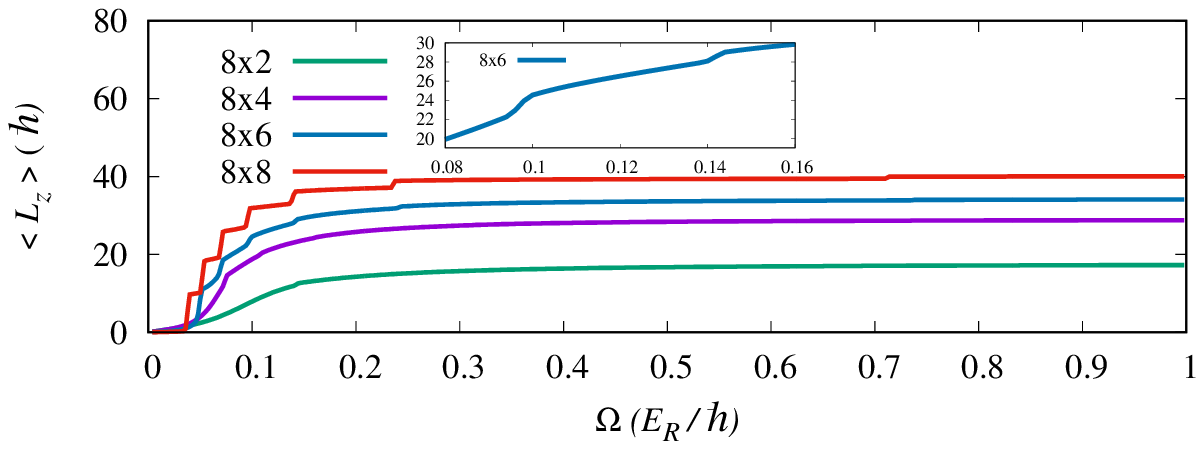}}\\
\vspace{-10pt}
     \subfloat[]{\includegraphics[scale=.7,keepaspectratio,trim={0 0cm 0 0cm},clip]{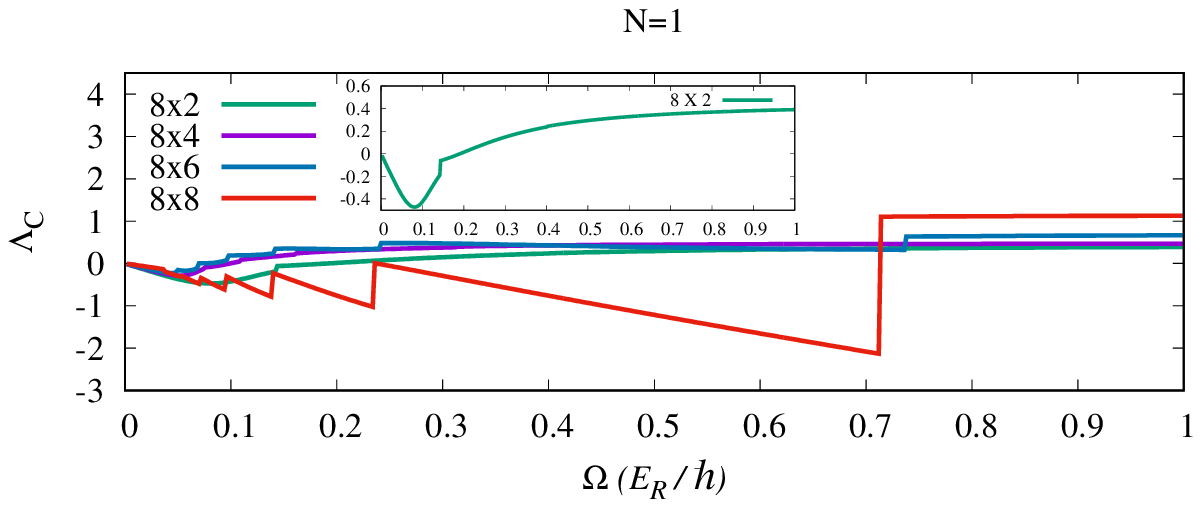}}\\
\vspace{-10pt}
     \subfloat[]{\includegraphics[scale=.7,keepaspectratio,trim={0 2cm 0 2.5cm},clip]{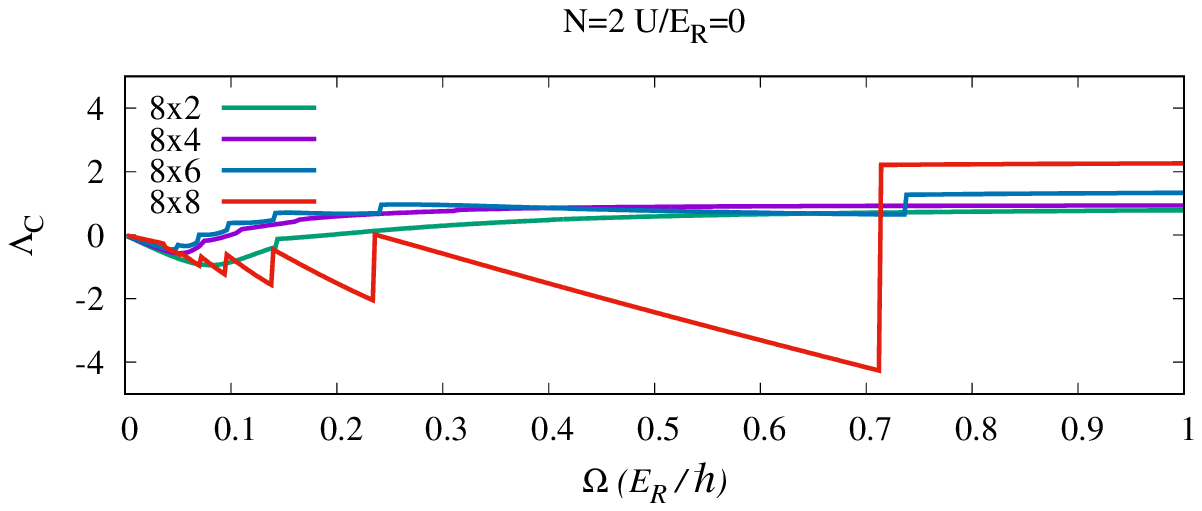}}
\vspace{-5pt}
      \caption{ Rectangular lattice with $L_X = 8$ and different values of $L_Y$ for $N=1,2$ without interaction $U / E_R =0$. (a) phase winding $\Theta_{cf}/2 \pi$, (b) average angular momentum $<L_z>$ for $N=1$, (c) sum of boundary current $\Lambda_C$ for $N=1$ and (d)  sum of boundary current $\Lambda_C$ for $N=2$ }\label{fig:3}
\end{figure}
As mentioned above, the vortex number, phase winding, boundary value current and phase difference between neighboring sites increases with the increase of the rotational frequency $\Omega$. For a square lattice of  $N_S= L \times L$ lattice sites, maximum  number of quantized vortices is $(L-1)$. The vortices enters the system one after another as the rotational frequency is increased which also increases the average value of the angular momentum $< L_z >$ of the system. This is shown in Fig. \ref{fig:2}(a) for a $4 \times 4$ square lattice for different particle number $N$. The abrupt change of the angular momentum for particular value of the rotational frequency shows the entry of a vortex in the system. The maximum phase difference between the nearest neighbor (NN) sites is $\pi/2$, and maximum phase winding is $2\pi \times (L-1)$ in the lowest Bloch band. This is shown in Fig. \ref{fig:2}(b). The boundary value current between the nearest-neighbor sites along the perimeter increases with increase in the strength of rotational frequency and it reaches the maximum value when the phase difference between the nearest-neighbor sites along the perimeter become $\pi/2$. For square lattice the maximum value of the boundary current ($\Lambda_C$) for one particle ($N=1$) is $\Lambda_C = 2$. For $N$ noninteracting particles ($N > 1$) the maximum boundary current is given by $\Lambda_c^{(N)} = N \Lambda_c^{(1)}$. This saturation of the boundary value current is shown in Fig. \ref{fig:2}(c). \\
\begin{figure}[!t]
         \subfloat[]{\includegraphics[scale=.8,keepaspectratio,trim={2cm 3.2cm 1cm 3.5cm},clip]{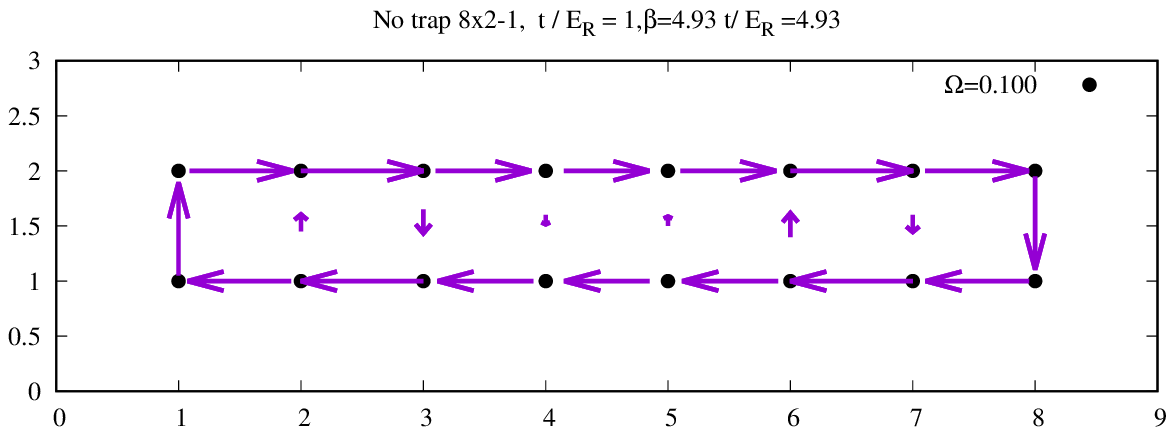}} \\
\vspace{-10pt}
    \subfloat[]{\includegraphics[scale=.8,keepaspectratio,trim={2cm 3.2cm 1cm 3.5cm},clip]{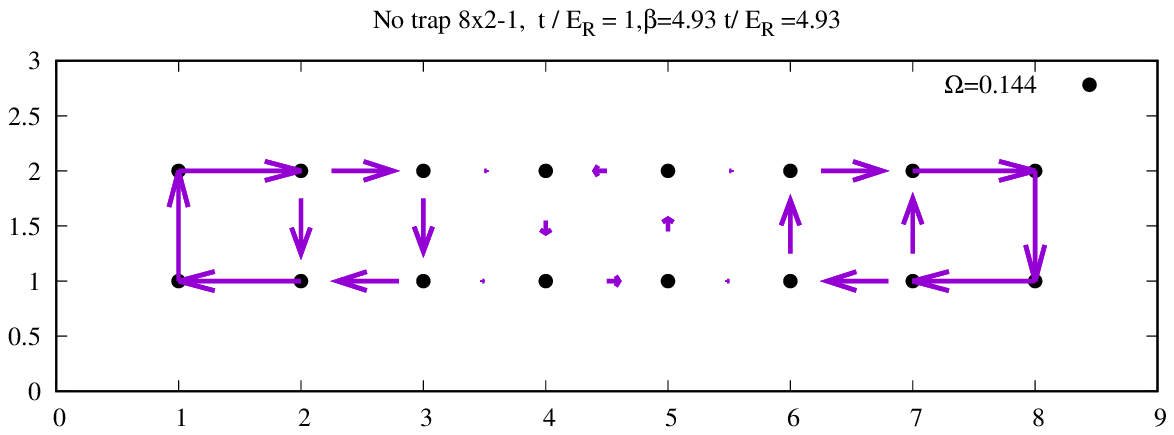}} \\
   \vspace{-10pt}
    \subfloat[]{\includegraphics[scale=.8,keepaspectratio,trim={2cm 3.2cm 1cm 3.5cm},clip]{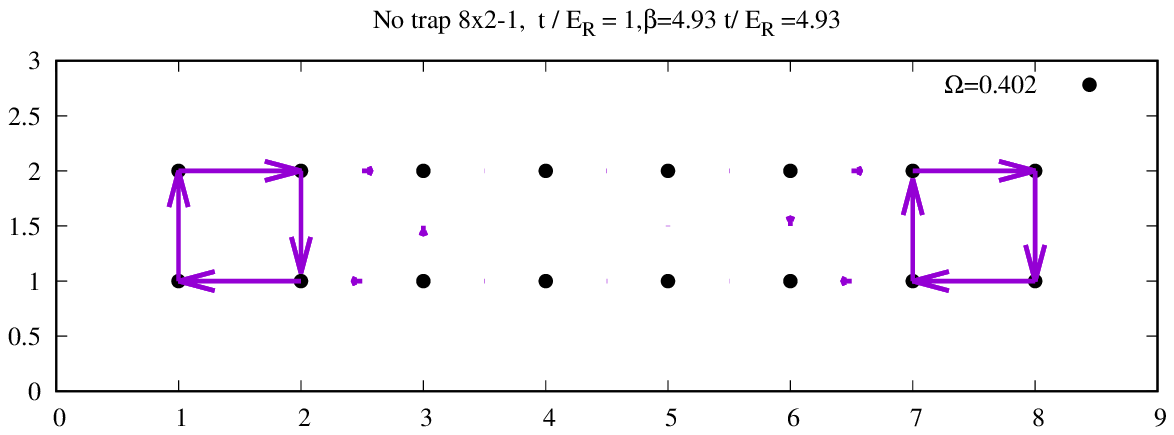}} \\
    \vspace{-10pt}
     \subfloat[]{\includegraphics[scale=.8,keepaspectratio,trim={2cm 3.2cm 1cm 3.5cm},clip]{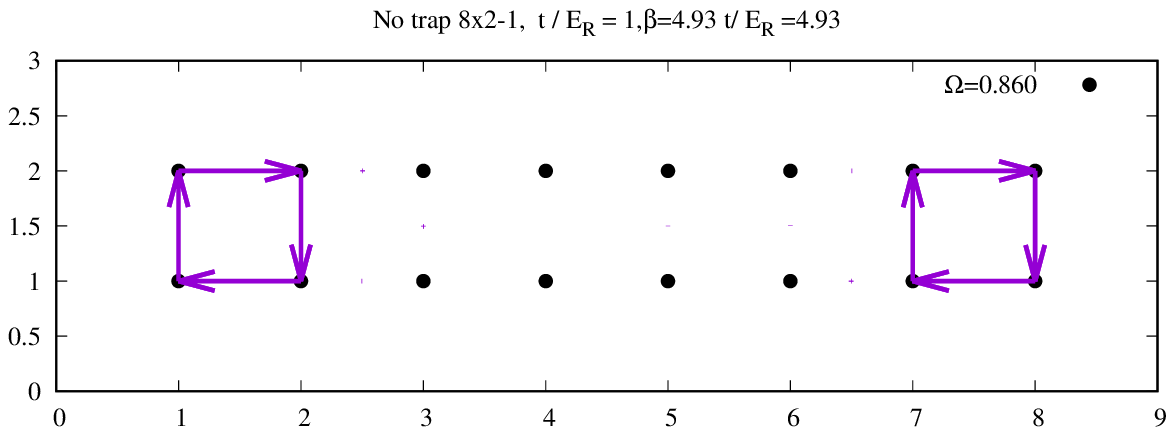}} 
     \vspace{-5pt}
     \caption{Current pattern for $8 \times 2$ rectangular lattice and $N=1$. 
Here length of arrow represent the current magnitude and arrow head represent the direction. For (a)  $\Omega =0.1$, (b) $\Omega =0.144$, (c) $\Omega =0.402$ and (d) $\Omega =0.860$.}\label{fig:4}
\end{figure}
For a rectangular lattice of $N_s = L_X \times L_Y$ lattice sites, where $L_X$ and $L_Y$ denote the number of lattice sites along x- and y-axis respectively, we define a asymmetry parameter $\epsilon = {L_Y \over L_X}$, such that, for $\epsilon = 1$ we get the square lattice ($L_X=L_Y=L$). We observe that for rectangular lattice, the maximum number of quantized vortices is nearest integer to $\sqrt{(L_X - 1) \times  (L_Y - 1)}$. For example, for the case of $L_X=8$ and $L_Y=2$ case, $\sqrt{(7) \times  (1)} = 2.64$ and  the maximum value of number of vortices is $3$.  Unlike the square lattice, the maximum phase difference between the neighboring sites is not ${\pi \over 2}$, but depends on the number of sites on the rectangular lattice. But the relation between the maximum value of the phase winding to the number of vortices, i.e. maximum value of the phase winding is $2\pi$ times the maximum value of number of vortices, as obtained for the square lattice, is also valid for the rectangular lattice. This is shown in Fig. \ref{fig:3}(a). Also, unlike the square lattice, the $< L_z >$ do not show sharp discontinuous jumps for the rectangular lattice. But as $\epsilon$ value increases the discrete jumps become sharper as shown in Fig. \ref{fig:3}(b). The inset of Fig. \ref{fig:3}(b) shows $< L_z >$ for $\epsilon = 0.25$. Unlike the square lattice, for the rectangular lattice there is no saturation of the boundary  current $\Lambda_C$ for the third vortex for small value of the parameter $\epsilon$. This is shown in the inset of Fig. \ref{fig:3}(c) for $\epsilon = 0.25$. The reason for the non-saturation of the boundary current with increasing angular frequency is due to the fact that the intersite currents in each of the two antivortices at the edges of the lattice (Fig. \ref{fig:4}(d)) also changes with change of angular frequency. In contrast, for the square lattice where the intersite current in each of the four anti-vortices within the perimeter or core of the third vortex almost remain constant with increasing rotational frequency \cite{bhat_hc2006}. As the value of the parameter $\epsilon$ increases we get the saturation of the boundary  current similar to that obtained for the square lattice. This is shown in Fig. \ref{fig:3}(c) for number of particle $N=1$. For $N$ number of non-interacting particles ($U = 0$), the maximum value of  $\Lambda_C^{(N)} = N \Lambda_C^{(1)}$, for all values of the parameter $\epsilon$, which is similar to that obtained for the square lattice. This is shown in Fig. \ref{fig:3}(d) for $N=2$ case. 
\begin{figure}[!ht]
         \subfloat[]{\includegraphics[scale=.75,keepaspectratio,trim={1.2cm 2.3cm 0cm 2.7cm},clip]{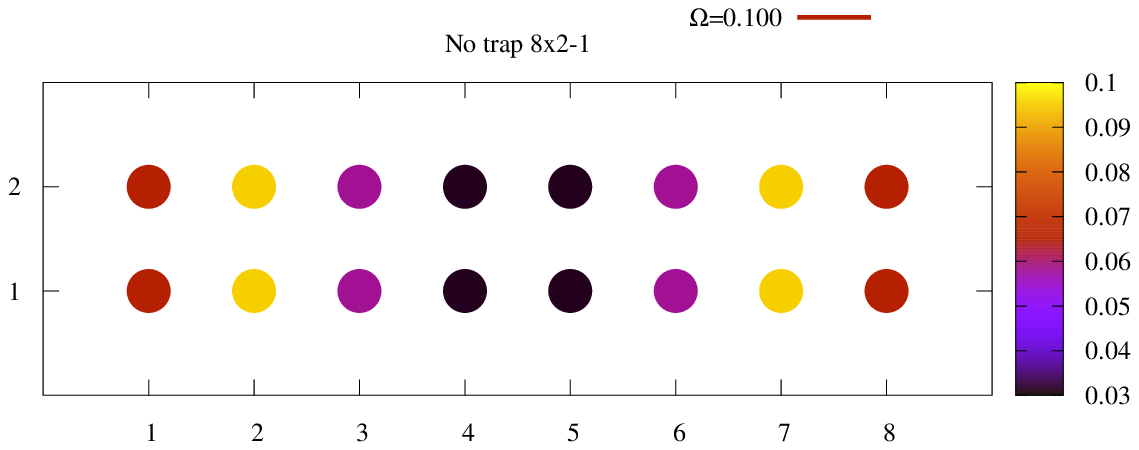}}\\
     \vspace{-10pt}         
    \subfloat[]{\includegraphics[scale=.75,keepaspectratio,trim={1.2cm 2.3cm 0cm 2.7cm},clip]{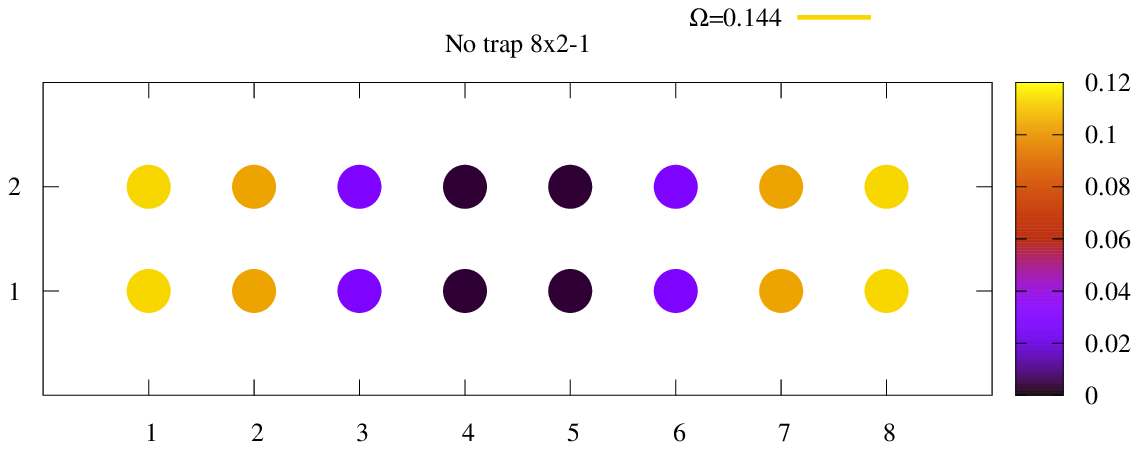}}\\
         \vspace{-10pt}
    \subfloat[]{\includegraphics[scale=.75,keepaspectratio,trim={1.2cm 2.3cm 0cm 2.7cm},clip]{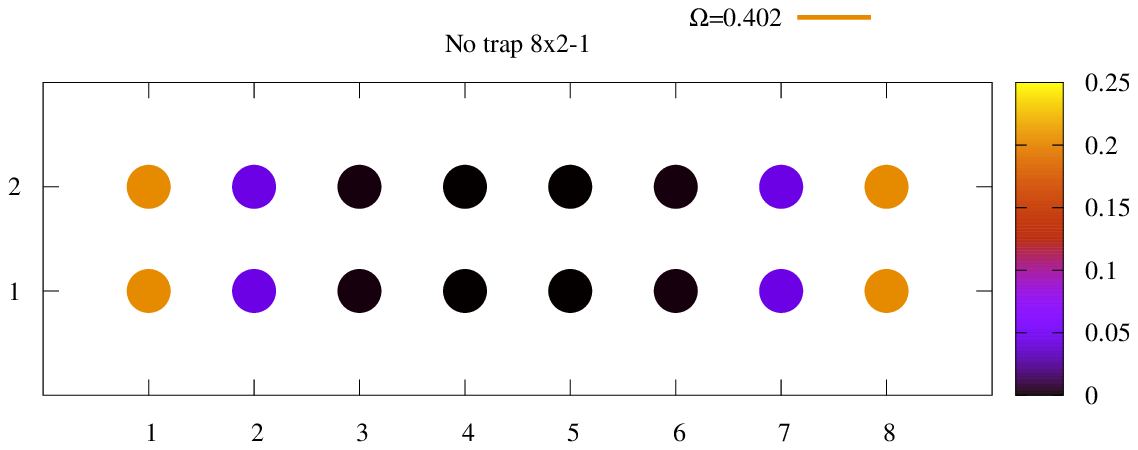}}\\
         \vspace{-10pt}
     \subfloat[]{\includegraphics[scale=.75,keepaspectratio,trim={1.2cm 2.3cm 0cm 2.7cm},clip]{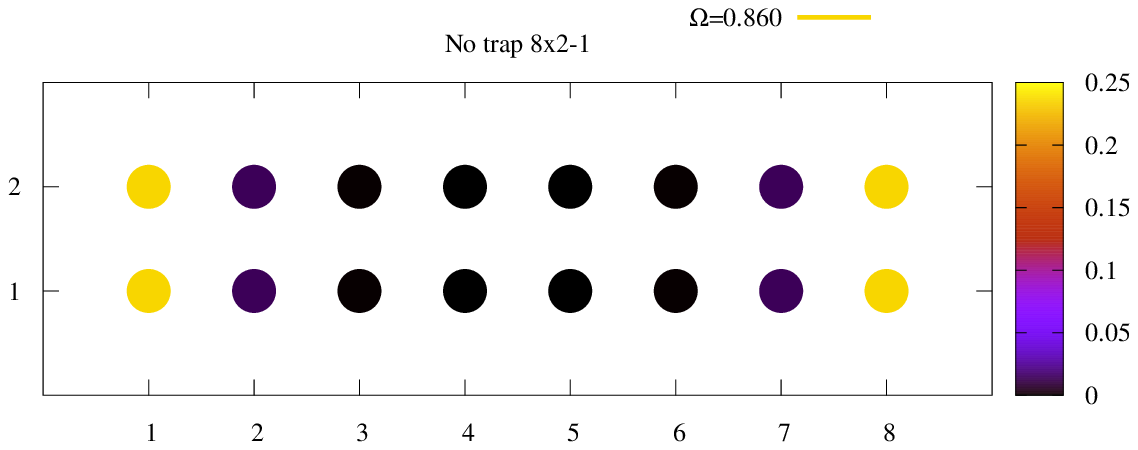}} 
          \vspace{-5pt}
     \caption{Site dependent density distribution for $8 \times 2$ lattice and $N=1$.  For (a)  $\Omega =0.1$, (b) $\Omega =0.144$, (c) $\Omega =0.402$ and (d) $\Omega =0.860$}\label{fig:5}
\end{figure}
The presence of interaction ($U \neq 0$) between the particles, however, changes this relations irrespective of the geometry of the lattice, as discussed in Section \ref{sec:l5} below. The current configurations or the circulation pattern for different vortices in a rectangular lattice is shown in Fig. \ref{fig:4}. In Fig. \ref{fig:4}(a) there is no vortex as the rotation ($\Omega = 0.1$) is not sufficient to create a vortex in the system. The current seems to flow backwards in the rotating frame (clockwise). As the rotation increases, a single vortex enters the system for $\Omega = 0.144$ as shown in Fig. \ref{fig:4}(b). The single vortex is located at the center of the lattice with central four sites defining the boundary of the vortex core. The current pattern in these four sites of the vortex is anti-clockwise with phase winding $\Theta_{cf}/2 \pi =1$. For further increase in rotation a second vortex enters the system ($\Omega = 0.402$)  which is  located at the central eight sites with the anti-clockwise current pattern and phase winding $\Theta_{cf}/2 \pi =2$ as shown in Fig. \ref{fig:4}(c). There are two anti-vortices with clockwise circulation at the two ends of the lattice. Fig. \ref{fig:4}(d) shows the current pattern for the third vortex for $\Omega = 0.860$ which is located at the central eight sites with similar current pattern as that of the second vortex but with $\Theta_{cf}/2 \pi =3$. The corresponding site dependent number density distribution $\hat{n_i}=<\hat{a_i}^\dagger \hat{a_i} >$ associated with each current pattern is shown in Fig. \ref{fig:5}. 
\begin{figure}[!ht]
     \subfloat[]{\includegraphics[scale=.7,keepaspectratio,trim={0 2cm 0cm 2.5cm},clip]{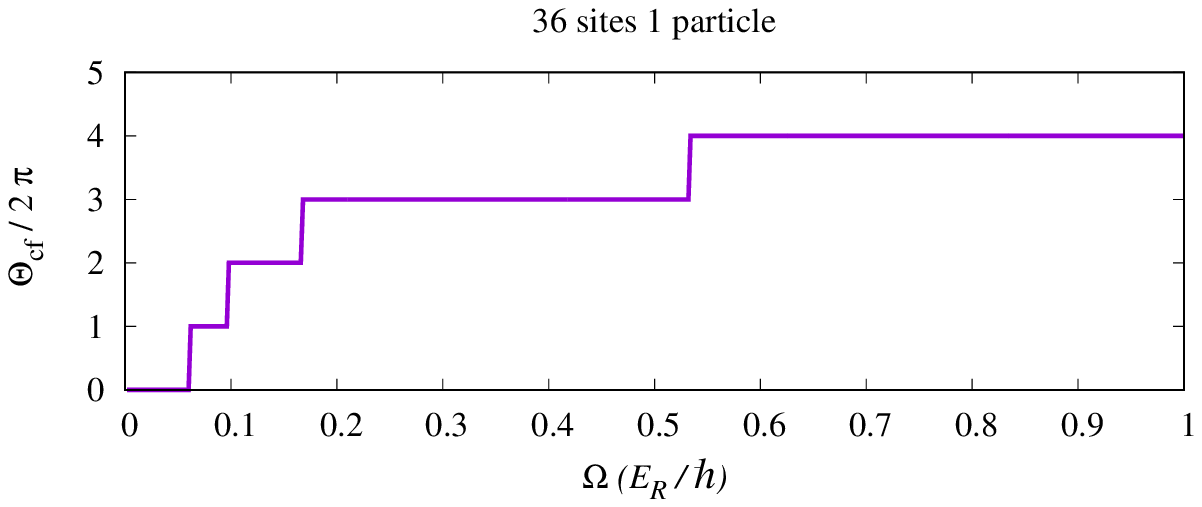}} \\  
     \vspace{-10pt}
     \subfloat[]{\includegraphics[scale=.7,keepaspectratio,trim={0 2cm 0cm 2.5cm},clip]{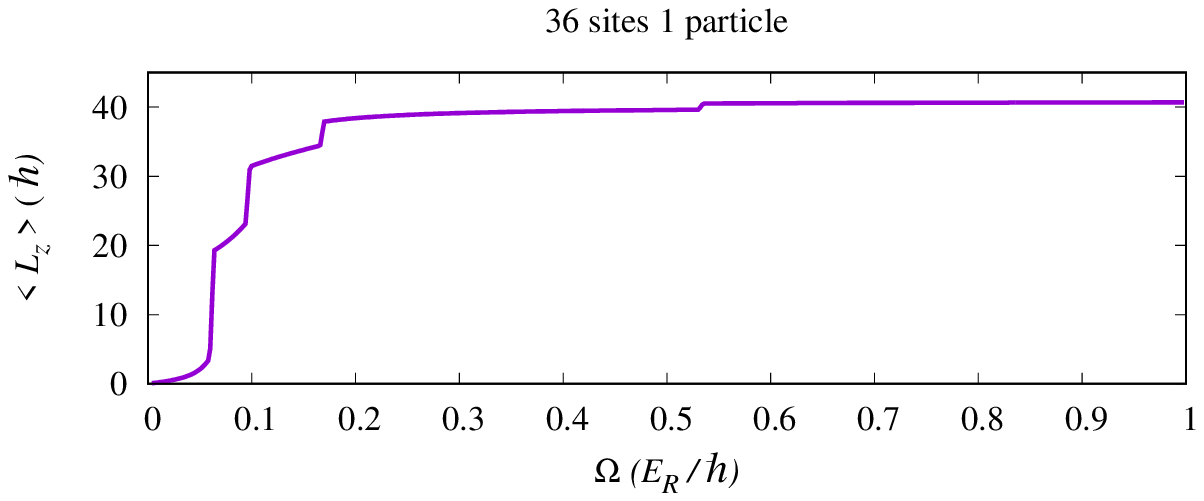}} \\
     \vspace{-10pt}
     \subfloat[]{\includegraphics[scale=.7,keepaspectratio,trim={0 2cm 0cm 2.5cm},clip]{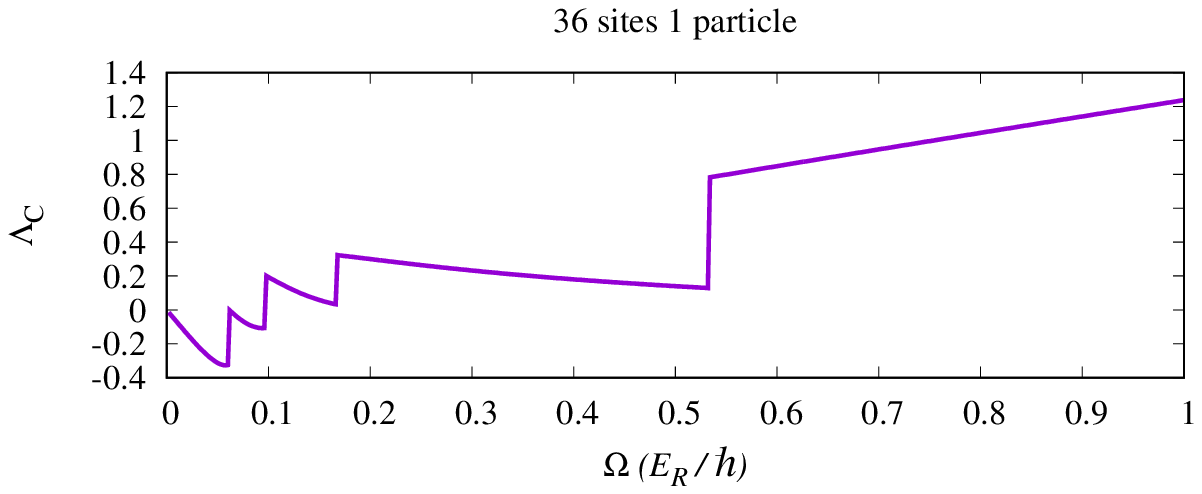}} \\     
     \vspace{-5pt}
     \caption{  Triangular lattice with $N_s = 36$ and $N=1$, (a) phase winding $\Theta_{cf}/2 \pi$,  (b) average angular momentum $<L_z>$, and (c) sum of boundary current $\Lambda_C $. }\label{fig:6}
     \end{figure}
In the figures each circle represent the site and color of each circle represent the magnitude of number density at that site. Fig. \ref{fig:5}(a) shows a no vortex state as the rotation in not enough to introduce the vortex in the system. There is no depletion of the site number density from the periphery towards the central region of the lattice. Fig. \ref{fig:5}(b) shows the presence of a vortex located at the center of the lattice with central four sites (black dots) defining the boundary of the vortex core. There is a continuous depletion of the site number density from the periphery towards the central region of the lattice. As expected, the site number density is minimum for the sites (black dots) on the boundary of the vortex core. Similarly, Fig. \ref{fig:5}(c) shows the density pattern for the  the second vortex having vortex core over central eight sites (black dots). Fig. \ref{fig:5}(d) shows similar site density distribution for the third vortex.\\
\begin{figure}[!t]
    \centering  
          \subfloat[]{\includegraphics[scale=.65,keepaspectratio,trim={3.2cm 1.25cm 3.2cm 2.3cm},clip]{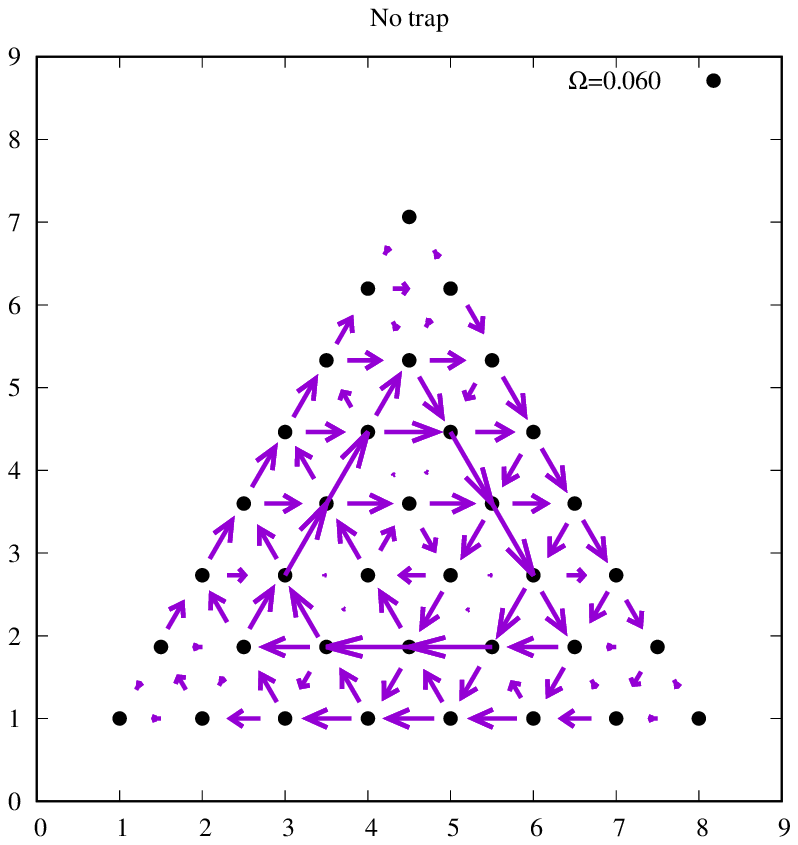}} 
          \subfloat[]{\includegraphics[scale=.65,keepaspectratio,trim={3.2cm 1.25cm 3.2cm 2.3cm},clip]{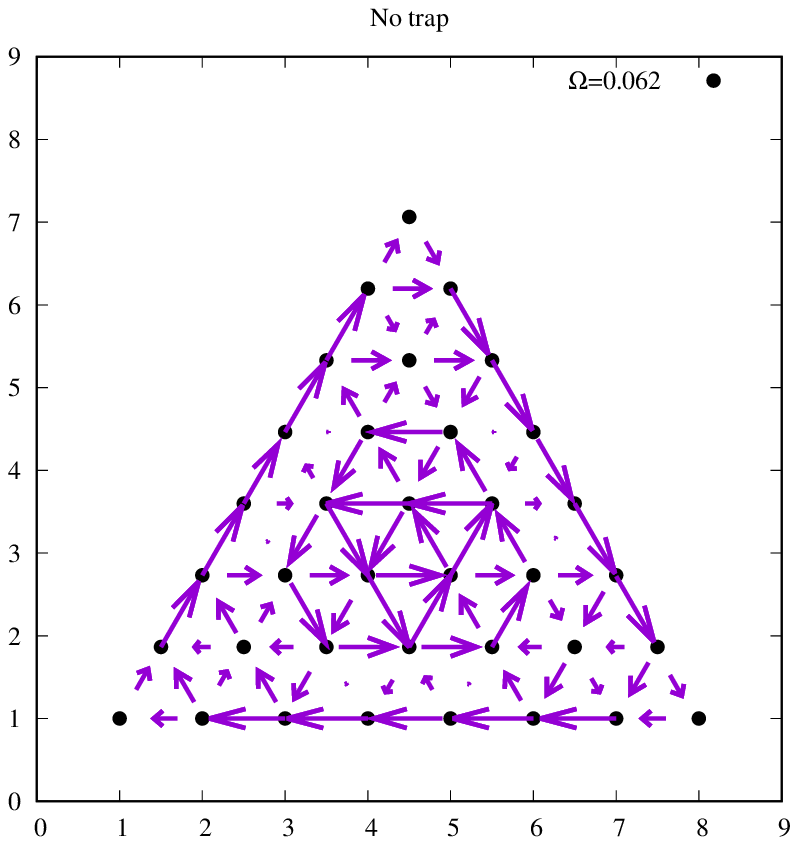}} \\
     \vspace{-5pt}          
          \subfloat[]{\includegraphics[scale=.65,keepaspectratio,trim={3.2cm 1.25cm 3.2cm 2.3cm},clip]{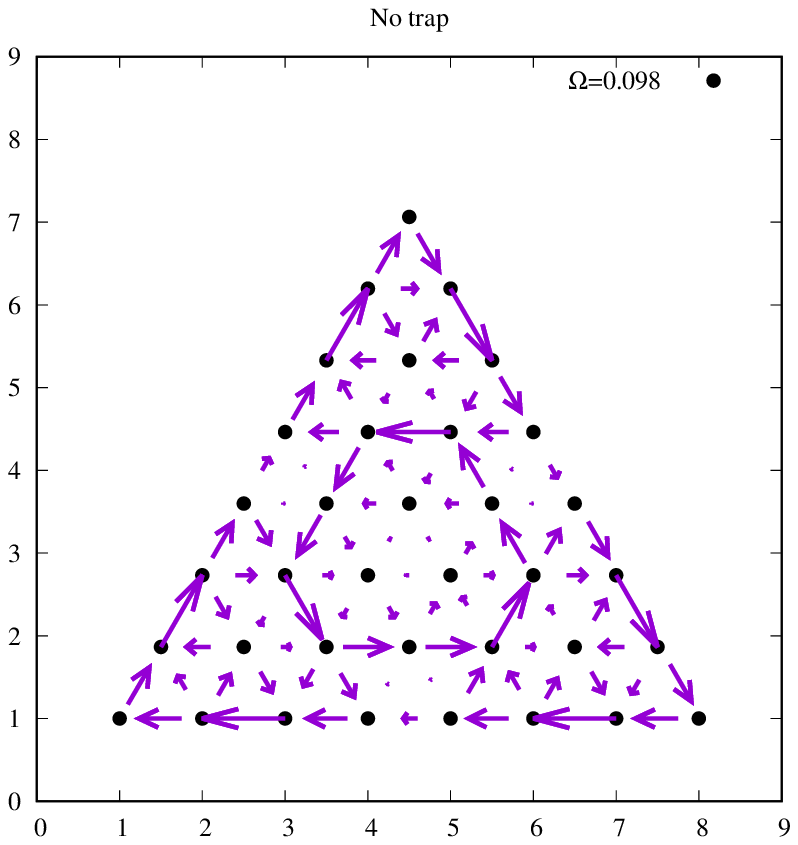}} 
          \subfloat[]{\includegraphics[scale=.65,keepaspectratio,trim={3.2cm 1.25cm 3.2cm 2.3cm},clip]{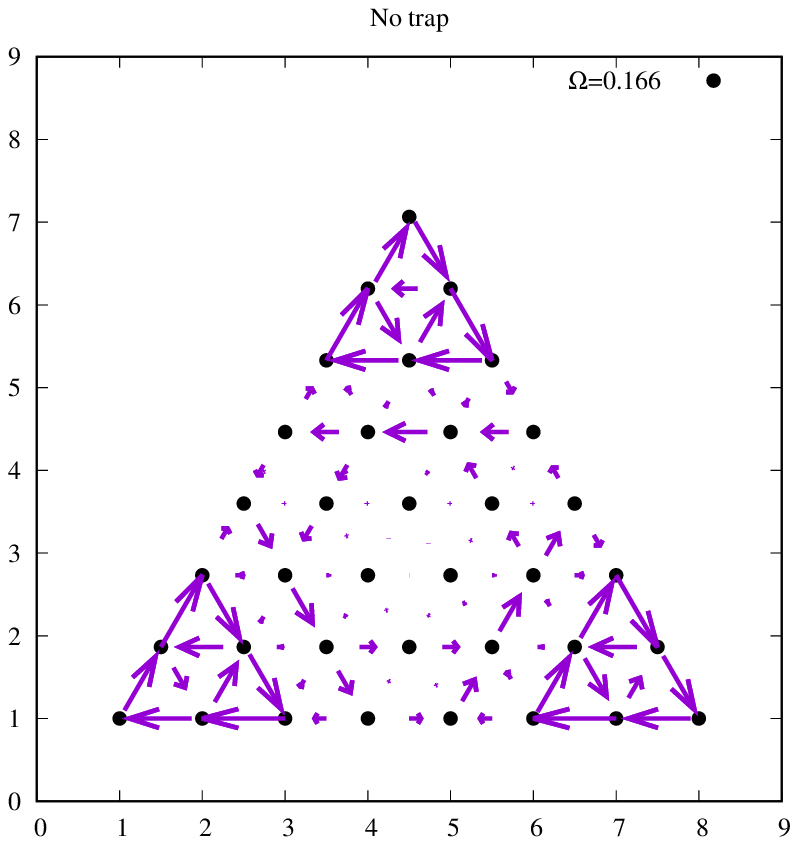}} \\
          \subfloat[]{\includegraphics[scale=.65,keepaspectratio,trim={3.2cm 1.25cm 3.2cm 2.3cm},clip]{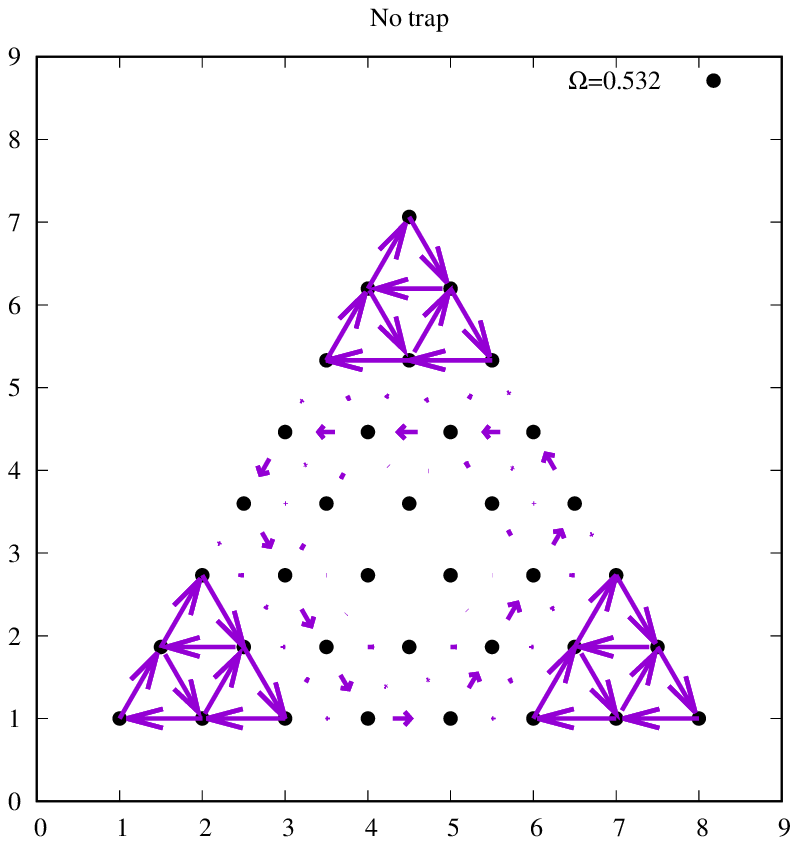}}           
               \vspace{-5pt}
 \caption{Current pattern for triangular lattice $N_s = 36, N=1$. For (a) $\Omega =0.06$, (b) $\Omega =0.062$, (c) $\Omega =0.098$, (d) $\Omega =0.168$ and (e) $\Omega =0.532$ }\label{fig:7}
\end{figure}   
\begin{figure}[!ht]
    \centering
         \subfloat[]{\includegraphics[scale=.52,keepaspectratio,trim={2.9cm 0.9cm 2cm 1.2cm},clip]{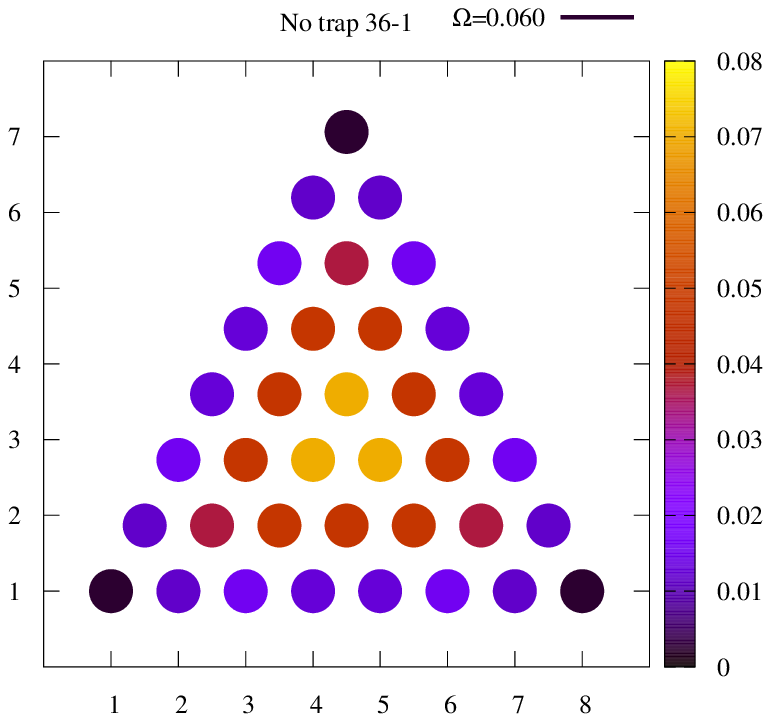}} 
    \subfloat[]{\includegraphics[scale=.52,keepaspectratio,trim={2.9cm 0.9cm 2cm 1.2cm},clip]{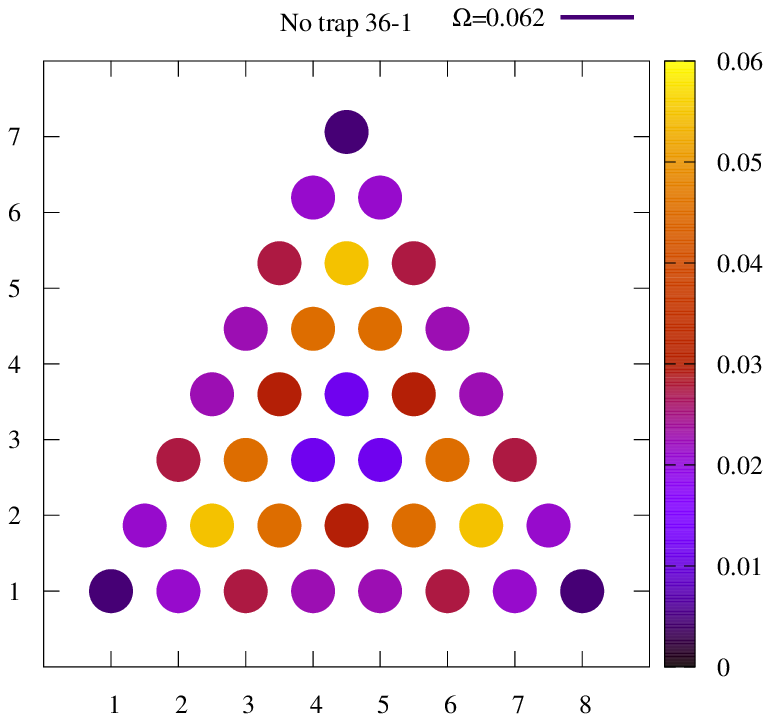}} 
\\
               \vspace{-5pt}
    \subfloat[]{\includegraphics[scale=.52,keepaspectratio,trim={2.9cm 0.9cm 2cm 1.2cm},clip]{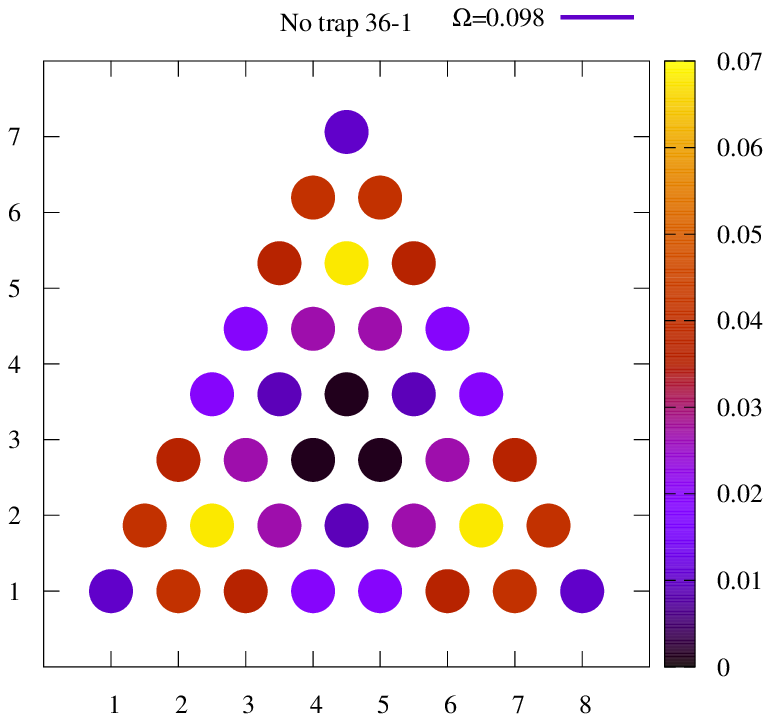}} 
     \subfloat[]{\includegraphics[scale=.52,keepaspectratio,trim={2.9cm 0.9cm 2cm 1.2cm},clip]{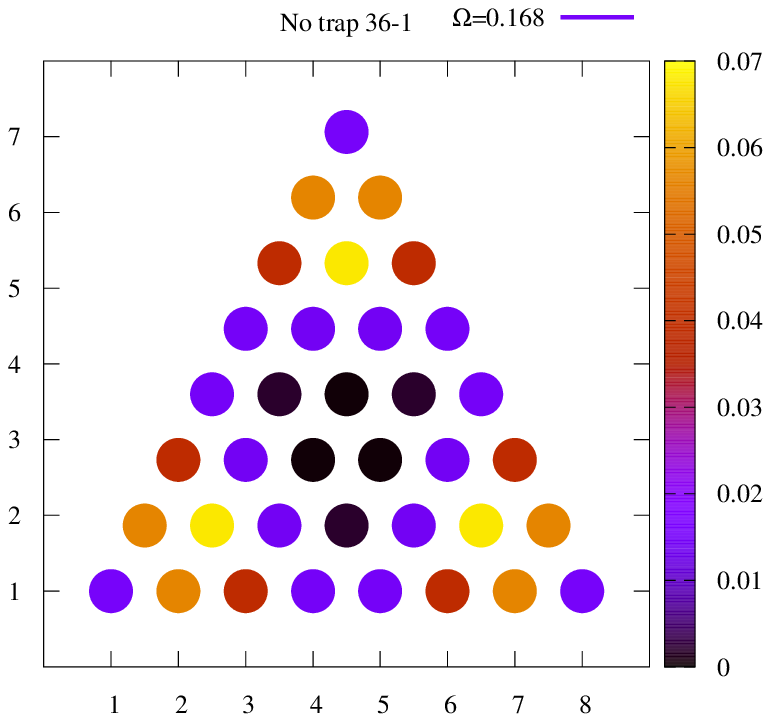}} 
\\     
     \subfloat[]{\includegraphics[scale=.52,keepaspectratio,trim={2.9cm 0.9cm 2cm 1.2cm},clip]{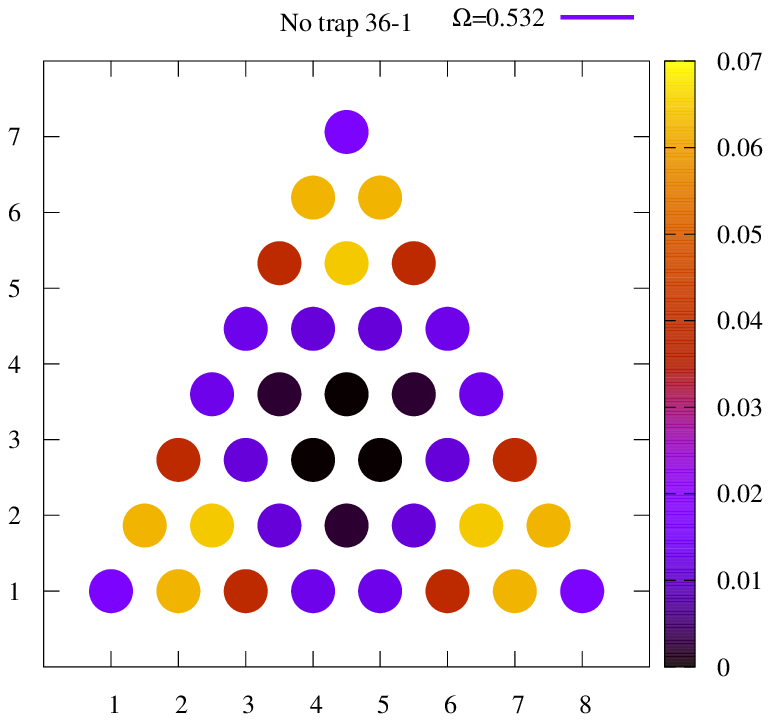}} 
      \vspace{-5pt}
     \caption{Density distribution for triangular lattice $N_s = 36, N=1$. For (a) $\Omega =0.06$, (b) $\Omega =0.062$, (c) $\Omega =0.098$, (d) $\Omega =0.168$ and (e) $\Omega =0.532$}\label{fig:8}
\end{figure}
For a triangular lattice of $N_s = {L(L + 1) \over 2}$ lattice sites, where $L$ denote the number of lattice sites in each side of the triangle, the maximum  number of quantized vortices is nearest integer to ${(L - 1)\over 2}$. For example, for a triangular lattice of $L=8$ sites per side, the number of vortices is $4$. This is shown in  Fig. \ref{fig:6} which shows four discrete jumps in phase winding (Fig. \ref{fig:6}(a)), the average angular momentum (Fig. \ref{fig:6}(b)) and the border current $\Lambda_C$ (Fig. \ref{fig:6}(c)) as the vortices enters the system with increasing rotational frequency. The maximum phase difference between the neighboring sites however depends on the number of sites. For example, for the unit cell of the triangular lattice ($L=2$), the  value of the phase difference between the neighboring sites is ${2\pi \over 3}$ which is expected due to the three fold rotational symmetry of the triangular lattice, but for larger lattice it is integer multiple of ${\pi \over 3}$, lowest value being ${\pi \over 3}$. This is because of the more complex current current configurations for the larger lattice which is shown in Fig. \ref{fig:7} for  lattice with $N_s = 36$. However, the maximum phase winding is $2\pi$ times the maximum value of number of vortices, similar to that of square and rectangular lattice as shown in Fig. \ref{fig:6}(a). Like the rectangular lattice, for the triangular lattice also the boundary current $\Lambda_C$ for the fourth vortex do not saturate with increasing rotational frequency. As in the rectangular case, for the triangular case also, the non-saturation behaviour in $\Lambda_C$ for the fourth vortex is due to the change in the intersite currents with increasing frequency in each of the three antivortices at the three edges of the triangle (Fig. \ref{fig:7}(e)). The corresponding number density distribution associated to each current pattern in  Fig. \ref {fig:7} is shown in Fig. \ref {fig:8}. In Fig. \ref{fig:8}(a) there is no vortex as can be seen from the density maximum at the three central sites. Fig. \ref{fig:8}(b) and Fig. \ref{fig:8}(c) shows the first and second vortices with three central sites (blue dots) and black dots respectively as the vortex cores. The total phase winding between the three central sites is $2\pi$ and $4\pi$ respectively. Similarly, Fig. \ref{fig:8}(d) and and Fig. \ref{fig:8}(e) shows the third and the fourth vortices with central six sites (black dots) as the vortex cores with total phase winding between the central six sites being $6\pi$ and $8\pi$ respectively. 
\section{\label{sec:l5}Effect of Interactions}
\subsection{\label{sec:l5-1} PARTICLE HOLE SYMMETRY}
\begin{figure}[!ht]
    {\includegraphics[scale=.5,keepaspectratio]{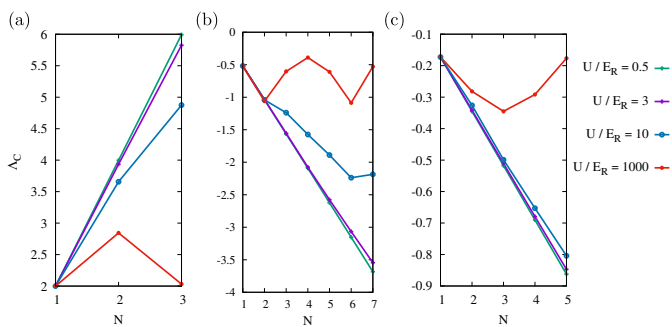}}
      \caption{Total boundary current $\Lambda_C$ for (a) square lattice, $2 \times 2, \Omega = 0.8$, (b) rectangular lattice, $4 \times 2, \Omega = 0.4$, and (c) triangular lattice $N_s = 6, \Omega = 0.5$.  The connecting lines here are for guide to the eyes.}\label{fig:9}
\end{figure}
Here we plot $\Lambda_C$ for different values of the number of particles $N$ for half filled case. For a square lattice of number of sites $N_s = 2\times 2$, we have maximum number of particles of $N=4$ for half filled case. We find that for strong two-body interaction ($U>>J$) and given $\Omega$,  $\Lambda_C$ for $N=1$ and $N=3$ matches exactly, showing the particle hole symmetry. Particle hole symmetry is also satisfied for rectangular and triangular lattice. For a rectangular lattice $N_s = 4\times 2$, we can have maximum $N=8$ for half filled case. We observed that for $U>>J$ and given $\Omega$, $\Lambda_C$ for $N=1$ and $N=7$ matches exactly, showing the particle hole symmetry for the rectangular lattice. Similarly for a triangular lattice of $N_s = 3$, we observed similar effect that $\Lambda_C$ for $N=1$ and $N=2$ matches exactly for strong two-body interaction. Fig. \ref{fig:9}(a), \ref{fig:9}(b) and \ref{fig:9}(c) shows the particle hole symmetry for square, rectangular and triangular lattice respectively.
\subsection{\label{sec:l5-2} ATTRACTIVE INTERACTION}
Here we consider interacting bosons on rotating optical lattice with two- and three-body interactions. First we will discuss the effect of attractive two-body interaction $U<0$.  We are interested to study effect of attractive interaction on the quantum vortex states in rotating square, rectangular and triangular lattices.\\
\begin{figure}[!tbp]
    \centering
     \subfloat[]{\includegraphics[scale=.7,keepaspectratio,trim={0 2cm 0cm 2.5cm},clip]{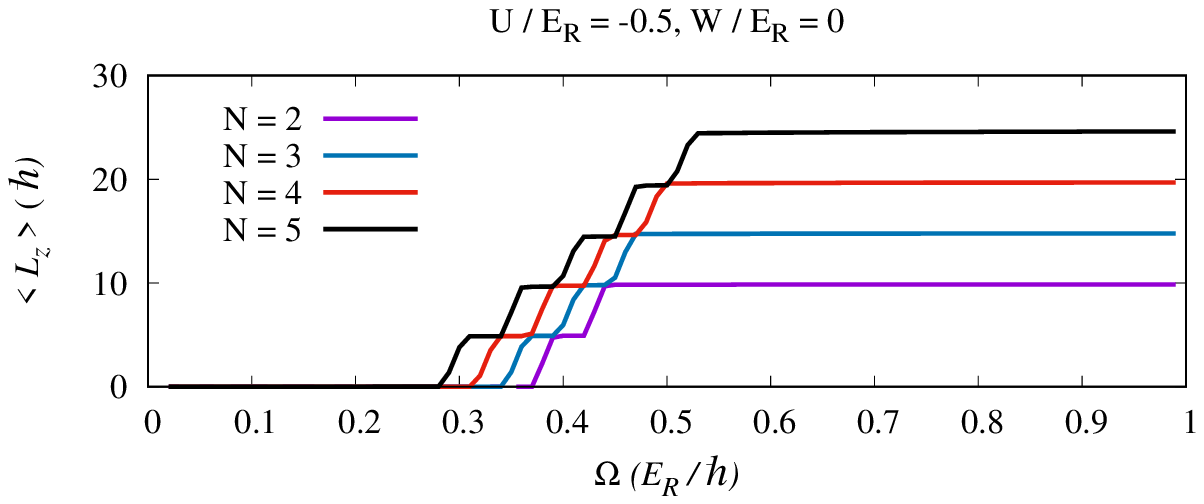}} \\
\vspace{-10 pt}
     \subfloat[]{\includegraphics[scale=.7,keepaspectratio,trim={0 2cm 0 2.5cm},clip]{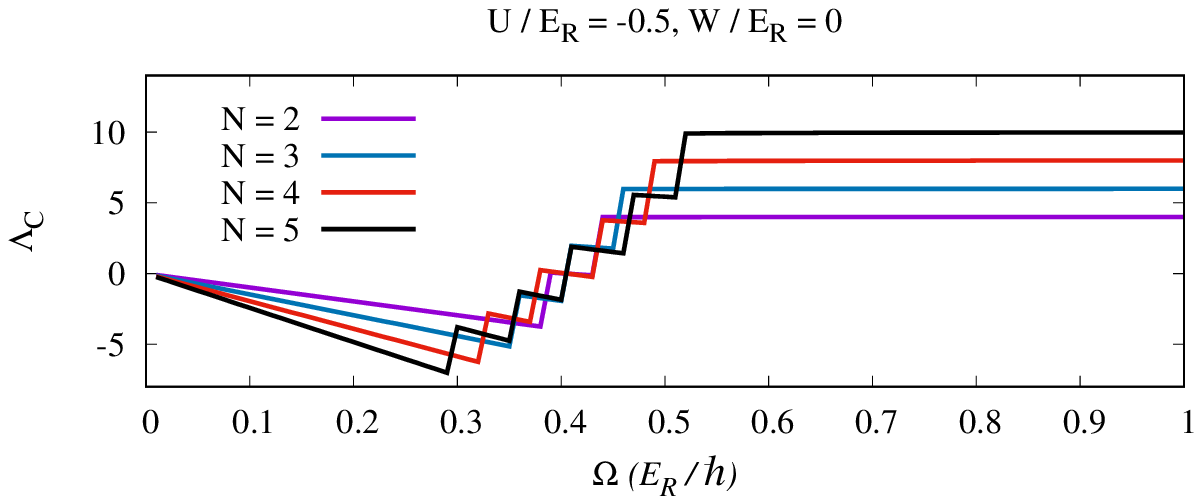}}\\
\vspace{-10 pt}
     \subfloat[]{\includegraphics[scale=.7,keepaspectratio,trim={0 2cm 0 2.5cm},clip]{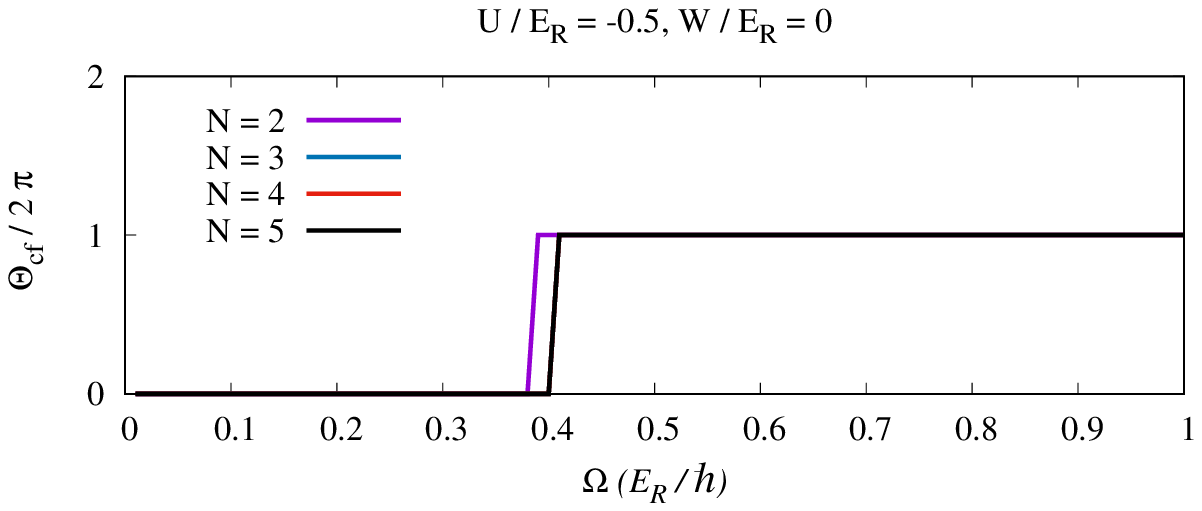}}
\vspace{-5 pt}
      \caption{$2 \times 2$ lattice for different values of $N$ and for $U / E_R =-0.5$. (a) average angular momentum $<L_z>$, (b) sum of boundary current $\Lambda_C $ and (c) phase winding $\Theta_{cf}/2 \pi$.}\label{fig:10}
\end{figure}
\begin{figure}[!ht]
    \centering
	    {\includegraphics[scale=.75,keepaspectratio,trim={0cm 0.3cm 0cm 0.75cm},clip]{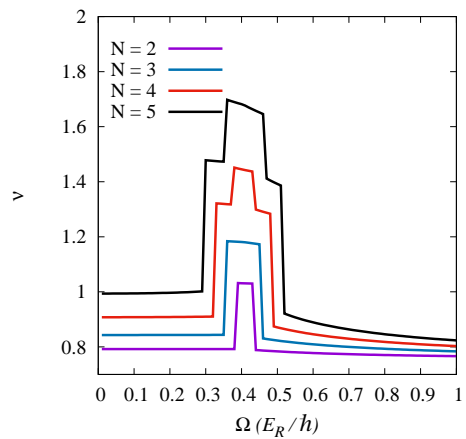}}
          \caption{ Normalized variance for $2 \times 2$ lattice for different values of $N$ and $U / E_R =-0.5$ as  function of $\Omega$}\label{fig:11}
\end{figure}
\begin{figure}[!ht]
    \centering
	    {\includegraphics[scale=.7,keepaspectratio,trim={0 2cm 0 2.5cm},clip]{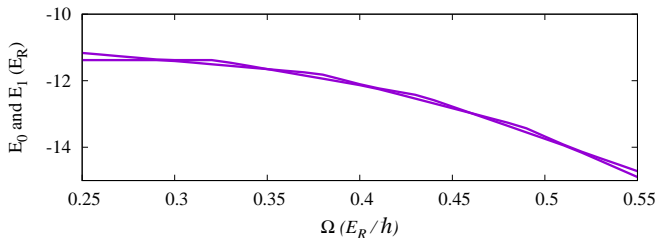}}
          \caption{ Lowest two energy eigen values of  $2 \times 2$ lattice for $N=5$ and  $U / E_R =-0.5$ as  function of $\Omega$}\label{fig:12}
\end{figure}
\begin{figure}[!hb]
      \subfloat[]{\includegraphics[scale=.7,keepaspectratio,trim={0 2cm 0 2.5cm},clip]{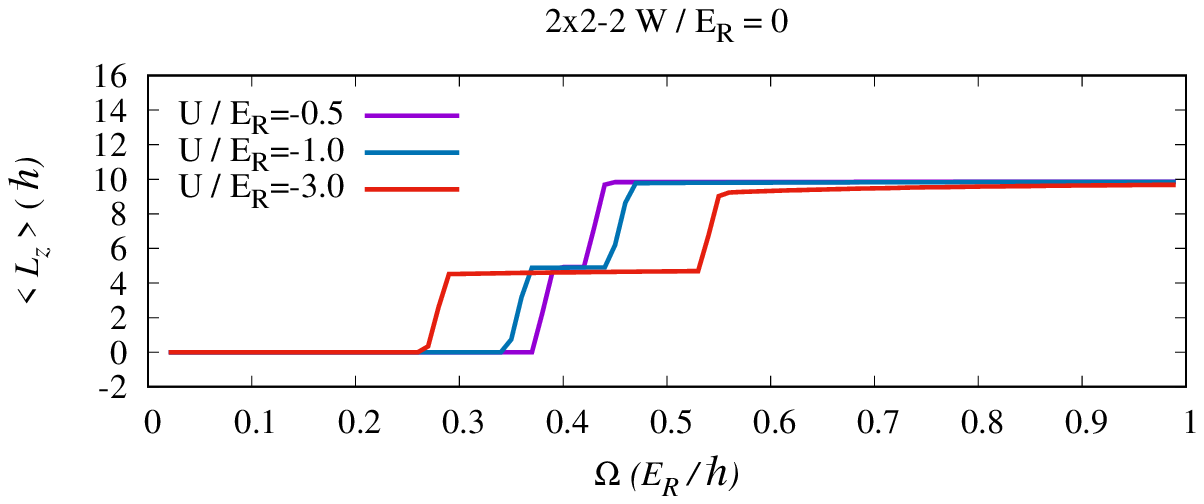}}\\
\vspace{-10 pt}
      \subfloat[]{\includegraphics[scale=.7,keepaspectratio,trim={0 2cm 0 2.5cm},clip]{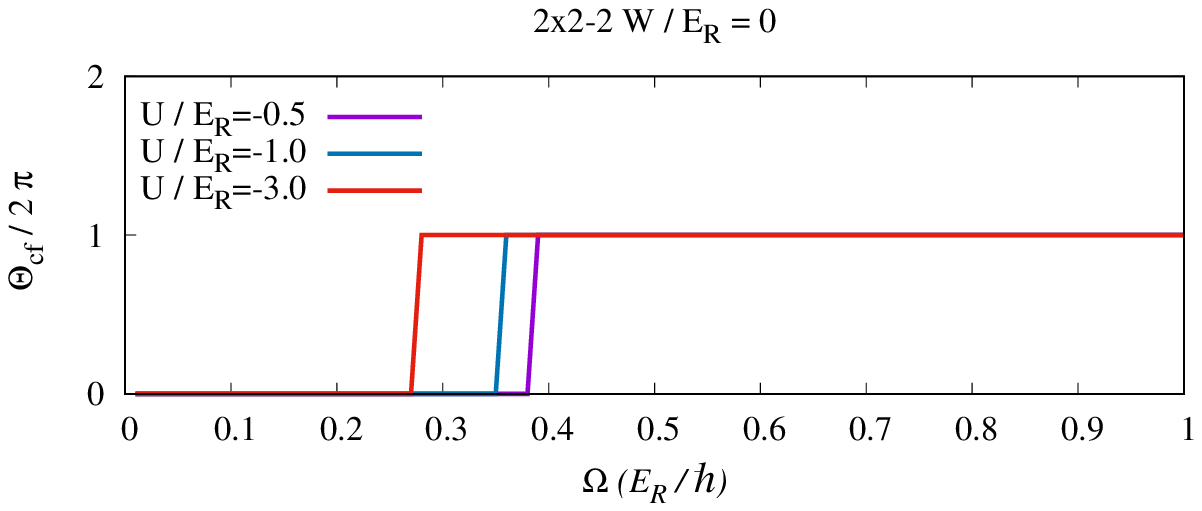}}
\vspace{-5 pt}
      \subfloat[]{\includegraphics[scale=.7,keepaspectratio,trim={2cm 0.3cm 0cm 0.8cm},clip]{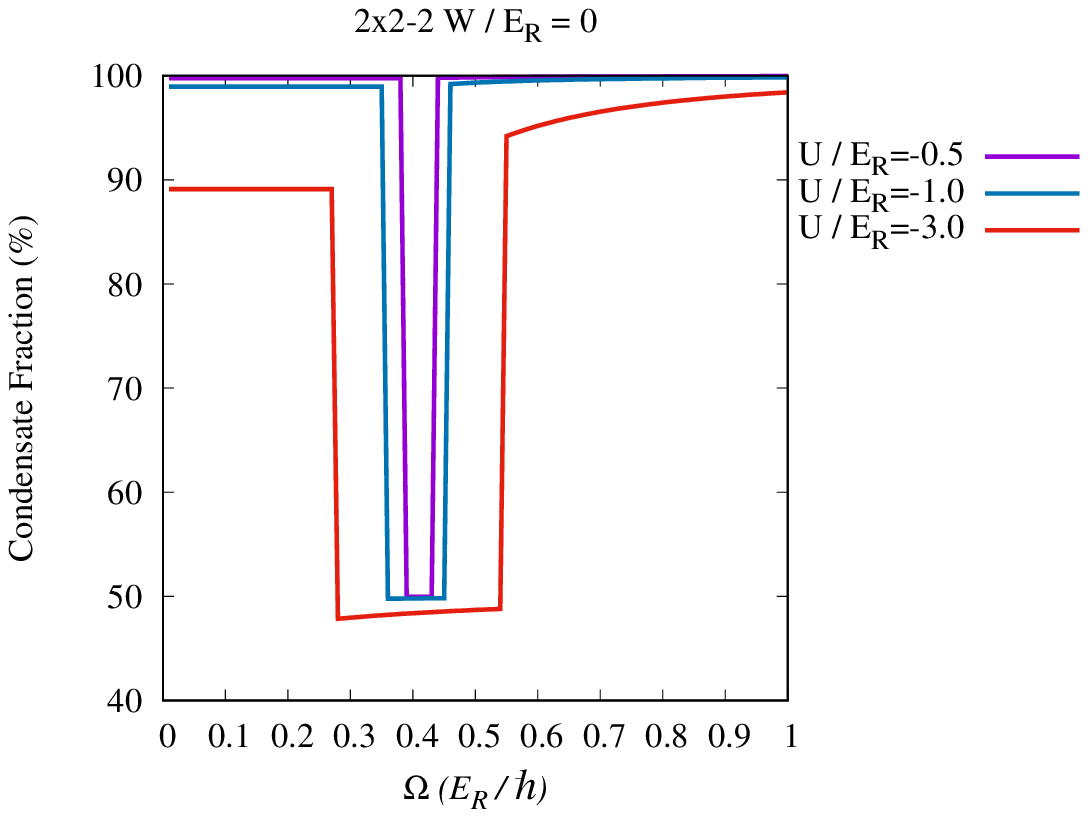}}
  \caption{  $2 \times 2$ lattice for $N=2$ and with $U / E_R =-0.5, -1.0, -3.0$ as  function of $\Omega$. (a) average angular momentum $<L_z>$, (b) phase winding $\Theta_{cf}/2 \pi$, and (c) condensate fraction for $N=2$.}\label{fig:13}
\end{figure}
At first we focus on the effect of attractive interaction between bosons on rotating square $L\times L$ lattice. As discussed above, for a $L \times L$ square lattice we get $(L-1)$ quantized vortices in the lowest Bloch band. Entry of each vortex is signaled by a jump in $<L_z>$, boundary current $\Lambda_C$ and phase winding similar to that for the non-interacting cases as discussed in Section \ref{sec:l4} above. On the other hand, for the attractive two-body interaction, we observed additional jumps in average angular momentum $<L_z>$ as well as boundary current $\Lambda_C$ in addition to the $L-1$ jumps associated with the entry of $L-1$ vortex. This is shown in Fig. \ref{fig:10}(a-b) for a $2 \times 2$ square lattice, for $N=2,3,4,5$. The same results are also obtained for larger lattice sizes. We have shown the figures for smaller lattice ($2\times 2$) since the corresponding figures for larger lattice become very cluttered due to large number of additional jumps in $<L_z>$ and $\Lambda_C$. However, the phase winding do not show any additional jump, thereby showing that the additional jumps in $<L_z>$ and $\Lambda_C$ do not correspond to entry of any additional vortices in the system. This is shown in Fig. \ref{fig:10}(c). The number of jumps increases with number of particles $N$ as shown in Fig. \ref{fig:10}(a-b). Each extra jump do not correspond to entry of a vortex but the normalized variance  $\nu$ [Eq.(\ref{bhm4})]  for such state is $\nu >1$ which denote a phase squeezed state. In contrast, the normalized variance is $\nu < 1$ for the repulsive two-body interaction \cite{bhat_hc2006}. The  normalized variance $\nu$ is shown in Fig. \ref {fig:11} for $U / E_R =-0.5$. The number of extra jumps in $<L_z>$ and $\Lambda_C$ are sensitive to number of particles in system $N$, in  particular for $N$ particles we get $N-1$ extra jumps as shown in Fig. \ref {fig:10}(a-b). For each jump in $<L_z>$ we observe crossing between ground state $E_0$ and first excited $E_1$ energies of the system as shown in Fig. \ref {fig:12} for a $2 \times 2$ lattice, $N=5$ and  $U / E_R =-0.5$. As strength of attractive interactions increases, the difference between the frequencies at which the  extra jumps occur also increases as shown in Fig. \ref{fig:13}(a). However, the required frequency strength for the entry of each vortex decreases  as shown in Fig. \ref{fig:13}(b). The condensate fraction also decreases with increasing strength of the attractive two-body interaction as shown in Fig. \ref{fig:13}(c).
\begin{figure}[!t]
    \centering
         \subfloat[]{\includegraphics[scale=.7,keepaspectratio,trim={0 2cm 0 2.5cm},clip]{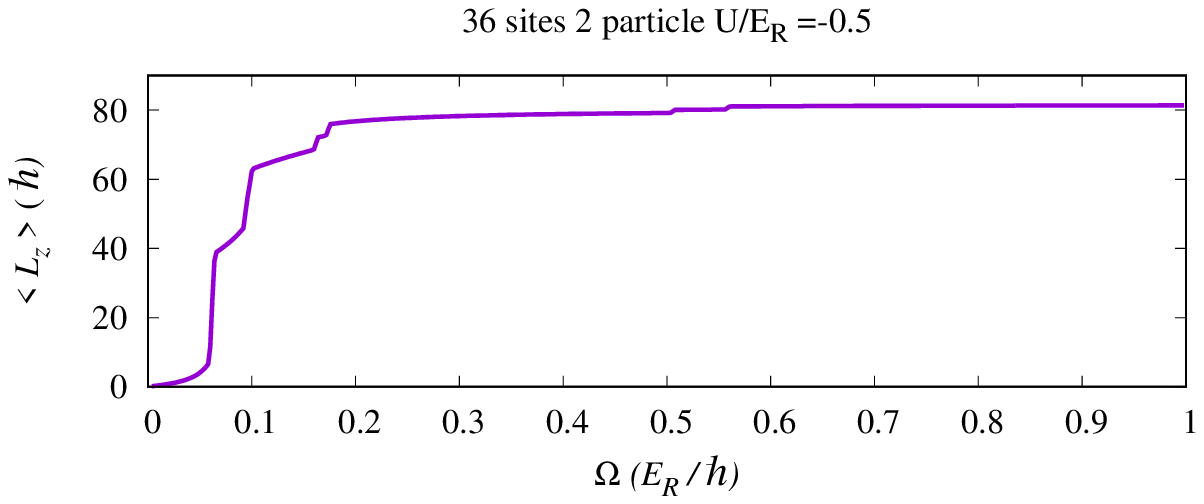}}\\
\vspace{-10 pt}     
      \subfloat[]{\includegraphics[scale=.7,keepaspectratio,trim={0 2cm 0 2.5cm},clip]{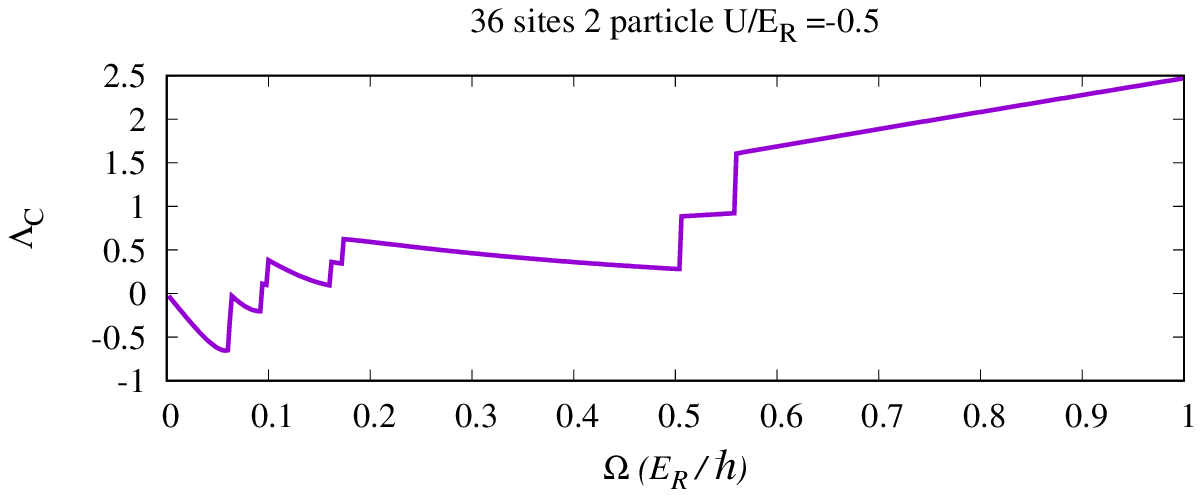}}\\
\vspace{-5pt}
  \caption{ Triangular lattice with $N_s = 36, U / E_R = -0.5, N=2$, (a) average angular momentum $<L_z>$, (d) sum of boundary current $\Lambda_C $.}\label{fig:14}
\end{figure}
\begin{figure}[!h]
    \centering
     \subfloat[]{\includegraphics[scale=.7,keepaspectratio,trim={0 2cm 0cm 2.5cm},clip]{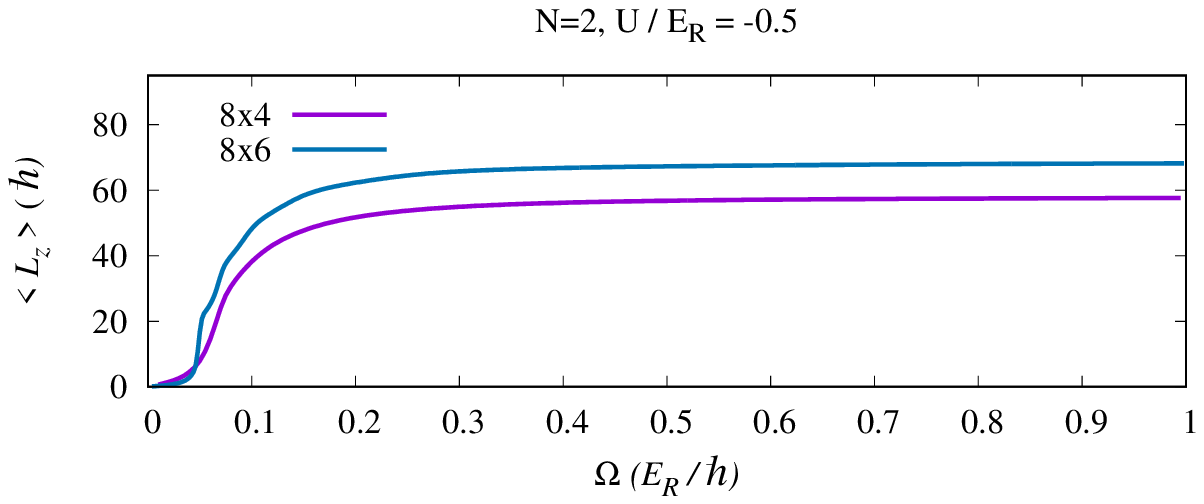}} \\
\vspace{-10 pt}
     \subfloat[]{\includegraphics[scale=.7,keepaspectratio,trim={0 2cm 0 2.5cm},clip]{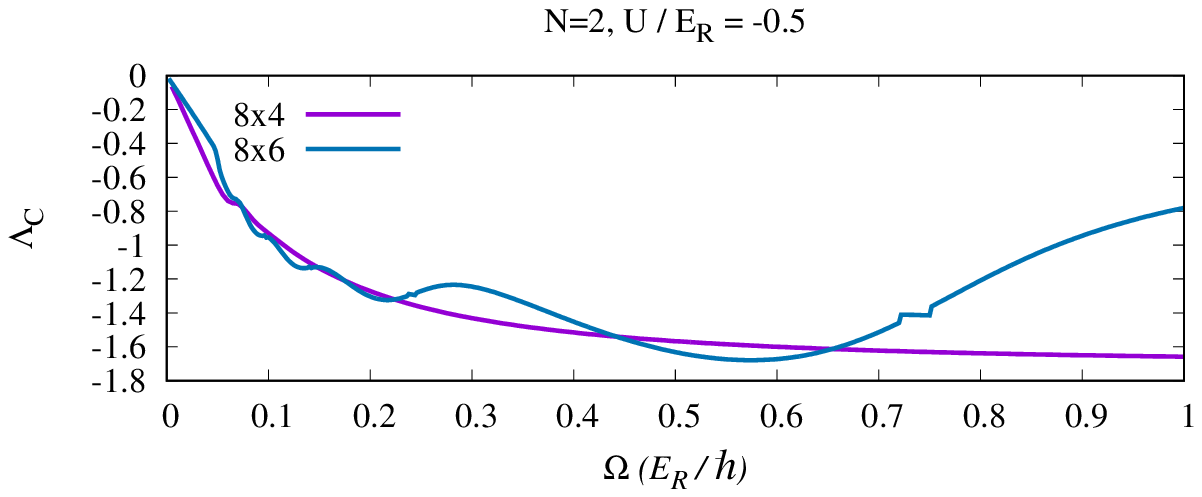}}\\
\vspace{-10 pt}
      \subfloat[]{\includegraphics[scale=.7,keepaspectratio,trim={0cm 0.3cm 0cm 0.8cm},clip]{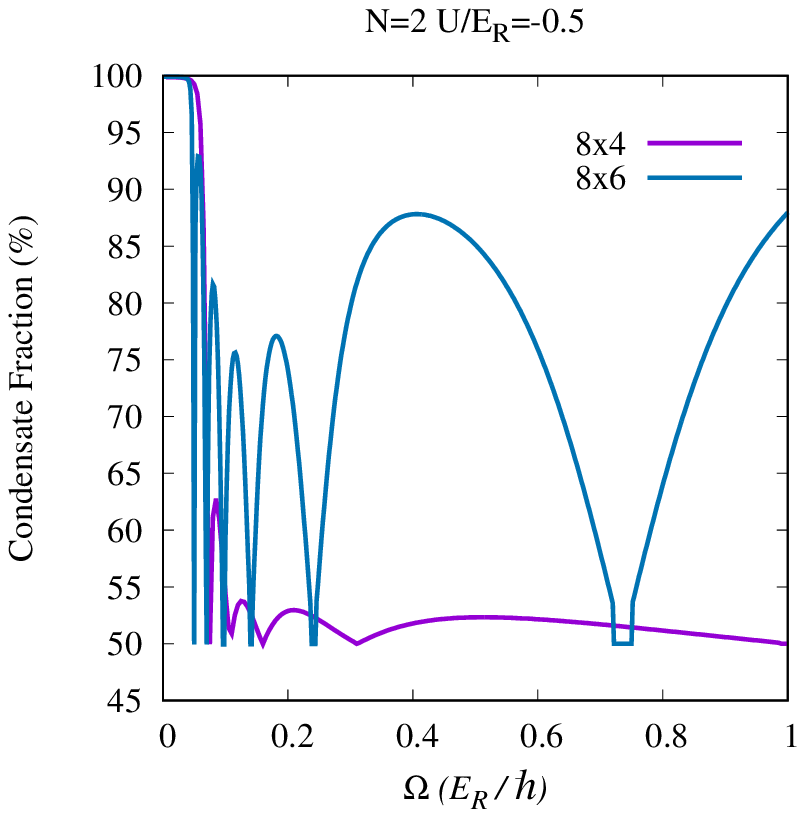}}\\
\vspace{-5 pt}
      \caption{ $U / E_R =-0.5, N=2$ for rectangular lattice, (a) average angular momentum $<L_z>$, (b) sum of boundary current $\Lambda_C $ for rectangular lattice and (c) condensate fraction for $N=2$ and $8 \times4, 8 \times6$ lattices.  }\label{fig:15}
\end{figure}
Similar behavior is also observed in presence of only three-body attractive interaction $W<0, \ U = 0$. However, it require more interaction strength as compared to two-body interaction. For example, for $2\times 2$ and $4\times 4$ square lattices, while we observed the above behavior for two-body interaction strength $U/E_R=-0.5$, for three-body interaction we observed similar behavior for $W /E_R = -2$. Similar behavior as that of the square lattice is also observed for the triangular lattice as shown in Fig. \ref{fig:14}(a-b). However, for rectangular lattice of $L_X \times L_Y$ sites, we observe that the effect of attractive interactions is sensitive to the value of parameter $\epsilon = L_Y / L_X$. For $\epsilon > 0.5$ extra jumps are observed for boundary current $\Lambda_C$ but not for average angular momentum $<L_z>$.  This is shown in Fig. \ref{fig:15}(a-b). Unlike the repulsive two-body interaction case, for the attractive two-body interaction the average angular momentum do not show any discontinuity (jump) but shows continuous behavior with increasing rotational frequency $\Omega$ as shown in Fig. \ref{fig:15}(b). Further, we also observe significant reduction ( $\sim 50\%$) of condensate fraction for $L_Y / L_X \leq 0.6$ as strength of rotation ($\Omega$) increases, and for $L_Y / L_X > 0.6$  the condensate fraction shows increase to all most $90 \%$. This is shown in Fig. \ref{fig:15}(c).  As the strength of the attractive interaction increases the condensate fraction decreases similar to the $L \times L$ square lattice case (Fig. \ref{fig:13}(c)). 
\subsection{\label{sec:l5-3} TWO- AND THREE-BODY INTERACTIONS}
\begin{figure}[!ht]
    \subfloat[]{\includegraphics[scale=.7,keepaspectratio,trim={0 2cm 0 2.5cm},clip]{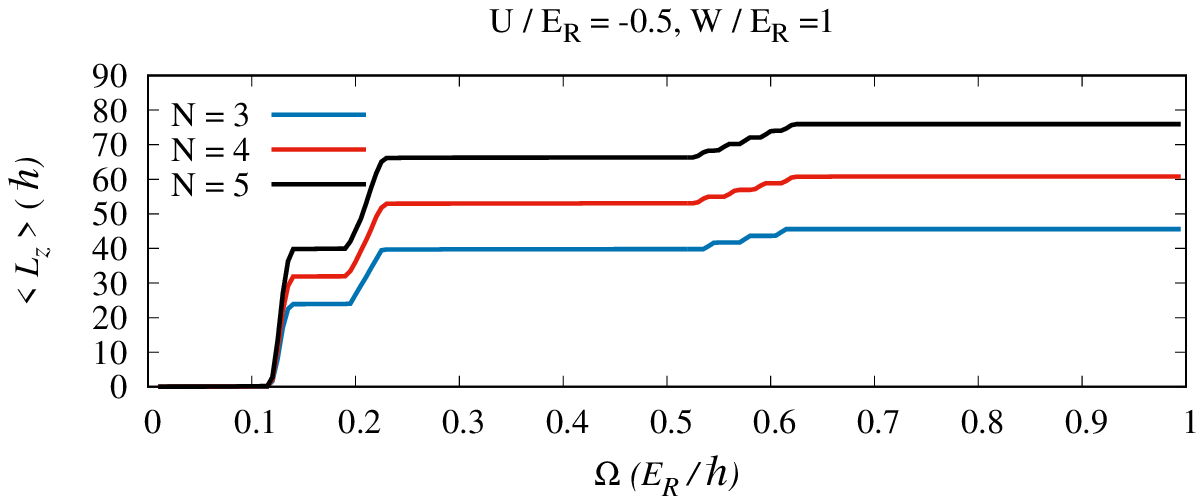}}\\
\vspace{-10pt}
   \subfloat[]{\includegraphics[scale=.7,keepaspectratio,trim={0 2cm 0 2.5cm},clip]{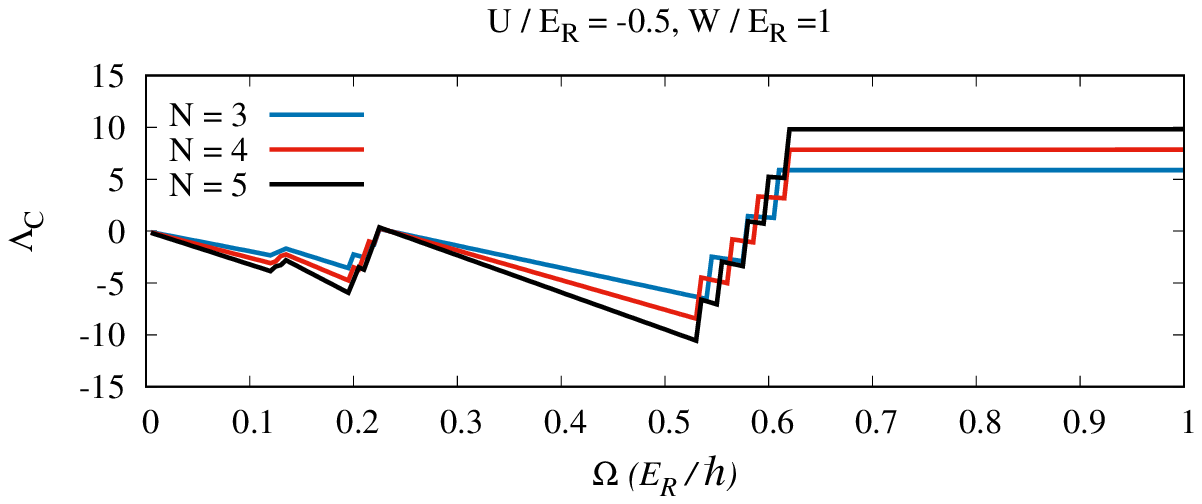}}\\
\vspace{-10pt}      
    \subfloat[]{\includegraphics[scale=.7,keepaspectratio,trim={0 2cm 0 2.5cm},clip]{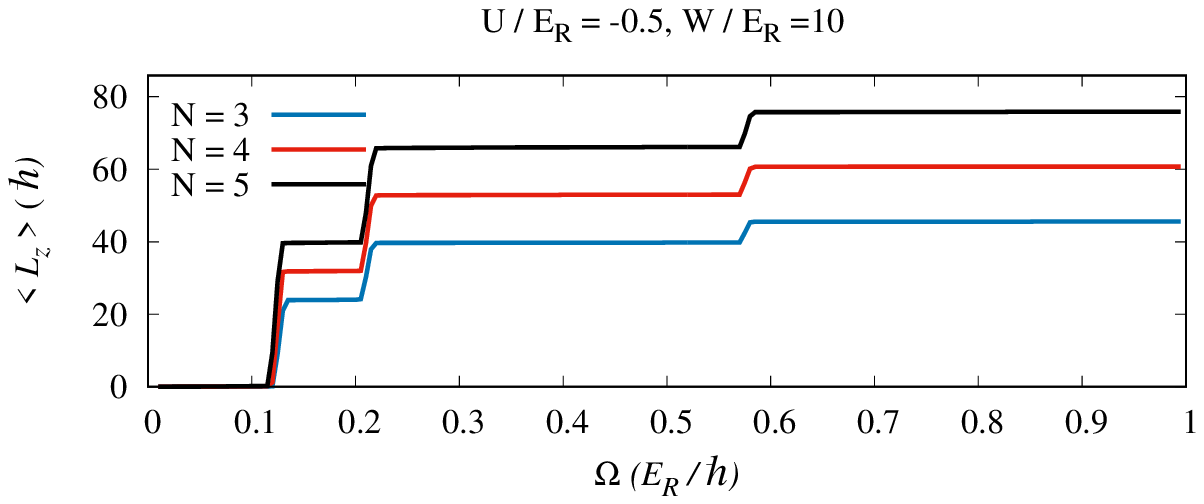}}\\
\vspace{-10pt}      
     \subfloat[]{\includegraphics[scale=.7,keepaspectratio,trim={0 2cm 0 2.5cm},clip]{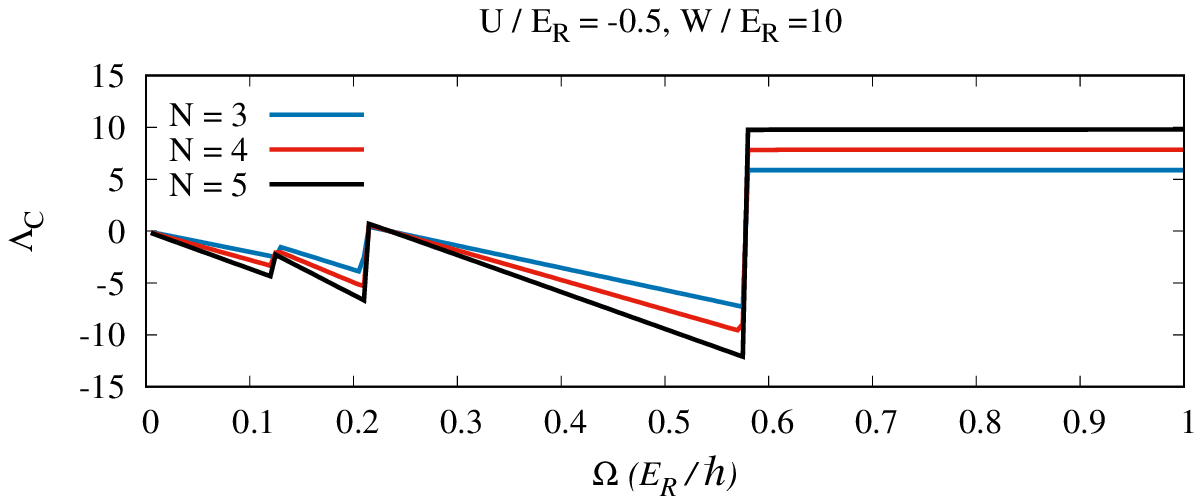}}     
\vspace{-5pt}
  \caption{ $4 \times 4$ lattice for $N=3,4,5$ and for $U / E_R =-0.5$ as a function of $\Omega$ for different values of repulsive three-body interaction. 
(a) average angular momentum $<L_z>$ for $W / E_R = 1$, (b) sum of boundary current $\Lambda_C $ for $W / E_R = 1$, (c) average angular momentum $<L_z>$ for $W / E_R = 10$ and (d) sum of boundary current $\Lambda_C $ for $W / E_R = 10$.}\label{fig:16}
\end{figure}
To study the effects of both two- and three-body interactions on the vortex formation, we consider the case when both two- and three-body interactions are repulsive and also the case when the two-body interaction is attractive but the three-body interaction is repulsive. We consider a $4 \times 4$ square lattice with number of particles $N=3, 4$ and $5$. We choose the strength of the two-body attractive interaction as $U/E_R = -0.5$ and that of the three-body repulsive interactions as $W/E_R = 1 \ {\rm and} \ 10$. When both the two- and three-body interactions are repulsive we get results similar to the case when only two-body repulsive interaction is present, except that the frequency ($\Omega$) at which the vortex enters the systems is lowered. On the other hand, for the case of attractive two-body interaction and weaker strength of the three-body interaction ($W/E_R = 1$) we observe extra discontinuous jumps in average angular momentum $<L_z>$ as well as boundary current $\Lambda_C$ similar to the case of attractive interactions as mentioned above. Similar to the attractive interaction case as mentioned above, the phase winding do not show any additional jump. But when the strength of the three-body interaction is stronger ($W/E_R = 10$) the additional jumps in $<L_z>$  and $\Lambda_C$ disappears (smooth out) and the behavior of $<L_z>$  and $\Lambda_C$ shows jumps only at the entry of each vortices similar to the cases of repulsive two and three-body interactions. This is shown in Fig. \ref{fig:16}.
\begin{figure}[!h]
 \subfloat[]{\includegraphics[scale=.7,keepaspectratio,trim={0 2cm 0cm 2.5cm},clip]{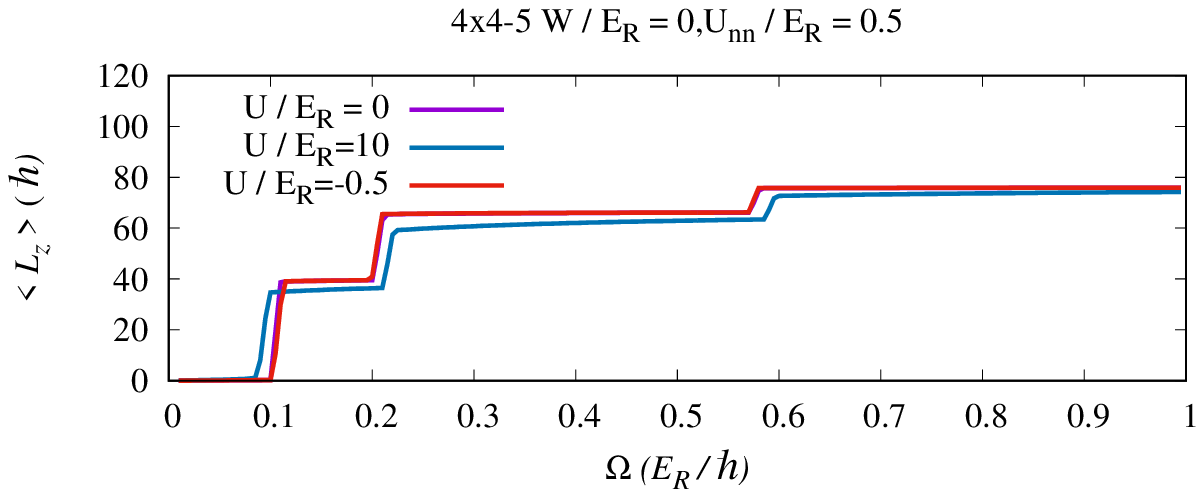}} \\
\vspace{-10pt}
 \subfloat[]{\includegraphics[scale=.7,keepaspectratio,trim={0 2cm 0 2.5cm},clip]{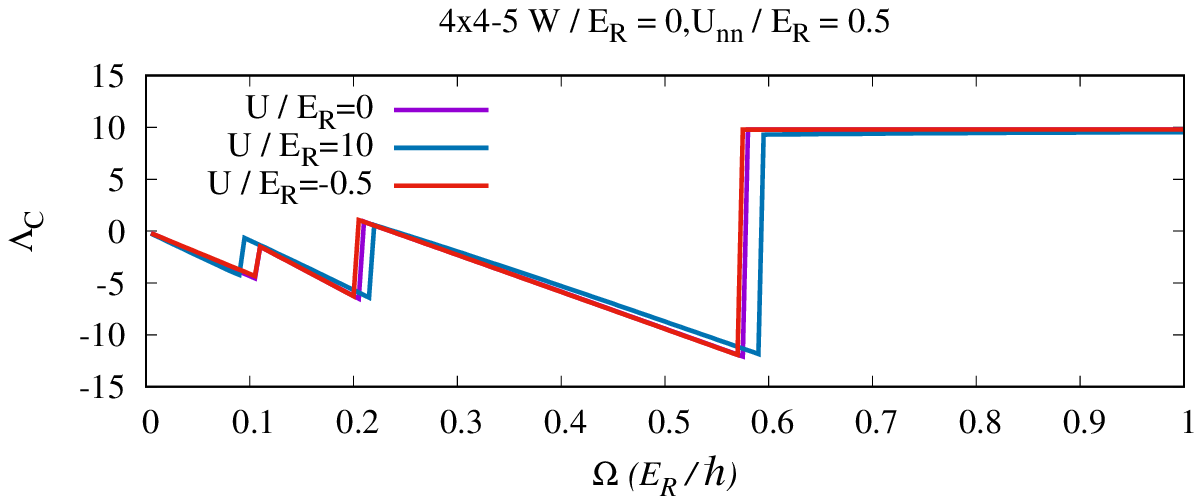}}\\
\vspace{-10pt}
 \subfloat[]{\includegraphics[scale=.7,keepaspectratio,trim={0 2cm 0 2.5cm},clip]{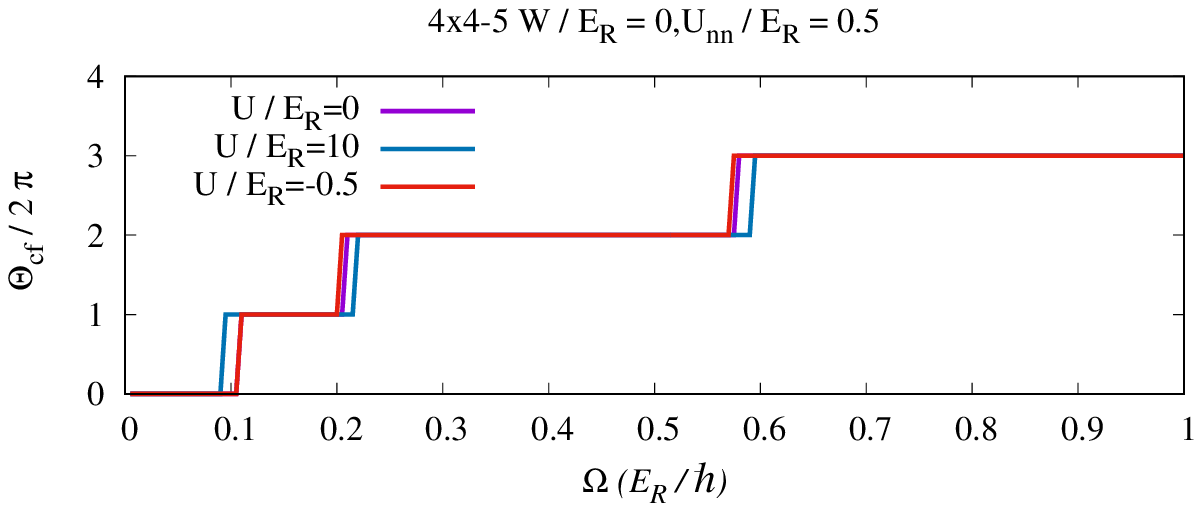}}
\vspace{-5pt}
      \caption{$4 \times 4$ lattice and $N=5$ for different values of two-body interactions for $U_{nn} / E_R =0.5$. (a) average angular momentum $<L_z>$, (b) sum of boundary current $\Lambda_C $ and (c) phase winding $\Theta_{cf}/2 \pi$.}\label{fig:17}
\end{figure}
\subsection{\label{sec:l5-4} NEAREST NEIGHBOR INTERACTIONS ($U_{nn}$)}

To study the effect of the long-range interaction we consider the cases of on site ($U$) two-body and the  nearest neighbor ($U_{nn}$) two-body interactions. We consider a $4 \times 4$ square lattice with number of particles $N=4$, strength of the nearest neighbor interaction as $U_{nn}/E_R = 0.5$ for three different strengths of the two-body interactions $U/E_R = 0$, $U/E_R = 0.5$ and  $U/E_R = 10$. We observe that the results are similar to that of repulsive two-body interactions, except that the frequencies at which the three vortices enters the system changes slightly. This is shown in Fig. \ref{fig:17}. Similarly, we have also checked that results in presence of two- and three-body along with combination of nearest neighbor interaction, and again all the results are consistent, i.e. in presence of attractive two- and three-body interactions and $U_{nn}$, we observe extra jump in $<L_z>$ and $\Lambda_C$ if $|U|>U_{nn}$ and the critical rotational strength require to entry each vortex is reduced. Similar behavior are also observed for the triangular and rectangular lattices.
\section{\label{sec:l6} EFFECT OF HARMONIC TRAP }
\begin{figure}[!h]
    \centering
         \subfloat[]{\includegraphics[scale=.52,keepaspectratio,trim={2.9cm 0.9cm 2cm 1.2cm},clip]{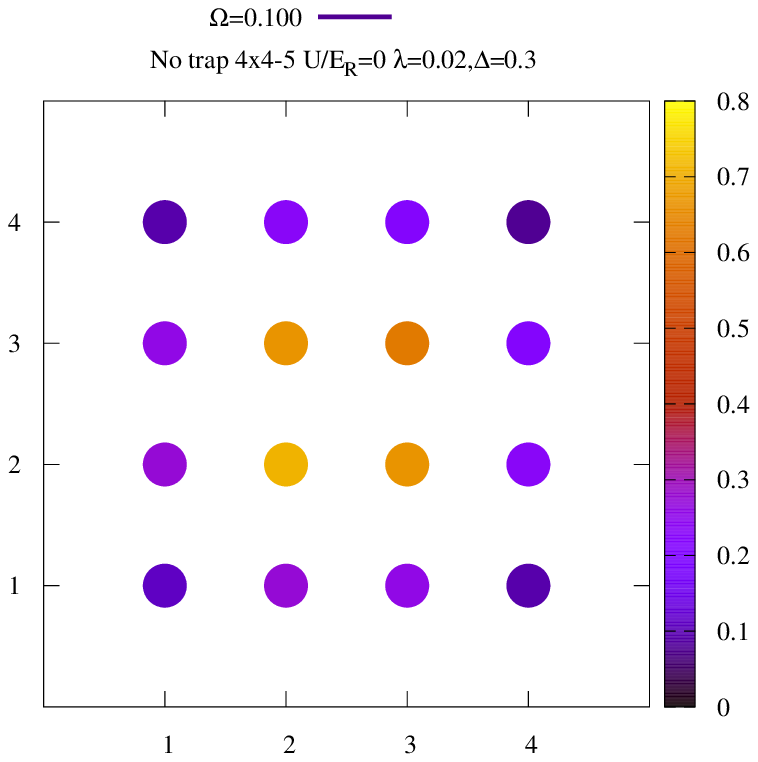}} 
    \subfloat[]{\includegraphics[scale=.52,keepaspectratio,trim={2.9cm 0.9cm 2cm 1.2cm},clip]{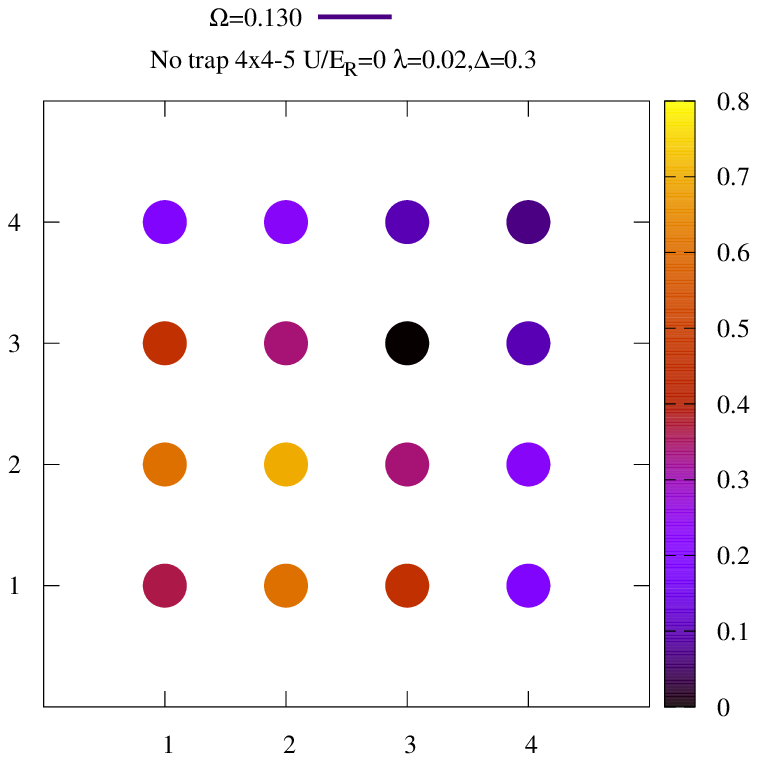}} 
\\
               \vspace{-10pt}
    \subfloat[]{\includegraphics[scale=.52,keepaspectratio,trim={2.9cm 0.9cm 2cm 1.2cm},clip]{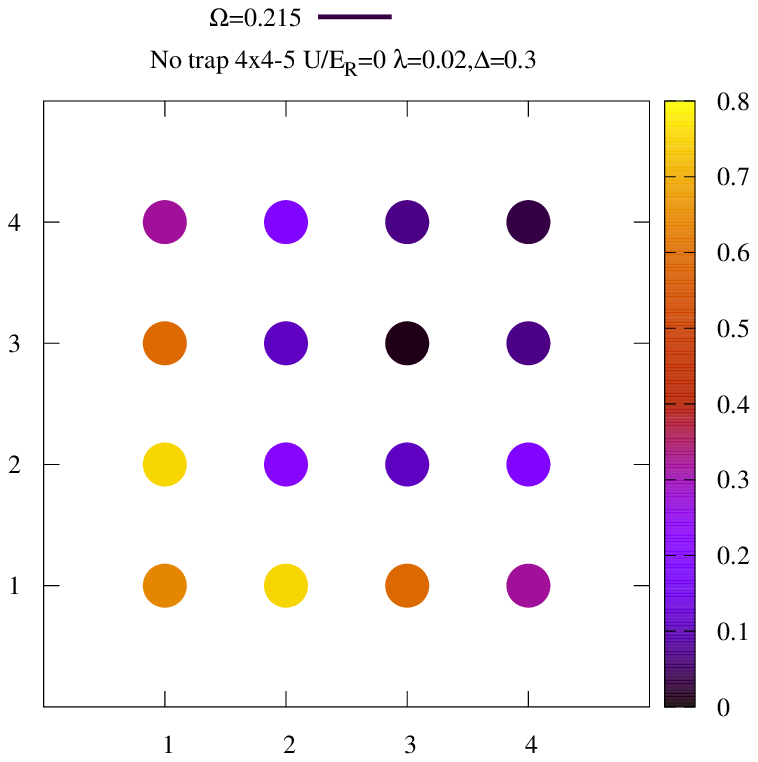}} 
     \subfloat[]{\includegraphics[scale=.52,keepaspectratio,trim={2.9cm 0.9cm 2cm 1.2cm},clip]{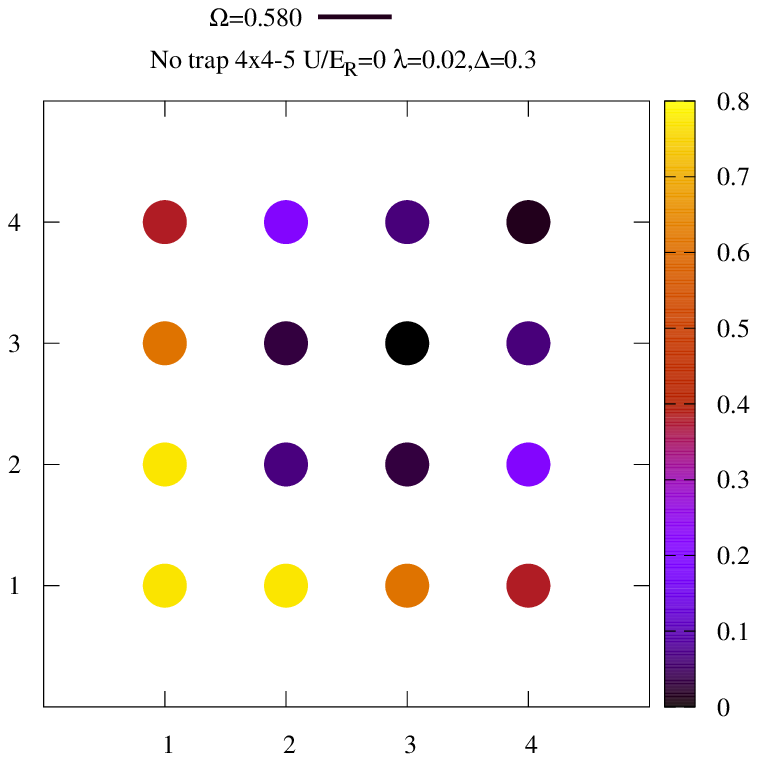}} 
      \vspace{-5pt}
     \caption{density distribution for harmonic trap $4 \times 4, N=5, \lambda =0.02 E_R$, $\Delta_x=\Delta_y=0.3$ and $U / E_R =0$. For (a) $\Omega =0.1$, (b) $\Omega =0.13$, (c) $\Omega =0.215$ and (d) $\Omega =0.58$.}\label{fig:18}
\end{figure}
 \begin{figure}
     \subfloat[]{\includegraphics[scale=.7,keepaspectratio,trim={0 2cm 0cm 2.5cm},clip]{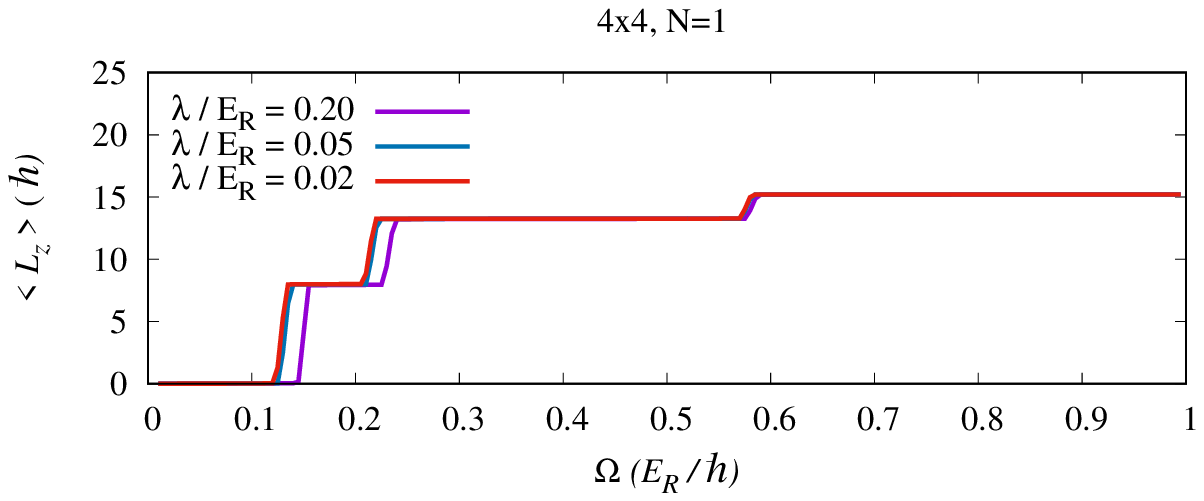}} \\
\vspace{-10pt}
     \subfloat[]{\includegraphics[scale=.7,keepaspectratio,trim={0 1.9cm 0 2.5cm},clip]{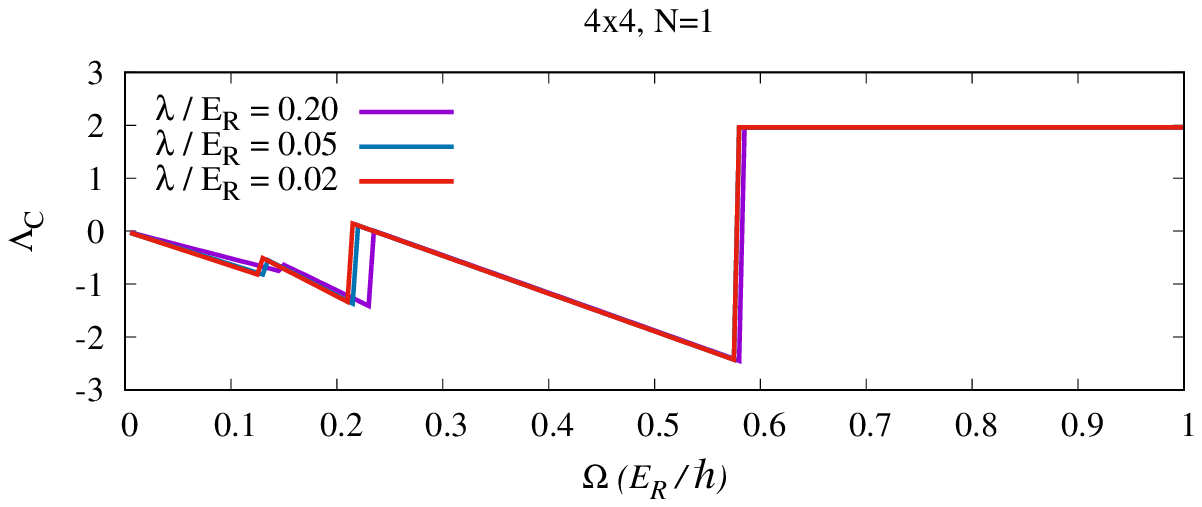}}\\
\vspace{-10pt}
     \subfloat[]{\includegraphics[scale=.7,keepaspectratio,trim={0 1.9cm 0 2.5cm},clip]{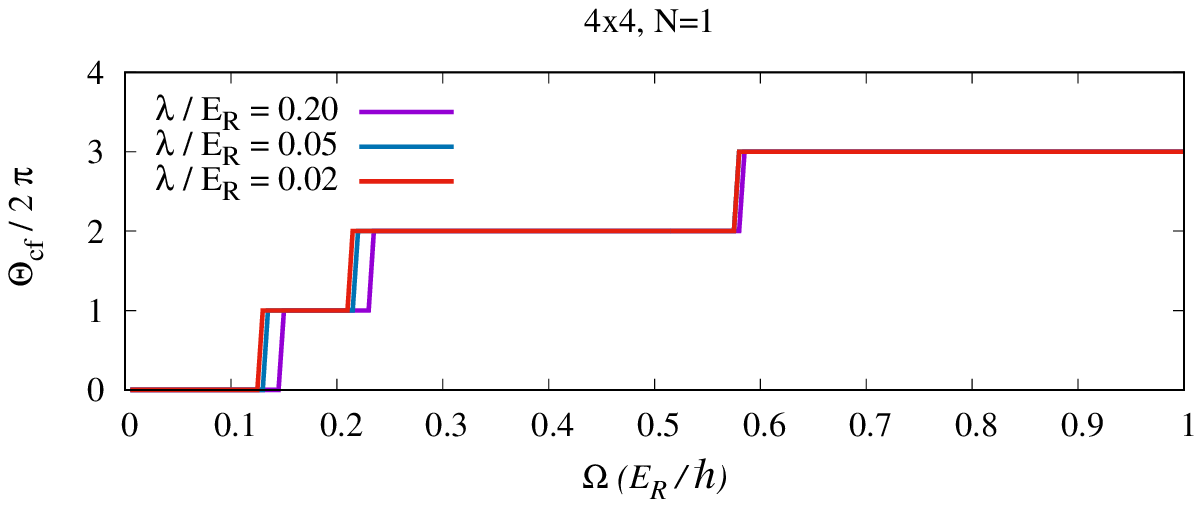}}
\vspace{-5pt}
      \caption{$4 \times 4$ lattice with $N=1$ for different values of $\lambda$. (a) average angular momentum $<L_z>$, (b) sum of boundary current $\Lambda_C $ and (c) phase winding $\Theta_{cf}/2 \pi$.}\label{fig:19}
\end{figure} 
We now consider the effects of the  external harmonic trap potential $\epsilon_i$ [Eq.(\ref{bhm2})] on the quantum vortex states. The harmonic trap  introduces spatial inhomogeneity in the lattice and the bosons at different lattice sites experience different energy offset due to the harmonic trap. For $\Delta =0$ the minimum of the harmonic trap potential coincide with the center of the lattice. 
  \begin{figure}[!h]
      \subfloat[]{\includegraphics[scale=.7,keepaspectratio,trim={0 1.9cm 0cm 2.5cm},clip]{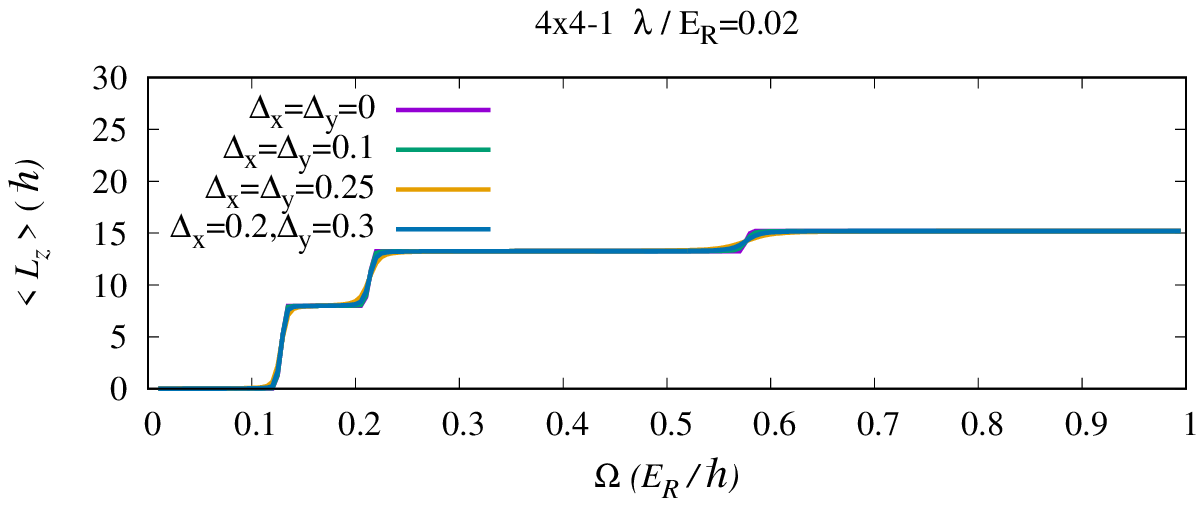}} \\
\vspace{-10pt}
     \subfloat[]{\includegraphics[scale=.7,keepaspectratio,trim={0 1.9cm 0 2.5cm},clip]{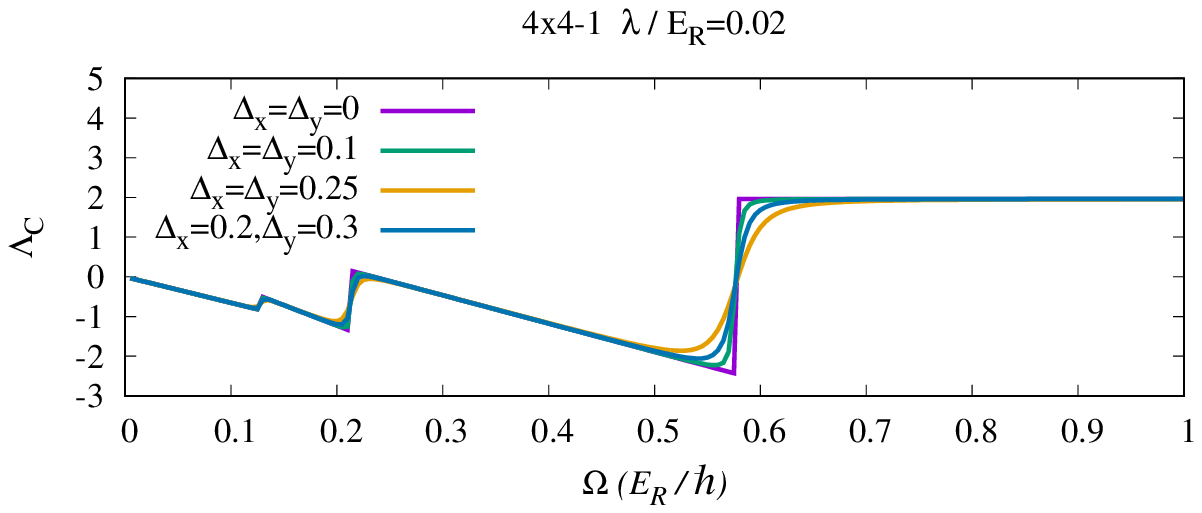}}\\
\vspace{-10pt}
     \subfloat[]{\includegraphics[scale=.7,keepaspectratio,trim={0 1.9cm 0 2.5cm},clip]{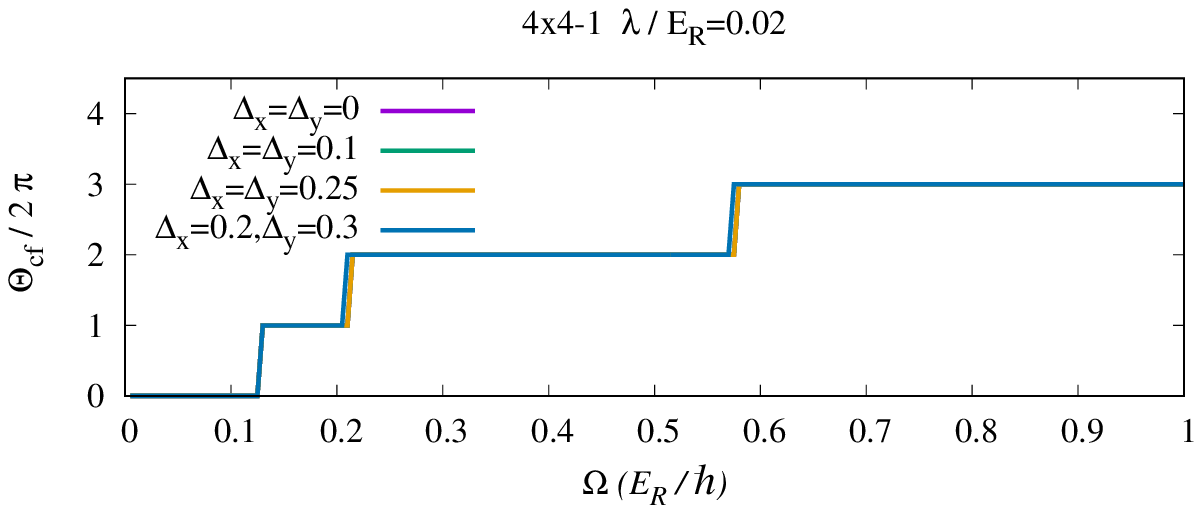}}
\vspace{-5pt}
      \caption{$4 \times 4$ lattice with $N=1, \lambda =0.02 E_R$ for different values of $\Delta$. (a) average angular momentum $<L_z>$, (b) sum of boundary current $\Lambda_C $ and (c) phase winding $\Theta_{cf}/2 \pi$.}\label{fig:20}
\end{figure} 
 We have taken the the axis of rotation perpendicular to the $(x,y)$ plane and passing through the center of lattice. We consider the center of the lattice to be off site and take $-0.5 \leq \Delta_x, \Delta_y \leq 0.5$, such that the minimum of the harmonic trap is also not on any lattice site. For $\Delta =0$, the harmonic trap potential $\epsilon_i$ have same value for all sites which are equidistant from the center of the lattice or the axis of rotation.  For $\Delta \neq 0$, $\epsilon_i$ depend on the position  of the minimum of the harmonic trap w.r.t. the center of lattice. In this case, some lattice sites will be favoured more than others for the particle occupation depending on $\epsilon_i$ and the density distribution will be asymmetric w.r.t. the center of the lattice. For the sites for which the harmonic trap potential $\epsilon_i$ have lower values, the corresponding site number densities $< n_i >$ will be higher \cite{khanore_bdey2015}.   For the choice of parameters $\Delta_x = \Delta_y = 0.30$, the minimum of the harmonic potential is on the left side (third quadrant) close to the center of the lattice  and  $< n_i >$ will be larger for sites closer to the minimum. This is shown in Fig. \ref{fig:18}(a) where we can see that the lattice site in the third quadrant nearest to the center of the lattice have higher site density as compared to the other three nearby sites. There is no vortex state in Fig. \ref{fig:18}(a) as the rotational frequency $\Omega$ is very small. As the frequency $\Omega$ increases, the first vortex enters for $\Omega = 0.13$ and rest at the center as shown in Fig. \ref{fig:18}(b). In contrast to the homogeneous case (no trap potential) where there are equal maximum currents and corresponding equal phase difference of $\pi\over 2$ between the adjacent four sites of the vortex core, for the inhomogeneous case (with trap)  the current between the  adjacent sites are all different due to site dependent occupation  number densities $< n_i >$. However, the total phase difference  between the central four sites is $2\pi$, showing the presence of a single vortex state. The total phase difference is also $2\pi$ between the adjacent sites at the perimeter of the lattice. 
As the rotational frequency increases further, $< n_i >$ changes due to hopping between the neighboring sites induced by rotation. The site density increases at sites in the perimeter of the lattice  and decreases in the center of the lattice. Again, the site density is higher at sites in the third quadrant from the center of the lattice due to imhomogeneity introduced by the harmonic trap potential. For $\Omega = 0.215$ a second vortex enters and sits in the center  with density dip in the four sites at the vortex core as shown in Fig. \ref{fig:18}(c). The total phase difference between the adjacent four sites at the vortex core and also between adjacent sites at the perimeter of the lattice is $4\pi$ showing the presence of the second vortex. Fig. \ref{fig:18}(d) shows the third vortex with similar site density distribution as the second vortex. 
\begin{figure}[!ht]
\includegraphics[scale=.7,keepaspectratio,trim={0 2cm 0 2.5cm},clip]{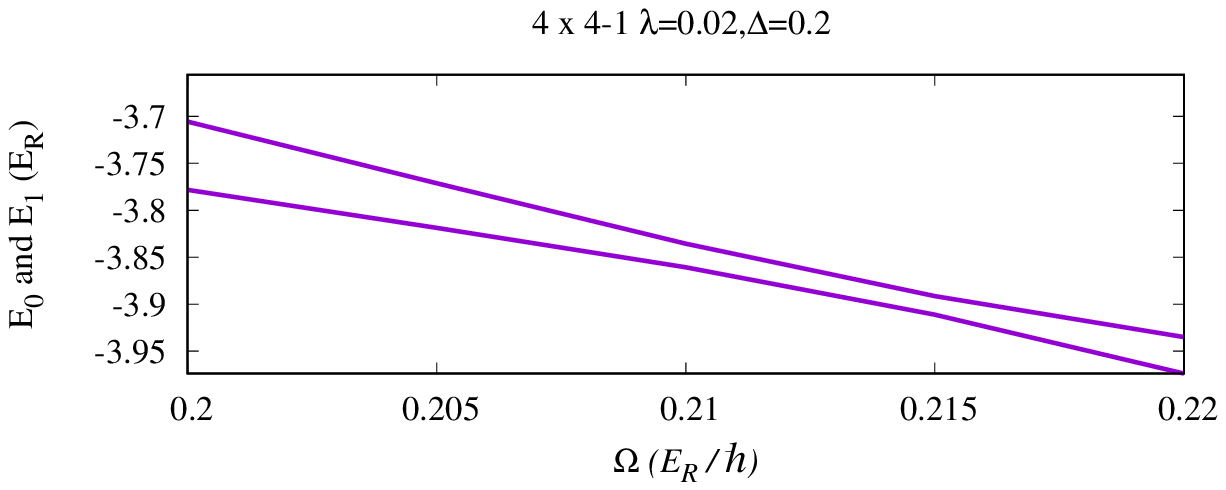}    
\caption{$E_0$ and $E_1$ for $2^{nd}$ vortex in $4 \times 4$ lattice with $N=1, \ \lambda =0.02 E_R$ and $\Delta _x=\Delta_y = 0.2$.}\label{fig:21}
 \end{figure}
The corresponding current distribution shows packing of four vortices with a antivortex at the center and total phase difference between adjacent sites is $6\pi$. There is no change in the number of vortices in presence of the harmonic trap. 
\begin{figure}[!ht]
     \subfloat[]{\includegraphics[scale=.7,keepaspectratio,trim={0 2cm 0cm 2.5cm},clip]{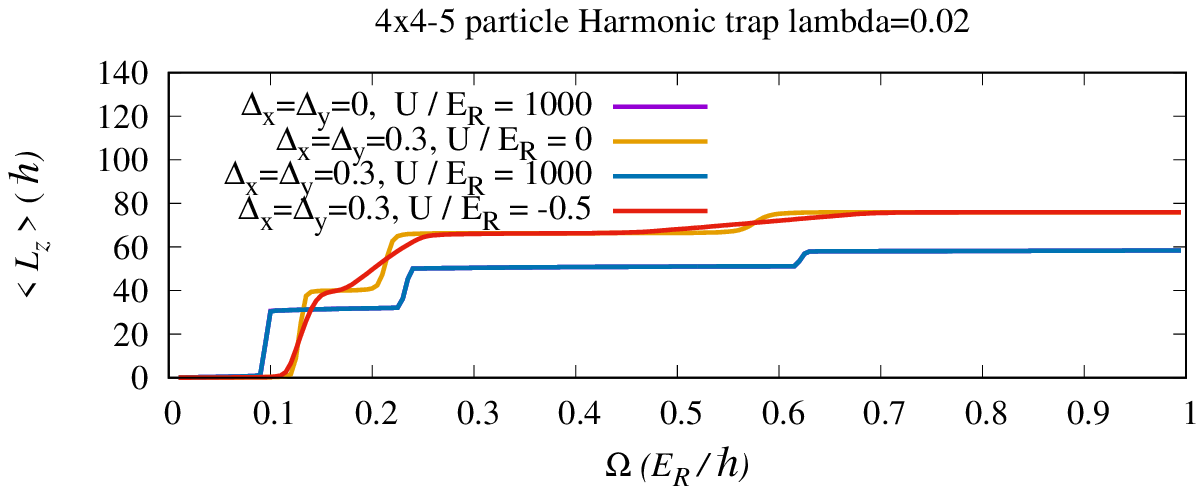}} \\
\vspace{-10pt}
     \subfloat[]{\includegraphics[scale=.7,keepaspectratio,trim={0 2cm 0 2.5cm},clip]{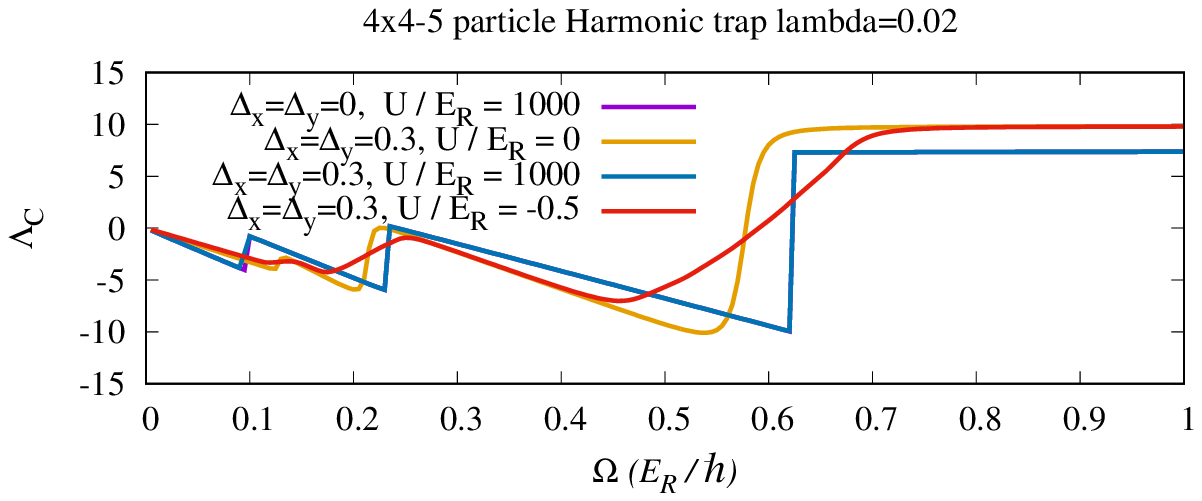}}\\
\vspace{-5pt}
      \caption{$4 \times 4$ lattice with $N=5, \lambda =0.02 E_R$ for different values of $\Delta$ and $U / E_R$.  (a) Average angular momentum $<L_z>$  and (b) sum of boundary current $\Lambda_C $.
}\label{fig:22}
\end{figure}
However,  We observe that for single particle case ($N = 1$) and for $\Delta=0$, the  discrete rotational frequencies at which the vortices enters the system increases with increase in $\lambda$ value as shown in Fig. \ref{fig:19}. The increase in rotational frequency is due to the fact that the harmonic trap acts as an additional pinning potential which adds up to the pinning potential due to the optical lattice. For fixed $\lambda$ and increasing  value of $\Delta$,  
\begin{figure}[!ht]
    \centering
         \subfloat[]{\includegraphics[scale=.52,keepaspectratio,trim={2.9cm 0.9cm 2cm 1.2cm},clip]{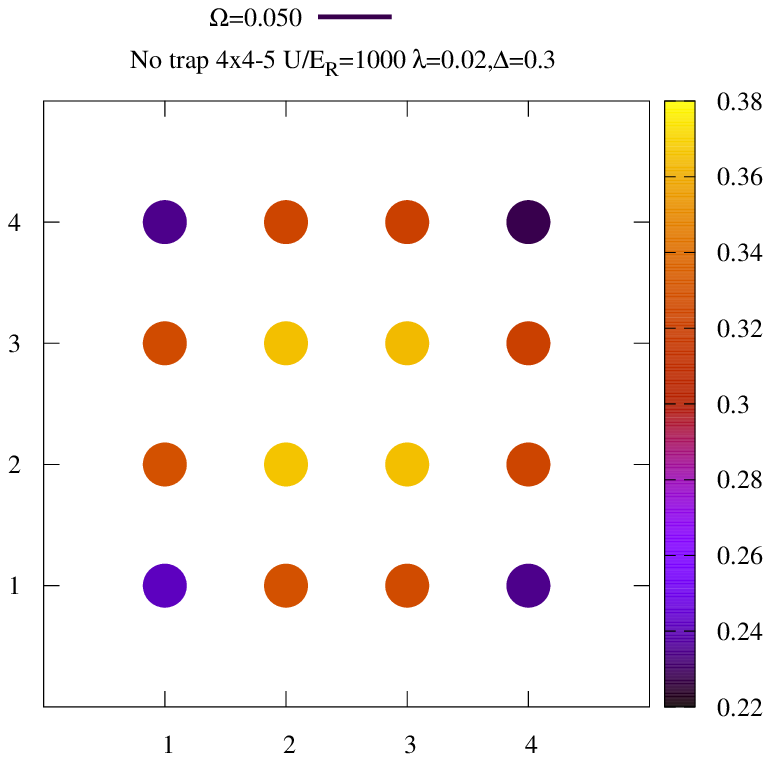}} 
    \subfloat[]{\includegraphics[scale=.52,keepaspectratio,trim={2.9cm 0.9cm 2cm 1.2cm},clip]{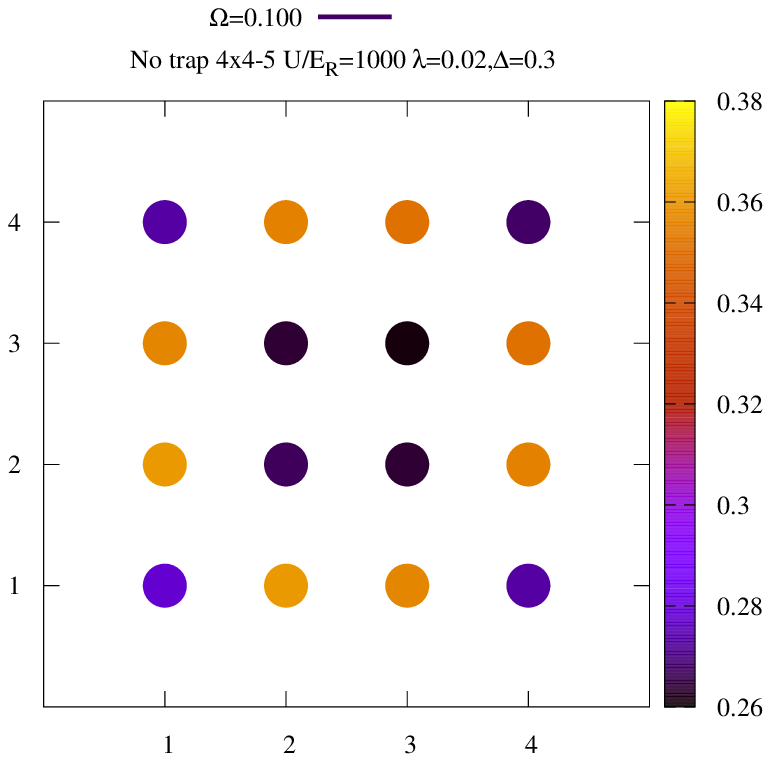}} 
\\
               \vspace{-10pt}
    \subfloat[]{\includegraphics[scale=.52,keepaspectratio,trim={2.9cm 0.9cm 2cm 1.2cm},clip]{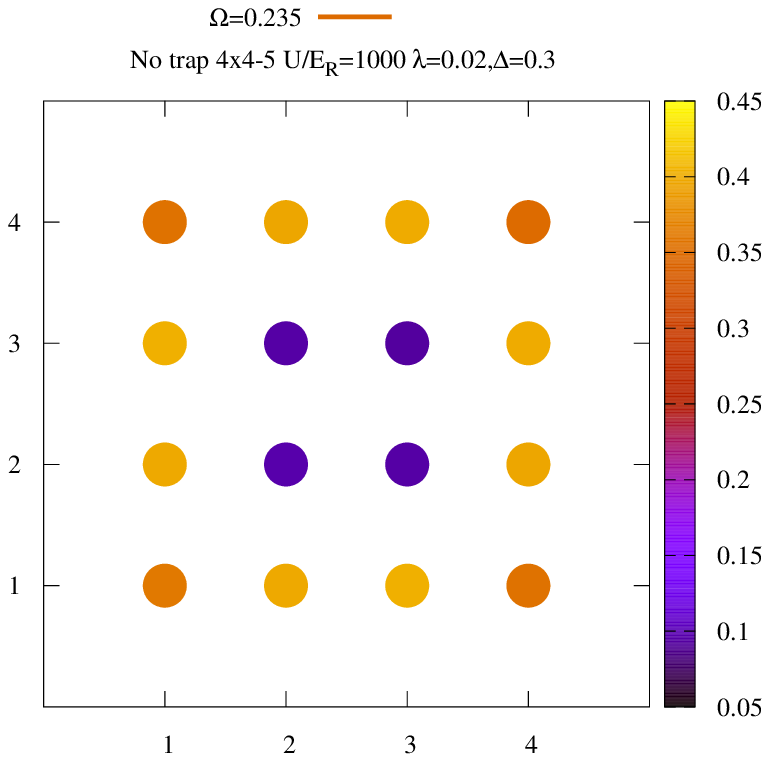}} 
     \subfloat[]{\includegraphics[scale=.52,keepaspectratio,trim={2.9cm 0.9cm 2cm 1.2cm},clip]{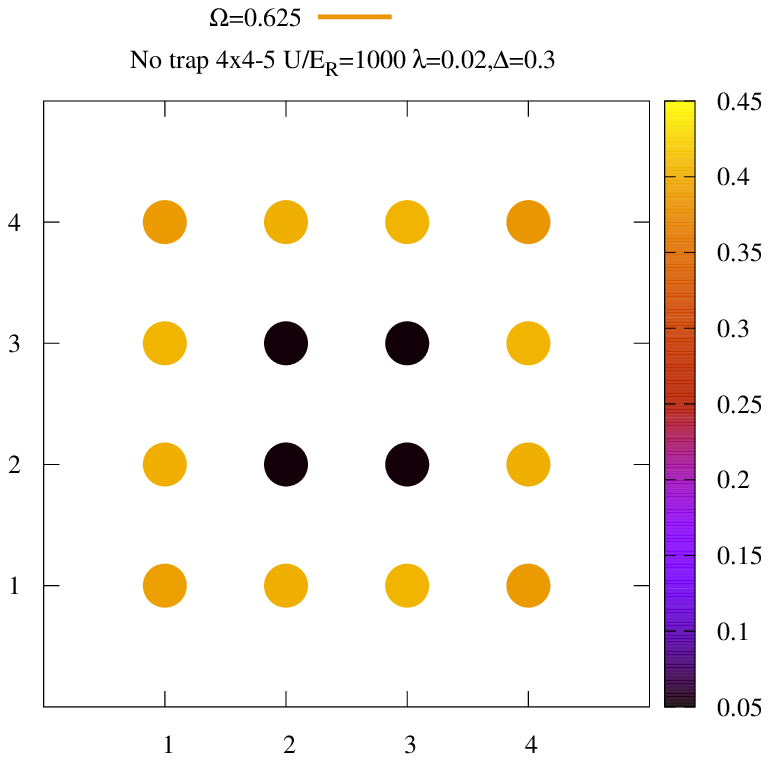}} 
      \vspace{-5pt}
     \caption{density distribution for harmonic trap $4 \times 4, N=5, \lambda =0.02 E_R$ and $\Delta_x=\Delta_y=0.3, U / E_R =1000$. For (a) $\Omega =0.05$, (b) $\Omega =0.1$, (c) $\Omega =0.235$ and (d) $\Omega =0.625$.}\label{fig:23}
\end{figure}
the observed sharp discontinuous jumps in average angular momentum $<L_z>$ and boundary current $\Lambda_C$  in absence of the harmonic trap, become smoother as shown in Fig. \ref{fig:20}. There is also avoided crossing between the ground state energy $E_0$ and the first excited state energy $E_1$ at all rotational frequencies as shown in Fig. \ref{fig:21}. \\
\begin{figure}[!t]
    \centering
         \subfloat[]{\includegraphics[scale=.52,keepaspectratio,trim={2.9cm 0.9cm 2cm 1.2cm},clip]{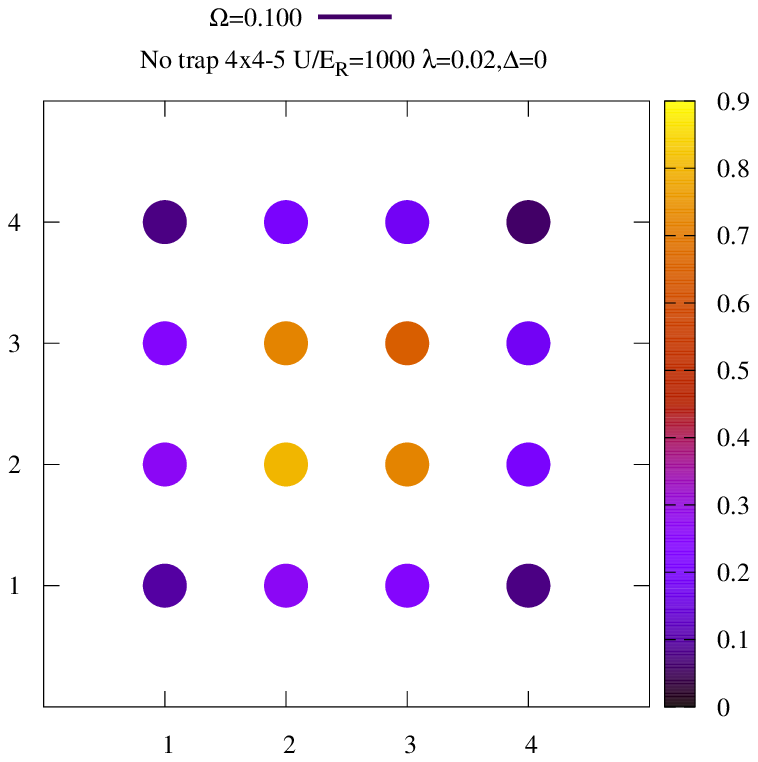}} 
    \subfloat[]{\includegraphics[scale=.52,keepaspectratio,trim={2.9cm 0.9cm 2cm 1.2cm},clip]{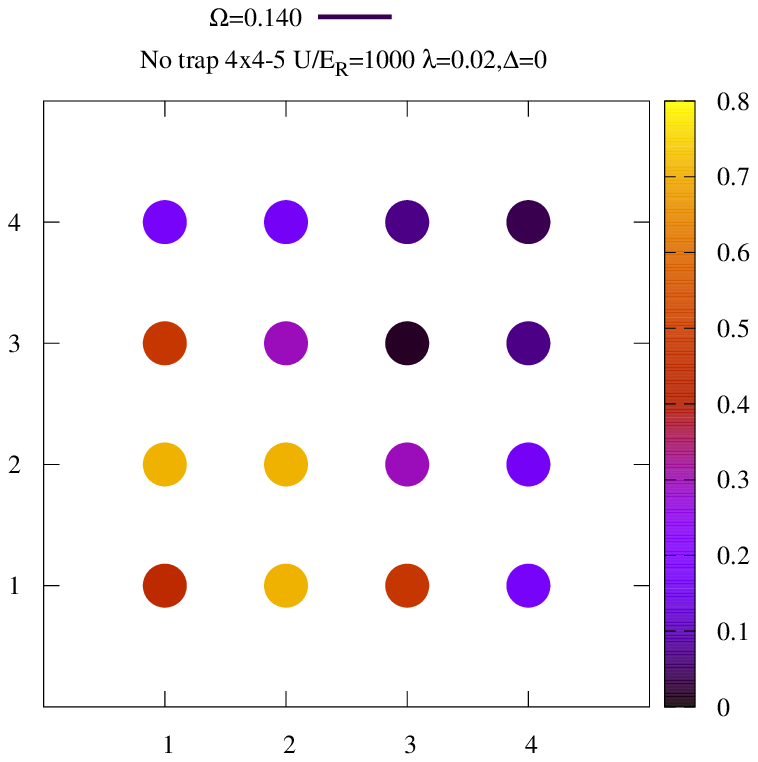}} 
\\
               \vspace{-10pt}
    \subfloat[]{\includegraphics[scale=.52,keepaspectratio,trim={2.9cm 0.9cm 2cm 1.2cm},clip]{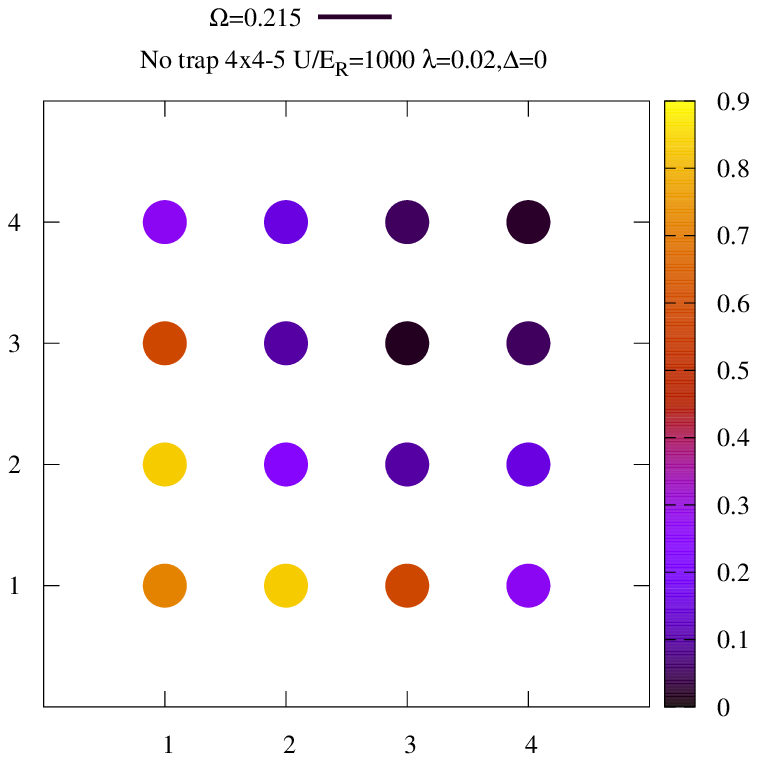}} 
     \subfloat[]{\includegraphics[scale=.52,keepaspectratio,trim={2.9cm 0.9cm 2cm 1.2cm},clip]{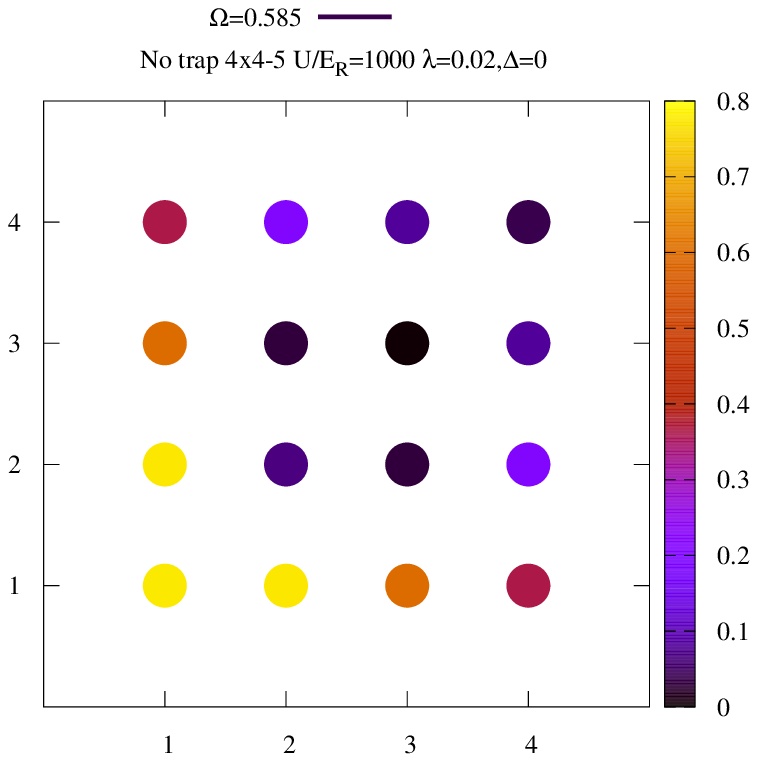}} 
      \vspace{-5pt}
     \caption{density distribution for harmonic trap $4 \times 4, N=5, \lambda =0.02 E_R$ and $\Delta_x=\Delta_y=0.3, U / E_R =-0.5$. For (a) $\Omega =0.1$, (b) $\Omega =0.14$, (c) $\Omega =0.215$ and (d) $\Omega =0.585$.}\label{fig:24}
\end{figure}
For many particles, the effect of two- and three-body interactions  produces vortex states which are similar to the homogeneous case for $\Delta = 0$. For $\Delta \neq 0$ and repulsive two- and three-body interactions, the avoided crossings between the energy levels are not observed even though the average angular momentum $<L_z>$ and boundary current $\Lambda_C$  become smoother as shown in Fig. \ref{fig:22}. Also, unlike the homogeneous case for attractive two- and three-body interactions, additional jumps in the $<L_z>$ and boundary current $\Lambda_C$  are not observed for the  inhomogenous case i.e. in presence of harmonic trap potential. The normalized invariance is however greater than one similar to the homogeneous case. The corresponding site density distribution is shown in Fig. \ref{fig:23} which shows vortex states similar to the homogeneous case for large two-body interactions.
Also,  avoided crossing between $E_0$ and $E_1$ is not observed for $U>0$ and $W>0$ case. However, for attractive interaction ($U<0$ and $W<0$),  we observe avoided crossing but no extra jumps in observable as in  homogeneous case as shown in Fig. \ref{fig:22}. Fig. \ref{fig:23} shows the effect of strong two-body interaction between the particles on the density and Fig. \ref{fig:24} shows the effect of attractive interaction between the particles on the density.\\ 
\section{\label{sec:l7} CONCLUSIONS}
In conclusion, we have studied the quantum vortex states of the strongly interacting bosons on  two-dimensional rotating optical lattice at zero temperature. In particular we have examined the effects of the  lattice geometry and the spatial inhomogeneity introduced by the additional harmonic trap potential on the characteristics of the quantum vortex states. We have shown that the rotation introduces quantum vortex states of different symmetry  at discrete rotation frequencies which are accompanied by jumps of $2\pi$ in the phase winding of the states. The transition between these states are indicated by the crossing of the energy levels. The maximum number of quantized vortices depends on the geometry of the  two-dimensional lattice. It is observed that the maximum phase difference between the neighboring sites depends on the number of sites and also on the lattice geometry. However, the maximum phase winding is $2\pi$ times the maximum number of vortices, irrespective of the lattice geometry, i.e.  square , rectangular or triangular. The variation of the average angular momentum $< L_z >$ with rotational frequency $\Omega$ also depends on the geometry of the lattice. Unlike the square lattice, the $< L_z >$ do not show sharp discontinuous jumps for the rectangular lattice and the triangular lattice. However, for the rectangular lattice, the discontinuous jumps become sharper with increasing value of the asymmetry parameter $\epsilon$ and for $\epsilon = 1$ (square lattice), it approaches the square lattice behaviour. The  effect of  the two- and three-body interactions  between the bosons  also depends on the geometry of the lattice as the current flow or the lattice current depends on the interactions. Particle-hole symmetry is observed for strong repulsive two-body interaction irrespective of the lattice geometry.  For attractive two- as well as three-body interactions, the average angular momentum $< L_z >$ and the total boundary current $\Lambda_C$ shows additional jumps in addition to the jumps associated with the entry of the vortices in the rotating lattice. However, the corresponding phase winding do not show any additional jumps showing that these jumps are not associated with entry of new vortices in the system. The corresponding normalized variance $\nu > 1$ shows that these are phase squeezed state. The number of extra jumps in $< L_z >$ and $\Lambda_C$ are sensitive to the number of particles. Similar behaviour is also observed for the triangular lattice. However, for the rectangular lattice, the effect of the attractive interactions is sensitive to the parameter $\epsilon$. For some value of the parameter $\epsilon$, the boundary current $\Lambda_C$ displays extra jumps but the average angular momentum $< L_z >$ shows continuous behaviour with increasing rotational frequency $\Omega$. The condensate fraction also shows dependence on the parameter $\epsilon$. In presence of both two- and three-body interactions, the results depends on the the nature of the interactions. When both the two- and three-body interactions are repulsive, we get results similar to the two-body repulsive interaction case, except that the rotational frequencies at which the vortices enters the system is lowered. However, when the two-body interaction is attractive and the three-body interaction is repulsive, then for weak three body repulsive interaction we get results similar to the attractive two-body interaction case. On the other hand, for stronger three-body repulsive interaction, the results are similar to the repulsive two-body interaction case. The presence of nearest neighbor interaction do not change the results except that the rotational frequencies at which the vortices enters the system changes slightly. These results are expected to have important implications for understanding the parallels between atoms in a rotating optical lattice and electrons in the presence of a magnetic field, in particular interpreting experiments on fractional quantum Hall effect for a 2D condensate in  confining  rotating optical lattice and co-rotating  harmonic trap potential.
\section{\label{sec:l8}Acknowledgment}
BD thanks SERB-DST, Government of India  for financial assistance through a research project, Grant No. EMR/2016/002627.
%
\bibliography{bhm_ref-mar-19}

\begin{thebibliography}{35}%
\makeatletter
\providecommand \@ifxundefined [1]{%
 \@ifx{#1\undefined}
}%
\providecommand \@ifnum [1]{%
 \ifnum #1\expandafter \@firstoftwo
 \else \expandafter \@secondoftwo
 \fi
}%
\providecommand \@ifx [1]{%
 \ifx #1\expandafter \@firstoftwo
 \else \expandafter \@secondoftwo
 \fi
}%
\providecommand \natexlab [1]{#1}%
\providecommand \enquote  [1]{``#1''}%
\providecommand \bibnamefont  [1]{#1}%
\providecommand \bibfnamefont [1]{#1}%
\providecommand \citenamefont [1]{#1}%
\providecommand \href@noop [0]{\@secondoftwo}%
\providecommand \href [0]{\begingroup \@sanitize@url \@href}%
\providecommand \@href[1]{\@@startlink{#1}\@@href}%
\providecommand \@@href[1]{\endgroup#1\@@endlink}%
\providecommand \@sanitize@url [0]{\catcode `\\12\catcode `\$12\catcode
  `\&12\catcode `\#12\catcode `\^12\catcode `\_12\catcode `\%12\relax}%
\providecommand \@@startlink[1]{}%
\providecommand \@@endlink[0]{}%
\providecommand \url  [0]{\begingroup\@sanitize@url \@url }%
\providecommand \@url [1]{\endgroup\@href {#1}{\urlprefix }}%
\providecommand \urlprefix  [0]{URL }%
\providecommand \Eprint [0]{\href }%
\providecommand \doibase [0]{http://dx.doi.org/}%
\providecommand \selectlanguage [0]{\@gobble}%
\providecommand \bibinfo  [0]{\@secondoftwo}%
\providecommand \bibfield  [0]{\@secondoftwo}%
\providecommand \translation [1]{[#1]}%
\providecommand \BibitemOpen [0]{}%
\providecommand \bibitemStop [0]{}%
\providecommand \bibitemNoStop [0]{.\EOS\space}%
\providecommand \EOS [0]{\spacefactor3000\relax}%
\providecommand \BibitemShut  [1]{\csname bibitem#1\endcsname}%
\let\auto@bib@innerbib\@empty
\bibitem [{\citenamefont {Fisher}\ \emph {et~al.}(1989)\citenamefont {Fisher},
  \citenamefont {Weichman}, \citenamefont {Grinstein},\ and\ \citenamefont
  {Fisher}}]{fisher1989}%
  \BibitemOpen
  \bibfield  {author} {\bibinfo {author} {\bibfnamefont {M.~P.~A.}\
  \bibnamefont {Fisher}}, \bibinfo {author} {\bibfnamefont {P.~B.}\
  \bibnamefont {Weichman}}, \bibinfo {author} {\bibfnamefont {G.}~\bibnamefont
  {Grinstein}}, \ and\ \bibinfo {author} {\bibfnamefont {D.~S.}\ \bibnamefont
  {Fisher}},\ }\href {\doibase 10.1103/PhysRevB.40.546} {\bibfield  {journal}
  {\bibinfo  {journal} {Phys. Rev. B}\ }\textbf {\bibinfo {volume} {40}},\
  \bibinfo {pages} {546} (\bibinfo {year} {1989})}\BibitemShut {NoStop}%
\bibitem [{\citenamefont {Greiner}\ \emph {et~al.}(2002)\citenamefont
  {Greiner}, \citenamefont {Mandel}, \citenamefont {Esslinger}, \citenamefont
  {H{\"{a}}nsch},\ and\ \citenamefont {Bloch}}]{Greiner2002}%
  \BibitemOpen
  \bibfield  {author} {\bibinfo {author} {\bibfnamefont {M.}~\bibnamefont
  {Greiner}}, \bibinfo {author} {\bibfnamefont {O.}~\bibnamefont {Mandel}},
  \bibinfo {author} {\bibfnamefont {T.}~\bibnamefont {Esslinger}}, \bibinfo
  {author} {\bibfnamefont {T.~W.}\ \bibnamefont {H{\"{a}}nsch}}, \ and\
  \bibinfo {author} {\bibfnamefont {I.}~\bibnamefont {Bloch}},\ }\href
  {\doibase 10.1038/415039a} {\bibfield  {journal} {\bibinfo  {journal}
  {Nature}\ }\textbf {\bibinfo {volume} {415}},\ \bibinfo {pages} {39}
  (\bibinfo {year} {2002})}\BibitemShut {NoStop}%
\bibitem [{\citenamefont {Sakmann}\ \emph {et~al.}(2010)\citenamefont
  {Sakmann}, \citenamefont {Streltsov}, \citenamefont {Alon},\ and\
  \citenamefont {Cederbaum}}]{sakmann_sacl2010}%
  \BibitemOpen
  \bibfield  {author} {\bibinfo {author} {\bibfnamefont {K.}~\bibnamefont
  {Sakmann}}, \bibinfo {author} {\bibfnamefont {A.~I.}\ \bibnamefont
  {Streltsov}}, \bibinfo {author} {\bibfnamefont {O.~E.}\ \bibnamefont {Alon}},
  \ and\ \bibinfo {author} {\bibfnamefont {L.~S.}\ \bibnamefont {Cederbaum}},\
  }\href {\doibase 10.1103/PhysRevA.82.013620} {\bibfield  {journal} {\bibinfo
  {journal} {Phys. Rev. A}\ }\textbf {\bibinfo {volume} {82}},\ \bibinfo
  {pages} {013620} (\bibinfo {year} {2010})}\BibitemShut {NoStop}%
\bibitem [{\citenamefont {Jack}\ and\ \citenamefont
  {Yamashita}(2005)}]{jack_y2005}%
  \BibitemOpen
  \bibfield  {author} {\bibinfo {author} {\bibfnamefont {M.~W.}\ \bibnamefont
  {Jack}}\ and\ \bibinfo {author} {\bibfnamefont {M.}~\bibnamefont
  {Yamashita}},\ }\href {\doibase 10.1103/PhysRevA.71.023610} {\bibfield
  {journal} {\bibinfo  {journal} {Phys. Rev. A}\ }\textbf {\bibinfo {volume}
  {71}},\ \bibinfo {pages} {023610} (\bibinfo {year} {2005})}\BibitemShut
  {NoStop}%
\bibitem [{\citenamefont {Khanore}\ and\ \citenamefont
  {Dey}(2014)}]{khanore_bdey2014}%
  \BibitemOpen
  \bibfield  {author} {\bibinfo {author} {\bibfnamefont {M.}~\bibnamefont
  {Khanore}}\ and\ \bibinfo {author} {\bibfnamefont {B.}~\bibnamefont {Dey}},\
  }\href
  {http://scitation.aip.org/content/aip/proceeding/aipcp/10.1063/1.4872521}
  {\bibfield  {journal} {\bibinfo  {journal} {AIP Conference Proceedings}\
  }\textbf {\bibinfo {volume} {1591}} (\bibinfo {year} {2014})}\BibitemShut
  {NoStop}%
\bibitem [{\citenamefont {Khanore}\ and\ \citenamefont
  {Dey}(2015)}]{khanore_bdey2015}%
  \BibitemOpen
  \bibfield  {author} {\bibinfo {author} {\bibfnamefont {M.}~\bibnamefont
  {Khanore}}\ and\ \bibinfo {author} {\bibfnamefont {B.}~\bibnamefont {Dey}},\
  }\href
  {http://scitation.aip.org/content/aip/proceeding/aipcp/10.1063/1.4917608}
  {\bibfield  {journal} {\bibinfo  {journal} {AIP Conference Proceedings}\
  }\textbf {\bibinfo {volume} {1665}},\ \bibinfo {eid} {030033} (\bibinfo
  {year} {2015})}\BibitemShut {NoStop}%
\bibitem [{\citenamefont {Tschischik}\ and\ \citenamefont
  {Haque}(2015)}]{wladimir_m2015}%
  \BibitemOpen
  \bibfield  {author} {\bibinfo {author} {\bibfnamefont {W.}~\bibnamefont
  {Tschischik}}\ and\ \bibinfo {author} {\bibfnamefont {M.}~\bibnamefont
  {Haque}},\ }\href {\doibase 10.1103/PhysRevA.91.053607} {\bibfield  {journal}
  {\bibinfo  {journal} {Phys. Rev. A}\ }\textbf {\bibinfo {volume} {91}},\
  \bibinfo {pages} {053607} (\bibinfo {year} {2015})}\BibitemShut {NoStop}%
\bibitem [{\citenamefont {Ziń}\ \emph {et~al.}(2008)\citenamefont {Ziń},
  \citenamefont {Chwedeńczuk}, \citenamefont {Oleś}, \citenamefont {Sacha},\
  and\ \citenamefont {Trippenbach}}]{zin_cost2008}%
  \BibitemOpen
  \bibfield  {author} {\bibinfo {author} {\bibfnamefont {P.}~\bibnamefont
  {Ziń}}, \bibinfo {author} {\bibfnamefont {J.}~\bibnamefont {Chwedeńczuk}},
  \bibinfo {author} {\bibfnamefont {B.}~\bibnamefont {Oleś}}, \bibinfo
  {author} {\bibfnamefont {K.}~\bibnamefont {Sacha}}, \ and\ \bibinfo {author}
  {\bibfnamefont {M.}~\bibnamefont {Trippenbach}},\ }\href
  {http://stacks.iop.org/0295-5075/83/i=6/a=64007} {\bibfield  {journal}
  {\bibinfo  {journal} {EPL}\ }\textbf {\bibinfo {volume} {83}},\ \bibinfo
  {pages} {64007} (\bibinfo {year} {2008})}\BibitemShut {NoStop}%
\bibitem [{\citenamefont {Mark}\ \emph {et~al.}(2012)\citenamefont {Mark},
  \citenamefont {Haller}, \citenamefont {Lauber}, \citenamefont {Danzl},
  \citenamefont {Janisch}, \citenamefont {B{\"u}chler}, \citenamefont {Daley},\
  and\ \citenamefont {N{\"a}gerl}}]{mark_hldjbdn2012}%
  \BibitemOpen
  \bibfield  {author} {\bibinfo {author} {\bibfnamefont {M.~J.}\ \bibnamefont
  {Mark}}, \bibinfo {author} {\bibfnamefont {E.}~\bibnamefont {Haller}},
  \bibinfo {author} {\bibfnamefont {K.}~\bibnamefont {Lauber}}, \bibinfo
  {author} {\bibfnamefont {J.~G.}\ \bibnamefont {Danzl}}, \bibinfo {author}
  {\bibfnamefont {A.}~\bibnamefont {Janisch}}, \bibinfo {author} {\bibfnamefont
  {H.~P.}\ \bibnamefont {B{\"u}chler}}, \bibinfo {author} {\bibfnamefont
  {A.~J.}\ \bibnamefont {Daley}}, \ and\ \bibinfo {author} {\bibfnamefont
  {H.-C.}\ \bibnamefont {N{\"a}gerl}},\ }\href {\doibase
  10.1103/PhysRevLett.108.215302} {\bibfield  {journal} {\bibinfo  {journal}
  {Phys. Rev. Lett.}\ }\textbf {\bibinfo {volume} {108}},\ \bibinfo {pages}
  {215302} (\bibinfo {year} {2012})}\BibitemShut {NoStop}%
\bibitem [{\citenamefont {Sowi{\' n}ski}\ \emph {et~al.}(2015)\citenamefont
  {Sowi{\' n}ski}, \citenamefont {Chhajlany}, \citenamefont {Dutta},
  \citenamefont {Tagliacozzo},\ and\ \citenamefont
  {Lewenstein}}]{sowinski_crdtlm2015}%
  \BibitemOpen
  \bibfield  {author} {\bibinfo {author} {\bibfnamefont {T.}~\bibnamefont
  {Sowi{\' n}ski}}, \bibinfo {author} {\bibfnamefont {R.~W.}\ \bibnamefont
  {Chhajlany}}, \bibinfo {author} {\bibfnamefont {O.}~\bibnamefont {Dutta}},
  \bibinfo {author} {\bibfnamefont {L.}~\bibnamefont {Tagliacozzo}}, \ and\
  \bibinfo {author} {\bibfnamefont {M.}~\bibnamefont {Lewenstein}},\ }\href
  {\doibase 10.1103/PhysRevA.92.043615} {\bibfield  {journal} {\bibinfo
  {journal} {Phys. Rev. A}\ }\textbf {\bibinfo {volume} {92}},\ \bibinfo
  {pages} {043615} (\bibinfo {year} {2015})}\BibitemShut {NoStop}%
\bibitem [{\citenamefont {Sowi{\' n}ski}(2012)}]{sowinski2012}%
  \BibitemOpen
  \bibfield  {author} {\bibinfo {author} {\bibfnamefont {T.}~\bibnamefont
  {Sowi{\' n}ski}},\ }\href {\doibase 10.1103/PhysRevA.85.065601} {\bibfield
  {journal} {\bibinfo  {journal} {Phys. Rev. A}\ }\textbf {\bibinfo {volume}
  {85}},\ \bibinfo {pages} {065601} (\bibinfo {year} {2012})}\BibitemShut
  {NoStop}%
\bibitem [{\citenamefont {Safavi-Naini}\ \emph {et~al.}(2012)\citenamefont
  {Safavi-Naini}, \citenamefont {von Stecher}, \citenamefont
  {Capogrosso-Sansone},\ and\ \citenamefont
  {Rittenhouse}}]{safavi-naini_scrs2012}%
  \BibitemOpen
  \bibfield  {author} {\bibinfo {author} {\bibfnamefont {A.}~\bibnamefont
  {Safavi-Naini}}, \bibinfo {author} {\bibfnamefont {J.}~\bibnamefont {von
  Stecher}}, \bibinfo {author} {\bibfnamefont {B.}~\bibnamefont
  {Capogrosso-Sansone}}, \ and\ \bibinfo {author} {\bibfnamefont {S.~T.}\
  \bibnamefont {Rittenhouse}},\ }\href {\doibase
  10.1103/PhysRevLett.109.135302} {\bibfield  {journal} {\bibinfo  {journal}
  {Phys. Rev. Lett.}\ }\textbf {\bibinfo {volume} {109}},\ \bibinfo {pages}
  {135302} (\bibinfo {year} {2012})}\BibitemShut {NoStop}%
\bibitem [{\citenamefont {Arwas}\ \emph {et~al.}(2014)\citenamefont {Arwas},
  \citenamefont {Vardi},\ and\ \citenamefont {Cohen}}]{arwas_vc2014}%
  \BibitemOpen
  \bibfield  {author} {\bibinfo {author} {\bibfnamefont {G.}~\bibnamefont
  {Arwas}}, \bibinfo {author} {\bibfnamefont {A.}~\bibnamefont {Vardi}}, \ and\
  \bibinfo {author} {\bibfnamefont {D.}~\bibnamefont {Cohen}},\ }\href
  {\doibase 10.1103/PhysRevA.89.013601} {\bibfield  {journal} {\bibinfo
  {journal} {Phys. Rev. A}\ }\textbf {\bibinfo {volume} {89}},\ \bibinfo
  {pages} {013601} (\bibinfo {year} {2014})}\BibitemShut {NoStop}%
\bibitem [{\citenamefont {Thomas}\ \emph {et~al.}(2017)\citenamefont {Thomas},
  \citenamefont {Barter}, \citenamefont {Leung}, \citenamefont {Okano},
  \citenamefont {Jo}, \citenamefont {Guzman}, \citenamefont {Kimchi},
  \citenamefont {Vishwanath},\ and\ \citenamefont {Stamper-Kurn}}]{thomas2017}%
  \BibitemOpen
  \bibfield  {author} {\bibinfo {author} {\bibfnamefont {C.~K.}\ \bibnamefont
  {Thomas}}, \bibinfo {author} {\bibfnamefont {T.~H.}\ \bibnamefont {Barter}},
  \bibinfo {author} {\bibfnamefont {T.-H.}\ \bibnamefont {Leung}}, \bibinfo
  {author} {\bibfnamefont {M.}~\bibnamefont {Okano}}, \bibinfo {author}
  {\bibfnamefont {G.-B.}\ \bibnamefont {Jo}}, \bibinfo {author} {\bibfnamefont
  {J.}~\bibnamefont {Guzman}}, \bibinfo {author} {\bibfnamefont
  {I.}~\bibnamefont {Kimchi}}, \bibinfo {author} {\bibfnamefont
  {A.}~\bibnamefont {Vishwanath}}, \ and\ \bibinfo {author} {\bibfnamefont
  {D.~M.}\ \bibnamefont {Stamper-Kurn}},\ }\href {\doibase
  10.1103/PhysRevLett.119.100402} {\bibfield  {journal} {\bibinfo  {journal}
  {Phys. Rev. Lett.}\ }\textbf {\bibinfo {volume} {119}},\ \bibinfo {pages}
  {100402} (\bibinfo {year} {2017})}\BibitemShut {NoStop}%
\bibitem [{\citenamefont {Jaksch}\ \emph {et~al.}(1998)\citenamefont {Jaksch},
  \citenamefont {Bruder}, \citenamefont {Cirac}, \citenamefont {Gardiner},\
  and\ \citenamefont {Zoller}}]{jaksch_bcgz1998}%
  \BibitemOpen
  \bibfield  {author} {\bibinfo {author} {\bibfnamefont {D.}~\bibnamefont
  {Jaksch}}, \bibinfo {author} {\bibfnamefont {C.}~\bibnamefont {Bruder}},
  \bibinfo {author} {\bibfnamefont {J.~I.}\ \bibnamefont {Cirac}}, \bibinfo
  {author} {\bibfnamefont {C.~W.}\ \bibnamefont {Gardiner}}, \ and\ \bibinfo
  {author} {\bibfnamefont {P.}~\bibnamefont {Zoller}},\ }\href {\doibase
  10.1103/PhysRevLett.81.3108} {\bibfield  {journal} {\bibinfo  {journal}
  {Phys. Rev. Lett.}\ }\textbf {\bibinfo {volume} {81}},\ \bibinfo {pages}
  {3108} (\bibinfo {year} {1998})}\BibitemShut {NoStop}%
\bibitem [{\citenamefont {Batrouni}\ \emph {et~al.}(2002)\citenamefont
  {Batrouni}, \citenamefont {Rousseau}, \citenamefont {Scalettar},
  \citenamefont {Rigol}, \citenamefont {Muramatsu}, \citenamefont {Denteneer},\
  and\ \citenamefont {Troyer}}]{batrouni_rsrmdt2002}%
  \BibitemOpen
  \bibfield  {author} {\bibinfo {author} {\bibfnamefont {G.~G.}\ \bibnamefont
  {Batrouni}}, \bibinfo {author} {\bibfnamefont {V.}~\bibnamefont {Rousseau}},
  \bibinfo {author} {\bibfnamefont {R.~T.}\ \bibnamefont {Scalettar}}, \bibinfo
  {author} {\bibfnamefont {M.}~\bibnamefont {Rigol}}, \bibinfo {author}
  {\bibfnamefont {A.}~\bibnamefont {Muramatsu}}, \bibinfo {author}
  {\bibfnamefont {P.~J.~H.}\ \bibnamefont {Denteneer}}, \ and\ \bibinfo
  {author} {\bibfnamefont {M.}~\bibnamefont {Troyer}},\ }\href {\doibase
  10.1103/PhysRevLett.89.117203} {\bibfield  {journal} {\bibinfo  {journal}
  {Phys. Rev. Lett.}\ }\textbf {\bibinfo {volume} {89}},\ \bibinfo {pages}
  {117203} (\bibinfo {year} {2002})}\BibitemShut {NoStop}%
\bibitem [{\citenamefont {Thonhauser}\ \emph {et~al.}(2007)\citenamefont
  {Thonhauser}, \citenamefont {Cooper}, \citenamefont {Li}, \citenamefont
  {Puzder}, \citenamefont {Hyldgaard},\ and\ \citenamefont
  {Langreth}}]{thonhauser2007}%
  \BibitemOpen
  \bibfield  {author} {\bibinfo {author} {\bibfnamefont {T.}~\bibnamefont
  {Thonhauser}}, \bibinfo {author} {\bibfnamefont {V.~R.}\ \bibnamefont
  {Cooper}}, \bibinfo {author} {\bibfnamefont {S.}~\bibnamefont {Li}}, \bibinfo
  {author} {\bibfnamefont {A.}~\bibnamefont {Puzder}}, \bibinfo {author}
  {\bibfnamefont {P.}~\bibnamefont {Hyldgaard}}, \ and\ \bibinfo {author}
  {\bibfnamefont {D.~C.}\ \bibnamefont {Langreth}},\ }\href {\doibase
  10.1103/PhysRevB.76.125112} {\bibfield  {journal} {\bibinfo  {journal} {Phys.
  Rev. B}\ }\textbf {\bibinfo {volume} {76}},\ \bibinfo {pages} {125112}
  (\bibinfo {year} {2007})}\BibitemShut {NoStop}%
\bibitem [{\citenamefont {Fetter}(2009)}]{fetter2009}%
  \BibitemOpen
  \bibfield  {author} {\bibinfo {author} {\bibfnamefont {A.~L.}\ \bibnamefont
  {Fetter}},\ }\href {\doibase 10.1103/RevModPhys.81.647} {\bibfield  {journal}
  {\bibinfo  {journal} {Rev. Mod. Phys.}\ }\textbf {\bibinfo {volume} {81}},\
  \bibinfo {pages} {647} (\bibinfo {year} {2009})}\BibitemShut {NoStop}%
\bibitem [{\citenamefont {Bhat}\ \emph {et~al.}(2007)\citenamefont {Bhat},
  \citenamefont {Kr\"amer}, \citenamefont {Cooper},\ and\ \citenamefont
  {Holland}}]{bhat_kch2007}%
  \BibitemOpen
  \bibfield  {author} {\bibinfo {author} {\bibfnamefont {R.}~\bibnamefont
  {Bhat}}, \bibinfo {author} {\bibfnamefont {M.}~\bibnamefont {Kr\"amer}},
  \bibinfo {author} {\bibfnamefont {J.}~\bibnamefont {Cooper}}, \ and\ \bibinfo
  {author} {\bibfnamefont {M.~J.}\ \bibnamefont {Holland}},\ }\href {\doibase
  10.1103/PhysRevA.76.043601} {\bibfield  {journal} {\bibinfo  {journal} {Phys.
  Rev. A}\ }\textbf {\bibinfo {volume} {76}},\ \bibinfo {pages} {043601}
  (\bibinfo {year} {2007})}\BibitemShut {NoStop}%
\bibitem [{\citenamefont {Mithun}\ \emph {et~al.}(2014)\citenamefont {Mithun},
  \citenamefont {Porsezian},\ and\ \citenamefont {Dey}}]{dey2014}%
  \BibitemOpen
  \bibfield  {author} {\bibinfo {author} {\bibfnamefont {T.}~\bibnamefont
  {Mithun}}, \bibinfo {author} {\bibfnamefont {K.}~\bibnamefont {Porsezian}}, \
  and\ \bibinfo {author} {\bibfnamefont {B.}~\bibnamefont {Dey}},\ }\href
  {\doibase 10.1103/PhysRevA.89.053625} {\bibfield  {journal} {\bibinfo
  {journal} {Phys. Rev. A}\ }\textbf {\bibinfo {volume} {89}},\ \bibinfo
  {pages} {053625} (\bibinfo {year} {2014})}\BibitemShut {NoStop}%
\bibitem [{\citenamefont {Mithun}\ \emph {et~al.}(2016)\citenamefont {Mithun},
  \citenamefont {Porsezian},\ and\ \citenamefont {Dey}}]{mithun2016}%
  \BibitemOpen
  \bibfield  {author} {\bibinfo {author} {\bibfnamefont {T.}~\bibnamefont
  {Mithun}}, \bibinfo {author} {\bibfnamefont {K.}~\bibnamefont {Porsezian}}, \
  and\ \bibinfo {author} {\bibfnamefont {B.}~\bibnamefont {Dey}},\ }\href
  {\doibase 10.1103/PhysRevA.93.013620} {\bibfield  {journal} {\bibinfo
  {journal} {Phys. Rev. A}\ }\textbf {\bibinfo {volume} {93}},\ \bibinfo
  {pages} {013620} (\bibinfo {year} {2016})}\BibitemShut {NoStop}%
\bibitem [{\citenamefont {Mithun}\ \emph {et~al.}(2018)\citenamefont {Mithun},
  \citenamefont {Ganguli}, \citenamefont {Raychaudhuri},\ and\ \citenamefont
  {Dey}}]{mithun2018}%
  \BibitemOpen
  \bibfield  {author} {\bibinfo {author} {\bibfnamefont {T.}~\bibnamefont
  {Mithun}}, \bibinfo {author} {\bibfnamefont {S.~C.}\ \bibnamefont {Ganguli}},
  \bibinfo {author} {\bibfnamefont {P.}~\bibnamefont {Raychaudhuri}}, \ and\
  \bibinfo {author} {\bibfnamefont {B.}~\bibnamefont {Dey}},\ }\href {\doibase
  10.1209/0295-5075/123/20004} {\bibfield  {journal} {\bibinfo  {journal}
  {{EPL} (Europhysics Letters)}\ }\textbf {\bibinfo {volume} {123}},\ \bibinfo
  {pages} {20004} (\bibinfo {year} {2018})}\BibitemShut {NoStop}%
\bibitem [{\citenamefont {Pastoriza}\ \emph {et~al.}(1994)\citenamefont
  {Pastoriza}, \citenamefont {Goffman}, \citenamefont {Arrib\'ere},\ and\
  \citenamefont {de~la Cruz}}]{pastoriza1994}%
  \BibitemOpen
  \bibfield  {author} {\bibinfo {author} {\bibfnamefont {H.}~\bibnamefont
  {Pastoriza}}, \bibinfo {author} {\bibfnamefont {M.~F.}\ \bibnamefont
  {Goffman}}, \bibinfo {author} {\bibfnamefont {A.}~\bibnamefont {Arrib\'ere}},
  \ and\ \bibinfo {author} {\bibfnamefont {F.}~\bibnamefont {de~la Cruz}},\
  }\href {\doibase 10.1103/PhysRevLett.72.2951} {\bibfield  {journal} {\bibinfo
   {journal} {Phys. Rev. Lett.}\ }\textbf {\bibinfo {volume} {72}},\ \bibinfo
  {pages} {2951} (\bibinfo {year} {1994})}\BibitemShut {NoStop}%
\bibitem [{\citenamefont {Paltiel}\ \emph {et~al.}(2000)\citenamefont
  {Paltiel}, \citenamefont {Zeldov}, \citenamefont {Myasoedov}, \citenamefont
  {Rappaport}, \citenamefont {Jung}, \citenamefont {Bhattacharya},
  \citenamefont {Higgins}, \citenamefont {Xiao}, \citenamefont {Andrei},
  \citenamefont {Gammel},\ and\ \citenamefont {Bishop}}]{paltiel2000}%
  \BibitemOpen
  \bibfield  {author} {\bibinfo {author} {\bibfnamefont {Y.}~\bibnamefont
  {Paltiel}}, \bibinfo {author} {\bibfnamefont {E.}~\bibnamefont {Zeldov}},
  \bibinfo {author} {\bibfnamefont {Y.}~\bibnamefont {Myasoedov}}, \bibinfo
  {author} {\bibfnamefont {M.~L.}\ \bibnamefont {Rappaport}}, \bibinfo {author}
  {\bibfnamefont {G.}~\bibnamefont {Jung}}, \bibinfo {author} {\bibfnamefont
  {S.}~\bibnamefont {Bhattacharya}}, \bibinfo {author} {\bibfnamefont {M.~J.}\
  \bibnamefont {Higgins}}, \bibinfo {author} {\bibfnamefont {Z.~L.}\
  \bibnamefont {Xiao}}, \bibinfo {author} {\bibfnamefont {E.~Y.}\ \bibnamefont
  {Andrei}}, \bibinfo {author} {\bibfnamefont {P.~L.}\ \bibnamefont {Gammel}},
  \ and\ \bibinfo {author} {\bibfnamefont {D.~J.}\ \bibnamefont {Bishop}},\
  }\href {\doibase 10.1103/PhysRevLett.85.3712} {\bibfield  {journal} {\bibinfo
   {journal} {Phys. Rev. Lett.}\ }\textbf {\bibinfo {volume} {85}},\ \bibinfo
  {pages} {3712} (\bibinfo {year} {2000})}\BibitemShut {NoStop}%
\bibitem [{\citenamefont {Kierfeld}\ and\ \citenamefont
  {Vinokur}(2004)}]{lindemann2004}%
  \BibitemOpen
  \bibfield  {author} {\bibinfo {author} {\bibfnamefont {J.}~\bibnamefont
  {Kierfeld}}\ and\ \bibinfo {author} {\bibfnamefont {V.}~\bibnamefont
  {Vinokur}},\ }\href {\doibase 10.1103/PhysRevB.69.024501} {\bibfield
  {journal} {\bibinfo  {journal} {Phys. Rev. B}\ }\textbf {\bibinfo {volume}
  {69}},\ \bibinfo {pages} {024501} (\bibinfo {year} {2004})}\BibitemShut
  {NoStop}%
\bibitem [{\citenamefont {Kierfeld}\ and\ \citenamefont
  {Vinokur}(2000)}]{kierfeld2000}%
  \BibitemOpen
  \bibfield  {author} {\bibinfo {author} {\bibfnamefont {J.}~\bibnamefont
  {Kierfeld}}\ and\ \bibinfo {author} {\bibfnamefont {V.}~\bibnamefont
  {Vinokur}},\ }\href {\doibase 10.1103/PhysRevB.61.R14928} {\bibfield
  {journal} {\bibinfo  {journal} {Phys. Rev. B}\ }\textbf {\bibinfo {volume}
  {61}},\ \bibinfo {pages} {R14928} (\bibinfo {year} {2000})}\BibitemShut
  {NoStop}%
\bibitem [{\citenamefont {Ganguli}\ \emph {et~al.}(2016)\citenamefont
  {Ganguli}, \citenamefont {Singh}, \citenamefont {Roy}, \citenamefont {Bagwe},
  \citenamefont {Bala}, \citenamefont {Thamizhavel},\ and\ \citenamefont
  {Raychaudhuri}}]{ganguli2016}%
  \BibitemOpen
  \bibfield  {author} {\bibinfo {author} {\bibfnamefont {S.~C.}\ \bibnamefont
  {Ganguli}}, \bibinfo {author} {\bibfnamefont {H.}~\bibnamefont {Singh}},
  \bibinfo {author} {\bibfnamefont {I.}~\bibnamefont {Roy}}, \bibinfo {author}
  {\bibfnamefont {V.}~\bibnamefont {Bagwe}}, \bibinfo {author} {\bibfnamefont
  {D.}~\bibnamefont {Bala}}, \bibinfo {author} {\bibfnamefont {A.}~\bibnamefont
  {Thamizhavel}}, \ and\ \bibinfo {author} {\bibfnamefont {P.}~\bibnamefont
  {Raychaudhuri}},\ }\href {\doibase 10.1103/PhysRevB.93.144503} {\bibfield
  {journal} {\bibinfo  {journal} {Phys. Rev. B}\ }\textbf {\bibinfo {volume}
  {93}},\ \bibinfo {pages} {144503} (\bibinfo {year} {2016})}\BibitemShut
  {NoStop}%
\bibitem [{\citenamefont {Ganguli}\ \emph {et~al.}(2015)\citenamefont
  {Ganguli}, \citenamefont {Singh}, \citenamefont {Saraswat}, \citenamefont
  {Ganguly}, \citenamefont {Bagwe}, \citenamefont {Shirage}, \citenamefont
  {Thamizhavel},\ and\ \citenamefont {Raychaudhuri}}]{ganguli2015}%
  \BibitemOpen
  \bibfield  {author} {\bibinfo {author} {\bibfnamefont {S.~C.}\ \bibnamefont
  {Ganguli}}, \bibinfo {author} {\bibfnamefont {H.}~\bibnamefont {Singh}},
  \bibinfo {author} {\bibfnamefont {G.}~\bibnamefont {Saraswat}}, \bibinfo
  {author} {\bibfnamefont {R.}~\bibnamefont {Ganguly}}, \bibinfo {author}
  {\bibfnamefont {V.}~\bibnamefont {Bagwe}}, \bibinfo {author} {\bibfnamefont
  {P.}~\bibnamefont {Shirage}}, \bibinfo {author} {\bibfnamefont
  {A.}~\bibnamefont {Thamizhavel}}, \ and\ \bibinfo {author} {\bibfnamefont
  {P.}~\bibnamefont {Raychaudhuri}},\ }\href {\doibase 10.1038/srep10613}
  {\bibfield  {journal} {\bibinfo  {journal} {Scientific Reports}\ }\textbf
  {\bibinfo {volume} {5}} (\bibinfo {year} {2015}),\
  10.1038/srep10613}\BibitemShut {NoStop}%
\bibitem [{\citenamefont {Wu}\ \emph {et~al.}(2004)\citenamefont {Wu},
  \citenamefont {Chen}, \citenamefont {Hu},\ and\ \citenamefont
  {Zhang}}]{wu_chjzs2004}%
  \BibitemOpen
  \bibfield  {author} {\bibinfo {author} {\bibfnamefont {C.}~\bibnamefont
  {Wu}}, \bibinfo {author} {\bibfnamefont {H.-d.}\ \bibnamefont {Chen}},
  \bibinfo {author} {\bibfnamefont {J.-p.}\ \bibnamefont {Hu}}, \ and\ \bibinfo
  {author} {\bibfnamefont {S.-C.}\ \bibnamefont {Zhang}},\ }\href {\doibase
  10.1103/PhysRevA.69.043609} {\bibfield  {journal} {\bibinfo  {journal} {Phys.
  Rev. A}\ }\textbf {\bibinfo {volume} {69}},\ \bibinfo {pages} {043609}
  (\bibinfo {year} {2004})}\BibitemShut {NoStop}%
\bibitem [{\citenamefont {Bhat}\ \emph
  {et~al.}(2006{\natexlab{a}})\citenamefont {Bhat}, \citenamefont {Holland},\
  and\ \citenamefont {Carr}}]{bhat_hc2006}%
  \BibitemOpen
  \bibfield  {author} {\bibinfo {author} {\bibfnamefont {R.}~\bibnamefont
  {Bhat}}, \bibinfo {author} {\bibfnamefont {M.~J.}\ \bibnamefont {Holland}}, \
  and\ \bibinfo {author} {\bibfnamefont {L.~D.}\ \bibnamefont {Carr}},\ }\href
  {\doibase 10.1103/PhysRevLett.96.060405} {\bibfield  {journal} {\bibinfo
  {journal} {Phys. Rev. Lett.}\ }\textbf {\bibinfo {volume} {96}},\ \bibinfo
  {pages} {060405} (\bibinfo {year} {2006}{\natexlab{a}})}\BibitemShut
  {NoStop}%
\bibitem [{\citenamefont {Bhat}\ \emph
  {et~al.}(2006{\natexlab{b}})\citenamefont {Bhat}, \citenamefont {Peden},
  \citenamefont {Seaman}, \citenamefont {Kr\"amer}, \citenamefont {Carr},\ and\
  \citenamefont {Holland}}]{bhat_ps2006}%
  \BibitemOpen
  \bibfield  {author} {\bibinfo {author} {\bibfnamefont {R.}~\bibnamefont
  {Bhat}}, \bibinfo {author} {\bibfnamefont {B.~M.}\ \bibnamefont {Peden}},
  \bibinfo {author} {\bibfnamefont {B.~T.}\ \bibnamefont {Seaman}}, \bibinfo
  {author} {\bibfnamefont {M.}~\bibnamefont {Kr\"amer}}, \bibinfo {author}
  {\bibfnamefont {L.~D.}\ \bibnamefont {Carr}}, \ and\ \bibinfo {author}
  {\bibfnamefont {M.~J.}\ \bibnamefont {Holland}},\ }\href {\doibase
  10.1103/PhysRevA.74.063606} {\bibfield  {journal} {\bibinfo  {journal} {Phys.
  Rev. A}\ }\textbf {\bibinfo {volume} {74}},\ \bibinfo {pages} {063606}
  (\bibinfo {year} {2006}{\natexlab{b}})}\BibitemShut {NoStop}%
\bibitem [{\citenamefont {Peden}\ \emph {et~al.}(2007)\citenamefont {Peden},
  \citenamefont {Bhat}, \citenamefont {Krämer},\ and\ \citenamefont
  {Holland}}]{peden2007}%
  \BibitemOpen
  \bibfield  {author} {\bibinfo {author} {\bibfnamefont {B.~M.}\ \bibnamefont
  {Peden}}, \bibinfo {author} {\bibfnamefont {R.}~\bibnamefont {Bhat}},
  \bibinfo {author} {\bibfnamefont {M.}~\bibnamefont {Krämer}}, \ and\
  \bibinfo {author} {\bibfnamefont {M.~J.}\ \bibnamefont {Holland}},\ }\href
  {\doibase 10.1088/0953-4075/40/18/012} {\bibfield  {journal} {\bibinfo
  {journal} {Journal of Physics B: Atomic, Molecular and Optical Physics}\
  }\textbf {\bibinfo {volume} {40}},\ \bibinfo {pages} {3725} (\bibinfo {year}
  {2007})}\BibitemShut {NoStop}%
\bibitem [{\citenamefont {Goldbaum}\ and\ \citenamefont
  {Mueller}(2008)}]{goldbaum2008}%
  \BibitemOpen
  \bibfield  {author} {\bibinfo {author} {\bibfnamefont {D.~S.}\ \bibnamefont
  {Goldbaum}}\ and\ \bibinfo {author} {\bibfnamefont {E.~J.}\ \bibnamefont
  {Mueller}},\ }\href {\doibase 10.1103/PhysRevA.77.033629} {\bibfield
  {journal} {\bibinfo  {journal} {Phys. Rev. A}\ }\textbf {\bibinfo {volume}
  {77}},\ \bibinfo {pages} {033629} (\bibinfo {year} {2008})}\BibitemShut
  {NoStop}%
\bibitem [{\citenamefont {Vignolo}\ \emph {et~al.}(2007)\citenamefont
  {Vignolo}, \citenamefont {Fazio},\ and\ \citenamefont {Tosi}}]{vignolo2007}%
  \BibitemOpen
  \bibfield  {author} {\bibinfo {author} {\bibfnamefont {P.}~\bibnamefont
  {Vignolo}}, \bibinfo {author} {\bibfnamefont {R.}~\bibnamefont {Fazio}}, \
  and\ \bibinfo {author} {\bibfnamefont {M.~P.}\ \bibnamefont {Tosi}},\ }\href
  {\doibase 10.1103/PhysRevA.76.023616} {\bibfield  {journal} {\bibinfo
  {journal} {Phys. Rev. A}\ }\textbf {\bibinfo {volume} {76}},\ \bibinfo
  {pages} {023616} (\bibinfo {year} {2007})}\BibitemShut {NoStop}%
\bibitem [{\citenamefont {Penrose}\ and\ \citenamefont
  {Onsager}(1956)}]{penrose1956}%
  \BibitemOpen
  \bibfield  {author} {\bibinfo {author} {\bibfnamefont {O.}~\bibnamefont
  {Penrose}}\ and\ \bibinfo {author} {\bibfnamefont {L.}~\bibnamefont
  {Onsager}},\ }\href {\doibase 10.1103/PhysRev.104.576} {\bibfield  {journal}
  {\bibinfo  {journal} {Phys. Rev.}\ }\textbf {\bibinfo {volume} {104}},\
  \bibinfo {pages} {576} (\bibinfo {year} {1956})}\BibitemShut {NoStop}%
\end{thebibliography}%
\end{document}